%% file: BPH-23-004_temp.tex
\documentclass[11pt,twoside,a4paper,cmspaper,final,collab]{cms-tdr}
\def\svnVersion{9370cfb}\def\svnDate{2026/05/07}\def\cmsCernNoTag{CERN-EP-2024-300}\def\cmsCernDate{\today}\def\cmsMessage{Published in Physical Review Letters as \href{http://dx.doi.org/10.1103/ctqx-vxrj}{\doi{10.1103/ctqx-vxrj}.}}
\begin{document}\cmsNoteHeader{BPH-23-004}

\newcommand{\setJpsiMu}{MT\xspace}
\newcommand{\setJpsiTrk}{ST\xspace}
\newcommand{\refCMSrunIIpartial}{\cite{CMS:2021136188}\xspace}
\newcommand{\refCMSrunI}{\cite{ref:cmsKhachatryan16plb}\xspace}
\newcommand{\refPDG}{\cite{ref:PDG}\xspace}
\newcommand{\refPhisSM}{\cite{ref:Charles11prd, ref:UTfit22}\xspace}
\newcommand{\sqabs}[1]{\abs{#1}^2}
\newcommand{\tenpow}[1]{10^{#1}}
\newcommand{\ltx}[1]{\ensuremath{<\!\tenpow{-#1}}}
\newcommand{\s}{\ensuremath{\mathrm{\PQs}}\xspace}
\newcommand{\Scap}{\ensuremath{\mathrm{S}}\xspace}
\newcommand{\f}{\ensuremath{\mathrm{f}}\xspace}
\newcommand{\xy}{\ensuremath{\mathrm{xy}}\xspace}
\newcommand{\phis}{\ensuremath{\phi_\s}\xspace}
\newcommand{\DGs}{\ensuremath{\Delta\Gamma_\s}\xspace}
\newcommand{\Gs}{\ensuremath{\Gamma_\s}\xspace}
\newcommand{\abslambda}{\ensuremath{\abs{\lambda}}\xspace}
\newcommand{\Dms}{\ensuremath{\Delta m_\s}\xspace}
\newcommand{\Dmd}{\ensuremath{\Delta m_{\PQd}}\xspace}
\newcommand{\Azero}{\ensuremath{A_0}\xspace}
\newcommand{\Aperp}{\ensuremath{A_{\perp}}\xspace}
\newcommand{\Apara}{\ensuremath{A_{\parallel}}\xspace}
\newcommand{\Aswav}{\ensuremath{A_\Scap}\xspace}
\newcommand{\dzero}{\ensuremath{\delta_0}\xspace}
\newcommand{\dperp}{\ensuremath{\delta_{\perp}}\xspace}
\newcommand{\dpara}{\ensuremath{\delta_{\parallel}}\xspace}
\newcommand{\dswav}{\ensuremath{\delta_\Scap}\xspace}
\newcommand{\dswpd}{\ensuremath{\delta_{\Scap\perp}}\xspace}
\newcommand{\sqAzero}{\ensuremath{\sqabs{\Azero}}\xspace}
\newcommand{\sqAperp}{\ensuremath{\sqabs{\Aperp}}\xspace}
\newcommand{\sqApara}{\ensuremath{\sqabs{\Apara}}\xspace}
\newcommand{\sqAswav}{\ensuremath{\sqabs{\Aswav}}\xspace}
\newcommand{\ct}{\ensuremath{ct}\xspace}
\newcommand{\sct}{\ensuremath{\sigma_{\ct}}\xspace}
\newcommand{\mBs}{\ensuremath{m_{\PGm\PGm KK}}\xspace}
\newcommand{\thetaT}{\ensuremath{\theta_{T}}\xspace}
\newcommand{\psiT}{\ensuremath{\psi_{T}}\xspace}
\newcommand{\phiT}{\ensuremath{\varphi_{T}}\xspace}
\newcommand{\cthetaT}{\ensuremath{\cos\thetaT}\xspace}
\newcommand{\cpsiT}{\ensuremath{\cos\psiT}\xspace}
\newcommand{\mrad}{\unit{mrad}}
\newcommand{\ips}{\ensuremath{\unit{ps}^{-1}}\xspace}
\newcommand{\hips}{\ensuremath{\hslash\ips}\xspace}
\newcommand{\Bs}{\ensuremath{\PBzs}\xspace}
\newcommand{\Bsbar}{\ensuremath{\PABzs}\xspace}
\newcommand{\Bu}{\ensuremath{\PBp}\xspace}
\newcommand{\Bm}{\ensuremath{\PBm}\xspace}
\newcommand{\Bd}{\ensuremath{\PBz}\xspace}
\newcommand{\jpsi}{\ensuremath{\PJGy}\xspace}
\newcommand{\kstar}{{\HepParticleResonanceFull{\PK}{}{*}{892}{}{0}}\xspace}
\newcommand{\mumu}{\ensuremath{\PGmp\PGmm}\xspace}
\newcommand{\KK}{\ensuremath{\PKp\PKm}\xspace}
\newcommand{\mumuKK}{\ensuremath{\mumu\,\KK}\xspace}
\newcommand{\phimes}{\ensuremath{\PGf(1020)}\xspace}
\newcommand{\jpsiphi}{\ensuremath{\jpsi\,\phimes}\xspace}
\newcommand{\jpsiphialt}{\ensuremath{\jpsi\,\PGf(1020)}\xspace}
\newcommand{\jpsimumu}{\ensuremath{\jpsi \to \mumu}\xspace}
\newcommand{\phiKK}{\ensuremath{\phimes \to \KK}\xspace}
\newcommand{\Bsjpsiphi}{\ensuremath{\Bs \to \jpsiphi }\xspace}
\newcommand{\BsjpsiKK}{\ensuremath{\Bs \to \jpsi\,\KK }\xspace}
\newcommand{\Bsjpsiphialt}{\ensuremath{\Bs \to \jpsiphialt }\xspace}
\newcommand{\BsjpsiphiMMKK}{\ensuremath{\Bs \to \jpsiphi \to \mumuKK}\xspace}
\newcommand{\BujpsiK}{\ensuremath{\Bu \to \jpsi\,\PKp}\xspace}
\newcommand{\Bdjpsikstar}{\ensuremath{\Bd \to \jpsi\,\kstar}\xspace}
\newcommand{\BdjpsikstarMMKP}{\ensuremath{\Bd \to \jpsi\,\kstar \to \mumu\,\PKp\Pgpm}\xspace}
\newcommand{\Lambdab}{\ensuremath{\PGLzb}\xspace}
\newcommand{\LambdabjpsiKpMMKp}{\ensuremath{\Lambdab \to \jpsi\,\Pp\PKm \to \mumu\,\Pp\PKm}\xspace}
\newcommand{\pp}{\ensuremath{\Pp\Pp}\xspace}
\newcommand{\BsBsbar}{\ensuremath{\Bs\text{--}\Bsbar}\xspace}
\newcommand{\mKK}{\ensuremath{m(\KK)}\xspace}
\newcommand{\betas}{\ensuremath{\beta_\s}\xspace}
\ifthenelse{\boolean{cms@external}}{\newcommand{\CP}{\ensuremath{C\mkern-3mu P}\xspace}}{\newcommand{\CP}{\ensuremath{CP}\xspace}}
\newcommand{\xitag}{\ensuremath{\xi_{\text{tag}}}\xspace}
\newcommand{\etag}{\ensuremath{\varepsilon_{\text{tag}}}\xspace}
\newcommand{\wtag}{\ensuremath{\omega_{\text{tag}}}\xspace}
\newcommand{\Ptag}{\ensuremath{P_{\text{tag}}}\xspace}
\newcommand{\Dtag}{\ensuremath{{\mathcal{D}_{\text{tag}}}}\xspace}
\newcommand{\sqDtag}{\ensuremath{{\mathcal{D}^2_{\text{tag}}}}\xspace}
\newcommand{\wmeas}{\ensuremath{\omega_{\text{meas}}}\xspace}
\newcommand{\ect}{\ensuremath{\varepsilon(\ct)}\xspace}
\newcommand{\NBsPtag}{\ensuremath{N_{\!\Bs} \Ptag}\xspace}
\newcommand{\Deltasyst}{\ensuremath{\Delta_\text{syst}}\xspace}
\newcommand{\tauBd}{\ensuremath{\tau^\text{PDG}_\Bd}\xspace}
\newcommand{\dR}{\ensuremath{\Delta R}\xspace}
\newcommand{\IPz}{\ensuremath{d_{\mathrm{z}}}\xspace}
\newcommand{\ksp}{\ensuremath{k_{\mathrm{SP}}}\xspace}
\newcommand{\cmeRunI}{\ensuremath{\sqrt{s} = 8\TeV}\xspace}
\newcommand{\cmeRunII}{\ensuremath{\sqrt{s} = 13\TeV}\xspace}
\newcommand{\phisRes}{\ensuremath{-73}\xspace}
\newcommand{\DGsRes}{\ensuremath{0.0761}\xspace}
\newcommand{\GsRes}{\ensuremath{0.6613}\xspace}
\newcommand{\DmsRes}{\ensuremath{17.757}\xspace}
\newcommand{\abslambdaRes}{\ensuremath{1.011}\xspace}
\newcommand{\sqAzeroRes}{\ensuremath{0.5300}\xspace}
\newcommand{\sqAperpRes}{\ensuremath{0.2409}\xspace}
\newcommand{\sqAswavRes}{\ensuremath{0.0067}\xspace}
\newcommand{\dparaRes}{\ensuremath{3.145}\xspace}
\newcommand{\dperpRes}{\ensuremath{2.931}\xspace}
\newcommand{\dswpdRes}{\ensuremath{0.48}\xspace}
\newcommand{\phisErrSym}{\ensuremath{22}\xspace}
\newcommand{\DGsErrSym}{\ensuremath{0.0040}\xspace}
\newcommand{\GsErrSym}{\ensuremath{0.0015}\xspace}
\newcommand{\DmsErrSym}{\ensuremath{0.040}\xspace}
\newcommand{\abslambdaErrSym}{\ensuremath{0.014}\xspace}
\newcommand{\sqAzeroErrSym}{\ensuremath{0.0015}\xspace}
\newcommand{\sqAperpErrSym}{\ensuremath{0.0020}\xspace}
\newcommand{\sqAswavErrSym}{\ensuremath{0.0034}\xspace}
\newcommand{\dparaErrSym}{\ensuremath{0.077}\xspace}
\newcommand{\dperpErrSym}{\ensuremath{0.096}\xspace}
\newcommand{\dswpdErrSym}{\ensuremath{0.16}\xspace}
\newcommand{\phisErr}{\ensuremath{\pm\ \phisErrSym}\xspace}
\newcommand{\DGsErr}{\ensuremath{\pm\ \DGsErrSym}\xspace}
\newcommand{\GsErr}{\ensuremath{\pm\ \GsErrSym}\xspace}
\newcommand{\DmsErr}{\ensuremath{\pm\ \DmsErrSym}\xspace}
\newcommand{\abslambdaErr}{\ensuremath{\pm\ \abslambdaErrSym}\xspace}
\newcommand{\sqAzeroErr}{\ensuremath{\pm\ \sqAzeroErrSym}\xspace}
\newcommand{\sqAperpErr}{\ensuremath{\pm\ \sqAperpErrSym}\xspace}
\newcommand{\sqAswavErr}{\ensuremath{^{+\hspace{0.5em} 0.0034}_{-\hspace{0.5em} 0.0027}}\xspace}
\newcommand{\dparaErr}{\ensuremath{\pm\ \dparaErrSym}\xspace}
\newcommand{\dperpErr}{\ensuremath{\pm\ \dperpErrSym}\xspace}
\newcommand{\dswpdErr}{\ensuremath{^{+\hspace{0.5em} 0.11}_{-\hspace{0.5em} 0.16}}\xspace}
\newcommand{\phisSystBias}{\ensuremath{4}\xspace}
\newcommand{\DGsSystBias}{\ensuremath{0.0011}\xspace}
\newcommand{\GsSystBias}{\ensuremath{0.0002}\xspace}
\newcommand{\DmsSystBias}{\ensuremath{0.004}\xspace}
\newcommand{\abslambdaSystBias}{\ensuremath{0.006}\xspace}
\newcommand{\sqAzeroSystBias}{\ensuremath{0.0012}\xspace}
\newcommand{\sqAperpSystBias}{\ensuremath{0.0022}\xspace}
\newcommand{\sqAswavSystBias}{\ensuremath{0.0006}\xspace}
\newcommand{\dparaSystBias}{\ensuremath{0.015}\xspace}
\newcommand{\dperpSystBias}{\ensuremath{0.017}\xspace}
\newcommand{\dswpdSystBias}{\ensuremath{0.03}\xspace}
\newcommand{\phisSystModel}{\NA\xspace}
\newcommand{\DGsSystModel}{\ensuremath{0.0015}\xspace}
\newcommand{\GsSystModel}{\ensuremath{0.0018}\xspace}
\newcommand{\DmsSystModel}{\NA\xspace}
\newcommand{\abslambdaSystModel}{\NA\xspace}
\newcommand{\sqAzeroSystModel}{\ensuremath{0.0009}\xspace}
\newcommand{\sqAperpSystModel}{\ensuremath{0.0009}\xspace}
\newcommand{\sqAswavSystModel}{\NA\xspace}
\newcommand{\dparaSystModel}{\NA\xspace}
\newcommand{\dperpSystModel}{\NA\xspace}
\newcommand{\dswpdSystModel}{\NA\xspace}
\newcommand{\phisSystBd}{\ensuremath{<\!1}\xspace}
\newcommand{\DGsSystBd}{\ensuremath{0.0002}\xspace}
\newcommand{\GsSystBd}{\ensuremath{0.0006}\xspace}
\newcommand{\DmsSystBd}{\ensuremath{\ltx{3}}\xspace}
\newcommand{\abslambdaSystBd}{\ensuremath{\ltx{3}}\xspace}
\newcommand{\sqAzeroSystBd}{\ensuremath{\ltx{4}}\xspace}
\newcommand{\sqAperpSystBd}{\ensuremath{\ltx{4}}\xspace}
\newcommand{\sqAswavSystBd}{\ensuremath{\ltx{4}}\xspace}
\newcommand{\dparaSystBd}{\ensuremath{\ltx{3}}\xspace}
\newcommand{\dperpSystBd}{\ensuremath{\ltx{3}}\xspace}
\newcommand{\dswpdSystBd}{\ensuremath{\ltx{2}}\xspace}
\newcommand{\phisSystLb}{\NA\xspace}
\newcommand{\DGsSystLb}{\NA\xspace}
\newcommand{\GsSystLb}{\ensuremath{0.0008}\xspace}
\newcommand{\DmsSystLb}{\NA\xspace}
\newcommand{\abslambdaSystLb}{\NA\xspace}
\newcommand{\sqAzeroSystLb}{\ensuremath{0.0008}\xspace}
\newcommand{\sqAperpSystLb}{\ensuremath{0.0006}\xspace}
\newcommand{\sqAswavSystLb}{\NA\xspace}
\newcommand{\dparaSystLb}{\NA\xspace}
\newcommand{\dperpSystLb}{\NA\xspace}
\newcommand{\dswpdSystLb}{\NA\xspace}
\newcommand{\phisSystKsp}{\ensuremath{<\!1}\xspace}
\newcommand{\DGsSystKsp}{\ensuremath{\ltx{4}}\xspace}
\newcommand{\GsSystKsp}{\ensuremath{\ltx{4}}\xspace}
\newcommand{\DmsSystKsp}{\ensuremath{\ltx{3}}\xspace}
\newcommand{\abslambdaSystKsp}{\ensuremath{\ltx{3}}\xspace}
\newcommand{\sqAzeroSystKsp}{\ensuremath{\ltx{4}}\xspace}
\newcommand{\sqAperpSystKsp}{\ensuremath{\ltx{4}}\xspace}
\newcommand{\sqAswavSystKsp}{\ensuremath{\ltx{4}}\xspace}
\newcommand{\dparaSystKsp}{\ensuremath{\ltx{3}}\xspace}
\newcommand{\dperpSystKsp}{\ensuremath{\ltx{3}}\xspace}
\newcommand{\dswpdSystKsp}{\ensuremath{\ltx{2}}\xspace}
\newcommand{\phisSystAng}{\ensuremath{4}\xspace}
\newcommand{\DGsSystAng}{\ensuremath{0.0002}\xspace}
\newcommand{\GsSystAng}{\ensuremath{\ltx{4}}\xspace}
\newcommand{\DmsSystAng}{\ensuremath{0.017}\xspace}
\newcommand{\abslambdaSystAng}{\ensuremath{0.013}\xspace}
\newcommand{\sqAzeroSystAng}{\ensuremath{0.0084}\xspace}
\newcommand{\sqAperpSystAng}{\ensuremath{0.0039}\xspace}
\newcommand{\sqAswavSystAng}{\ensuremath{0.0001}\xspace}
\newcommand{\dparaSystAng}{\ensuremath{0.017}\xspace}
\newcommand{\dperpSystAng}{\ensuremath{0.051}\xspace}
\newcommand{\dswpdSystAng}{\ensuremath{0.02}\xspace}
\newcommand{\phisSystCt}{\ensuremath{<\!1}\xspace}
\newcommand{\DGsSystCt}{\ensuremath{0.0014}\xspace}
\newcommand{\GsSystCt}{\ensuremath{0.0026}\xspace}
\newcommand{\DmsSystCt}{\ensuremath{\ltx{3}}\xspace}
\newcommand{\abslambdaSystCt}{\ensuremath{\ltx{3}}\xspace}
\newcommand{\sqAzeroSystCt}{\ensuremath{0.0004}\xspace}
\newcommand{\sqAperpSystCt}{\ensuremath{0.0005}\xspace}
\newcommand{\sqAswavSystCt}{\ensuremath{\ltx{4}}\xspace}
\newcommand{\dparaSystCt}{\ensuremath{0.001}\xspace}
\newcommand{\dperpSystCt}{\ensuremath{0.002}\xspace}
\newcommand{\dswpdSystCt}{\ensuremath{\ltx{2}}\xspace}
\newcommand{\phisSystReso}{\ensuremath{<\!1}\xspace}
\newcommand{\DGsSystReso}{\ensuremath{\ltx{4}}\xspace}
\newcommand{\GsSystReso}{\ensuremath{\ltx{4}}\xspace}
\newcommand{\DmsSystReso}{\ensuremath{\ltx{3}}\xspace}
\newcommand{\abslambdaSystReso}{\ensuremath{\ltx{3}}\xspace}
\newcommand{\sqAzeroSystReso}{\ensuremath{\ltx{4}}\xspace}
\newcommand{\sqAperpSystReso}{\ensuremath{\ltx{4}}\xspace}
\newcommand{\sqAswavSystReso}{\ensuremath{\ltx{4}}\xspace}
\newcommand{\dparaSystReso}{\ensuremath{\ltx{3}}\xspace}
\newcommand{\dperpSystReso}{\ensuremath{0.001}\xspace}
\newcommand{\dswpdSystReso}{\ensuremath{\ltx{3}}\xspace}
\newcommand{\phisSystTag}{\ensuremath{8}\xspace}
\newcommand{\DGsSystTag}{\ensuremath{\ltx{4}}\xspace}
\newcommand{\GsSystTag}{\ensuremath{0.0010}\xspace}
\newcommand{\DmsSystTag}{\ensuremath{0.015}\xspace}
\newcommand{\abslambdaSystTag}{\ensuremath{0.003}\xspace}
\newcommand{\sqAzeroSystTag}{\ensuremath{\ltx{4}}\xspace}
\newcommand{\sqAperpSystTag}{\ensuremath{\ltx{4}}\xspace}
\newcommand{\sqAswavSystTag}{\ensuremath{0.0012}\xspace}
\newcommand{\dparaSystTag}{\ensuremath{0.024}\xspace}
\newcommand{\dperpSystTag}{\ensuremath{0.033}\xspace}
\newcommand{\dswpdSystTag}{\ensuremath{0.06}\xspace}
\newcommand{\phisSystFix}{\ensuremath{<\!1}\xspace}
\newcommand{\DGsSystFix}{\ensuremath{0.0002}\xspace}
\newcommand{\GsSystFix}{\ensuremath{0.0003}\xspace}
\newcommand{\DmsSystFix}{\ensuremath{\ltx{3}}\xspace}
\newcommand{\abslambdaSystFix}{\ensuremath{\ltx{3}}\xspace}
\newcommand{\sqAzeroSystFix}{\ensuremath{0.0001}\xspace}
\newcommand{\sqAperpSystFix}{\ensuremath{0.0001}\xspace}
\newcommand{\sqAswavSystFix}{\ensuremath{\ltx{4}}\xspace}
\newcommand{\dparaSystFix}{\ensuremath{\ltx{3}}\xspace}
\newcommand{\dperpSystFix}{\ensuremath{\ltx{3}}\xspace}
\newcommand{\dswpdSystFix}{\ensuremath{\ltx{2}}\xspace}
\newcommand{\phisSyst}{\ensuremath{10}\xspace}
\newcommand{\DGsSyst}{\ensuremath{0.0024}\xspace}
\newcommand{\GsSyst}{\ensuremath{0.0034}\xspace}
\newcommand{\DmsSyst}{\ensuremath{0.023}\xspace}
\newcommand{\abslambdaSyst}{\ensuremath{0.015}\xspace}
\newcommand{\sqAzeroSyst}{\ensuremath{0.0086}\xspace}
\newcommand{\sqAperpSyst}{\ensuremath{0.0047}\xspace}
\newcommand{\sqAswavSyst}{\ensuremath{0.0013}\xspace}
\newcommand{\dparaSyst}{\ensuremath{0.033}\xspace}
\newcommand{\dperpSyst}{\ensuremath{0.063}\xspace}
\newcommand{\dswpdSyst}{\ensuremath{0.07}\xspace}
\newcommand{\phisResComb}{\ensuremath{-75}\xspace}
\newcommand{\DGsResComb}{\ensuremath{0.0778}\xspace}
\newcommand{\GsResComb}{\ensuremath{0.6639}\xspace}
\newcommand{\DmsResComb}{\ensuremath{17.763}\xspace}
\newcommand{\abslambdaResComb}{\ensuremath{1.012}\xspace}
\newcommand{\sqAzeroResComb}{\ensuremath{0.5222}\xspace}
\newcommand{\sqAperpResComb}{\ensuremath{0.2441}\xspace}
\newcommand{\sqAswavResComb}{\ensuremath{0.0075}\xspace}
\newcommand{\dparaResComb}{\ensuremath{3.147}\xspace}
\newcommand{\dperpResComb}{\ensuremath{2.94}\xspace}
\newcommand{\dswpdResComb}{\ensuremath{0.44}\xspace}
\newcommand{\phisErrComb}{\ensuremath{23}\xspace}
\newcommand{\DGsErrComb}{\ensuremath{0.0046}\xspace}
\newcommand{\GsErrComb}{\ensuremath{0.0032}\xspace}
\newcommand{\DmsErrComb}{\ensuremath{0.044}\xspace}
\newcommand{\abslambdaErrComb}{\ensuremath{0.019}\xspace}
\newcommand{\sqAzeroErrComb}{\ensuremath{0.0070}\xspace}
\newcommand{\sqAperpErrComb}{\ensuremath{0.0045}\xspace}
\newcommand{\sqAswavErrComb}{\ensuremath{0.0034}\xspace}
\newcommand{\dparaErrComb}{\ensuremath{0.083}\xspace}
\newcommand{\dperpErrComb}{\ensuremath{0.11}\xspace}
\newcommand{\dswpdErrComb}{\ensuremath{0.15}\xspace}
\newcommand{\phisSM}{\ensuremath{-37 \pm 1}\xspace}
\newcommand{\intLnum}{\ensuremath{96.5}\xspace}
\newcommand{\intL}{\ensuremath{\intLnum\fbinv}\xspace}
\newcommand{\nBs}{\ensuremath{491\,270 \pm 950}\xspace}
\newcommand{\nEvtHltMu}{\ensuremath{67\,908}\xspace}
\newcommand{\nEvtHltTrk}{\ensuremath{555\,213}\xspace}
\newcommand{\nBootstrap}{\ensuremath{1300}\xspace}
\newcommand{\nBsEff}{\ensuremath{27\,500}\xspace}
\newcommand\varpm{\mathbin{\vcenter{\hbox{%
  \oalign{\hfil\ensuremath{\scriptstyle+}\hfil\cr
          \noalign{\kern-.3ex}
          \ensuremath{\scriptscriptstyle({-})}\cr}%
}}}}
\newcommand{\splot}{\ensuremath{{}_{s}\mathcal{P}lot}\xspace}

\ifthenelse{\boolean{cms@external}}{\providecommand{\cmsTable}[1]{#1}}{\providecommand{\cmsTable}[1]{\resizebox{\textwidth}{!}{#1}}}
\newlength\cmsTabSkip\setlength{\cmsTabSkip}{1ex}
\ifthenelse{\boolean{cms@external}}{\providecommand{\cmsLeft}{upper\xspace}}{\providecommand{\cmsLeft}{left\xspace}}
\ifthenelse{\boolean{cms@external}}{\providecommand{\cmsRight}{lower\xspace}}{\providecommand{\cmsRight}{right\xspace}}
\ifthenelse{\boolean{cms@external}}
{\providecommand{\suppMaterial}{the supplemental material~\cite{suppMatBib}}}
{\providecommand{\suppMaterial}{Appendix~\ref{app:suppMat}\nocite{suppMatBib}}}

\cmsNoteHeader{BPH-23-004}

\title{First evidence for mixing-induced \texorpdfstring{\CP}{CP} violation in \texorpdfstring{\Bsjpsiphi}{Bs to J/Psi phi(1020)} decays in \texorpdfstring{\pp}{pp} collisions at \texorpdfstring{\cmeRunII}{sqrt(s) = 13 TeV}}
\date{\today}

\abstract{A novel machine-learning-based flavor-tagging algorithm combining same-side and opposite-side tagging is used to obtain the equivalent of \nBsEff tagged \Bsjpsiphi decays from \pp collisions at \cmeRunII, collected by the CMS experiment and corresponding to an integrated luminosity of \intL. A time- and flavor-dependent angular analysis of the \mumuKK final state, consistent with a \phiKK decay, is used to measure parameters of the \BsBsbar system. The weak phase is measured to be $\phis = \phisRes \pm \phisErrSym\stat \pm \phisSyst\syst$\mrad, which, combined with the \cmeRunI CMS result, gives $\phis = \phisResComb \pm \phisErrComb$\mrad. This value differs from zero by 3.2 standard deviations, providing the first evidence for mixing-induced \CP violation in \Bsjpsiphi decays. All measured physics parameters are found to agree with standard model predictions where available.
}

\hypersetup{%
pdfauthor={CMS Collaboration},%
pdftitle={First evidence for mixing-induced CP violation in B0s to J/Psi Phi(1020) decays in pp collisions at sqrt(s) = 13 TeV},%
pdfsubject={CMS},%
pdfkeywords={CMS, B physics, CP violation, flavor oscillation, mixing}}

\maketitle

The decays of \Bs mesons provide powerful ways to test the standard model (SM) predictions for \CP violation (CPV) in the quark sector and search for new physics effects.
In the SM, the weak phase \phis, which arises from CPV in the interference between decays with and without mixing of \Bs mesons to a \CP eigenstate, is related to the elements of the Cabibbo--Kobayashi--Maskawa quark mixing matrix~\cite{PhysRevLett.10.531, Kobayashi:1973fv} by $\phis \approx -2\betas = -2\arg(-V_{\PQt\PQs} V_{\PQt\PQb}^{*} / V_{\PQc\PQs} V_{\PQc\PQb}^{*})$, ignoring small contributions from penguin diagrams~\cite{Barel_2021}.
Global fits to measurements of \PQb-hadron and kaon decays in the context of the SM predict $-2\betas \approx \phisSM$\mrad~\refPhisSM.
This precise determination makes \phis an excellent probe for beyond-the-SM physics, as its value can be modified by the contribution of new particles to \Bs mixing~\cite{CDF:2006imy, Chiang:2009ev, ref:Artuso2015swg}.
Several experiments have measured \phis via $\PQb \to \PQc\PAQc\PQs$ transitions in \Bs decays~\cite{ref:d0Abazov08prl, ref:d0Abazov12prd, ref:cdfAaltonen08prl, ref:cdfAaltonen12prd, ref:cdfAaltonen12prl, ref:cmsKhachatryan16plb, CMS:2021136188, ref:lhcbAaij12plb, ref:lhcbAaij13prd, ref:lhcbAaij14plb, ref:lhcbAaij15prl,PhysRevLett.132.051802, ref:lhcbAaij14prl, ref:lhcbAaij16plb, ref:atlasAad12jhep, ref:atlasAad14prd, ref:atlasAad16jhep, Aad2021}, with no evident deviation from SM-based predictions.
This Letter presents a study of the \Bsjpsiphialt decay to the \mumuKK final state using proton-proton collisions collected by the CMS experiment~\cite{CMS:2008xjf, CMS:2023gfb} at \cmeRunII during 2017--2018, corresponding to an integrated luminosity of \intL~\cite{CMS-LUM-17-003, CMS-PAS-LUM-17-004, CMS-PAS-LUM-18-002}.
Data collected in 2016 are not used as they were collected with a different inner tracker detector and trigger strategies.
This analysis not only introduces a pioneering flavor-tagging algorithm and incorporates additional trigger paths compared to Ref.~\refCMSrunIIpartial, but also features a comprehensively reworked and improved analysis procedure, ultimately yielding the largest effective number of tagged signal candidates ever used for a single measurement of \phis.
A combination of this result with the one obtained by CMS at \cmeRunI~\refCMSrunI is also presented.
In this Letter, wherever a particle or its decay mode is specified, the charge conjugate is also implied unless otherwise indicated.

\begin{figure*}[tb!]
  \centering
  \includegraphics[width=0.5\textwidth]{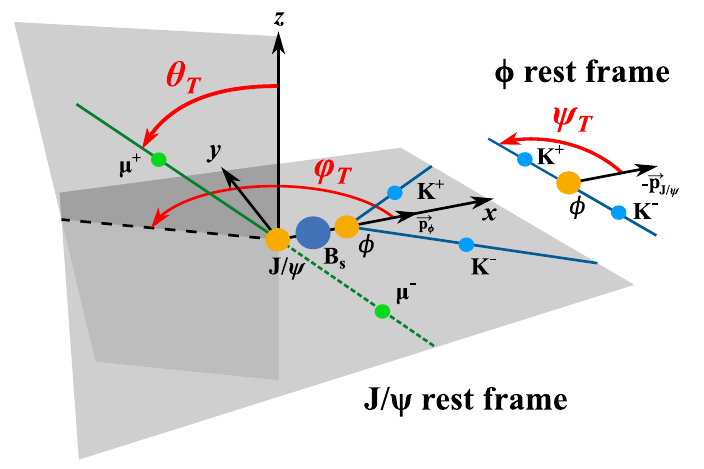} \hfill
  \includegraphics[width=0.35\textwidth]{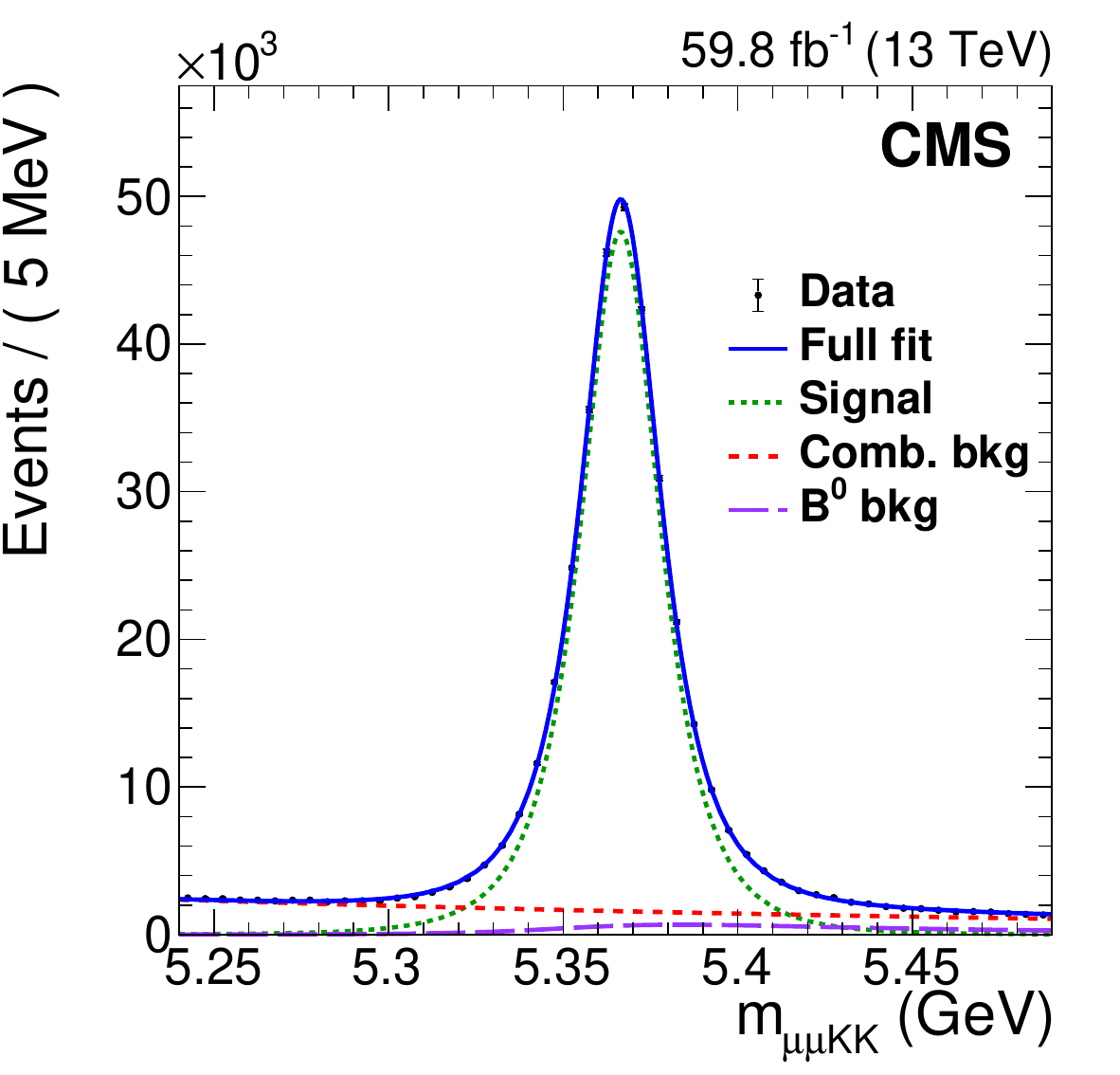}
  \caption{Left: Definition of the three decay angles \thetaT, \psiT, and \phiT. Right: Invariant mass distribution and fit projection of the selected candidates for the \setJpsiTrk trigger category (2018 data).}
  \label{fig:fig1}
\end{figure*}

The \Bsjpsiphi decay proceeds through a mixture of \CP eigenstates, which requires an amplitude analysis to separate the \CP-odd and \CP-even components.
A time-dependent angular analysis is performed in the transversity basis~\cite{ref:Dighe99epjc} by measuring the decay angles of the final-state particles $\Theta = (\thetaT,\, \psiT,\, \phiT)$ and \ct, which is the proper decay time of the reconstructed \Bs meson candidate, as described in Refs.~\cite{ref:PDG, CMS:2021136188}.
In the transversity basis, a reference frame is chosen such that the $x$ axis aligns with the \phimes momentum, and the $x$-$y$ plane is the \phiKK decay plane.
The polar and azimuthal angles of the \PGmp in the \jpsi rest frame are denoted \thetaT and \phiT, respectively.
The helicity angle \psiT, computed in the \phimes rest frame, is the angle between the  \PKp momentum and the direction of negative \jpsi momentum.
The three decay angles are illustrated in Fig.~\ref{fig:fig1} (left).
The decay rate formalism used to describe the distribution of final-state particles is the same as in previous CMS results~\cite{ref:cmsKhachatryan16plb, CMS:2021136188}.
The CPV in the \BsBsbar system is described by the complex parameter $\lambda = (q/p) (\overline{A}_\f/A_\f)$, where $A_\f$ ($\overline{A}_\f)$ is the decay amplitude of the \Bs (\Bsbar) meson to the final state $f$, and the parameters $p$ and $q$ relate the heavy and light mass eigenstates to the flavor eigenstates through $\PB^{\mathrm{L(H)}}_{\s} = p |\Bs\rangle \varpm q|\Bsbar\rangle$~\cite{ref:Branco99irmp}.
The parameters of interest (POIs) describing the decay rate are the weak phase \phis; the decay width and absolute mass differences between the \Bs mass eigenstates \DGs and \Dms; their average decay width \Gs; the direct CPV observable \abslambda; the longitudinal, parallel, and perpendicular transversity amplitudes \Azero, \Apara, and \Aperp; and the corresponding (\CP conserving) strong phases \dzero, \dpara, and \dperp.
Possible contributions from $\Bs \to \jpsi \, \PfDzP{980}$ and nonresonant \BsjpsiKK decays are taken into account by including an additional $S$-wave term in the decay model, parametrized with amplitude \Aswav and phase \dswav.
An effective coupling factor is required to model the $S$-$P$ wave interference due to the lack of an \mKK parametrization in the fit model.
Within the \mKK range used in this analysis (1009.5--1029.5\MeV), we calculate this correction to be $\ksp = 0.538 \pm 0.004$ by integrating the mass lineshape interference term.
A relativistic Breit--Wigner distribution and a constant distribution are assumed to describe the $P$- and $S$-wave, respectively, with a Flatt\'{e} distribution~\cite{Flatte:1976xv} used as an alternative assumption to estimate the \ksp uncertainty.
The strong phase \dzero is selected as the reference phase and set to zero, following a general convention~\cite{ref:d0Abazov12prd, ref:cdfAaltonen12prd}.
The value of \DGs is constrained to be positive, based on Ref.~\cite{ref:Aaij2012eq}.
To reduce correlations, $\dswpd \equiv \dswav-\dperp$ is used instead of \dswav.
Possible \BsBsbar production asymmetries are neglected, as the effect is expected to be small and below the sensitivity of the analysis, consistent with what is typically done in measurements of this type.

\begin{figure*}[tb!]
  \centering
  \includegraphics[width=0.32\textwidth]{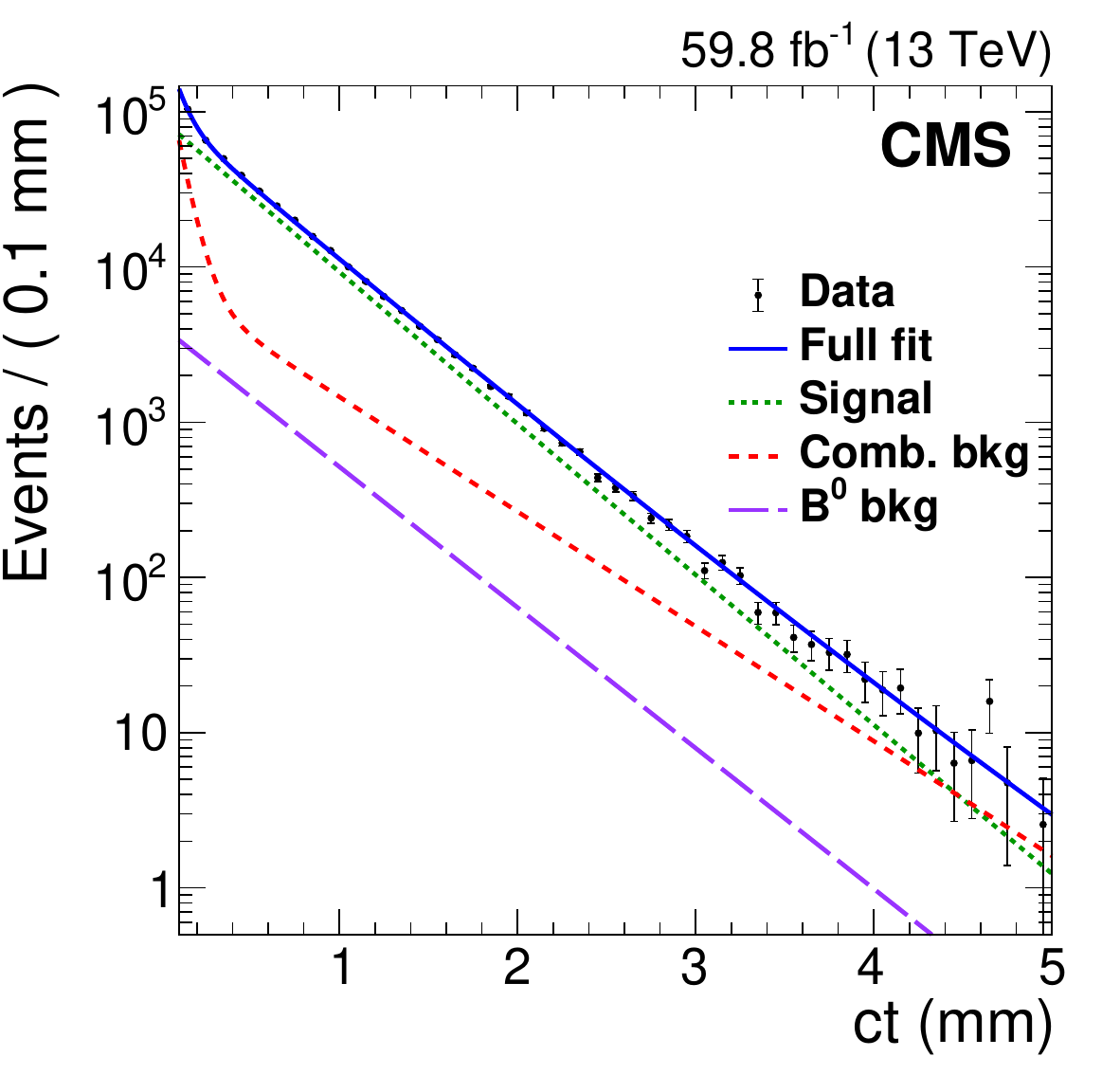} 
  \includegraphics[width=0.32\textwidth]{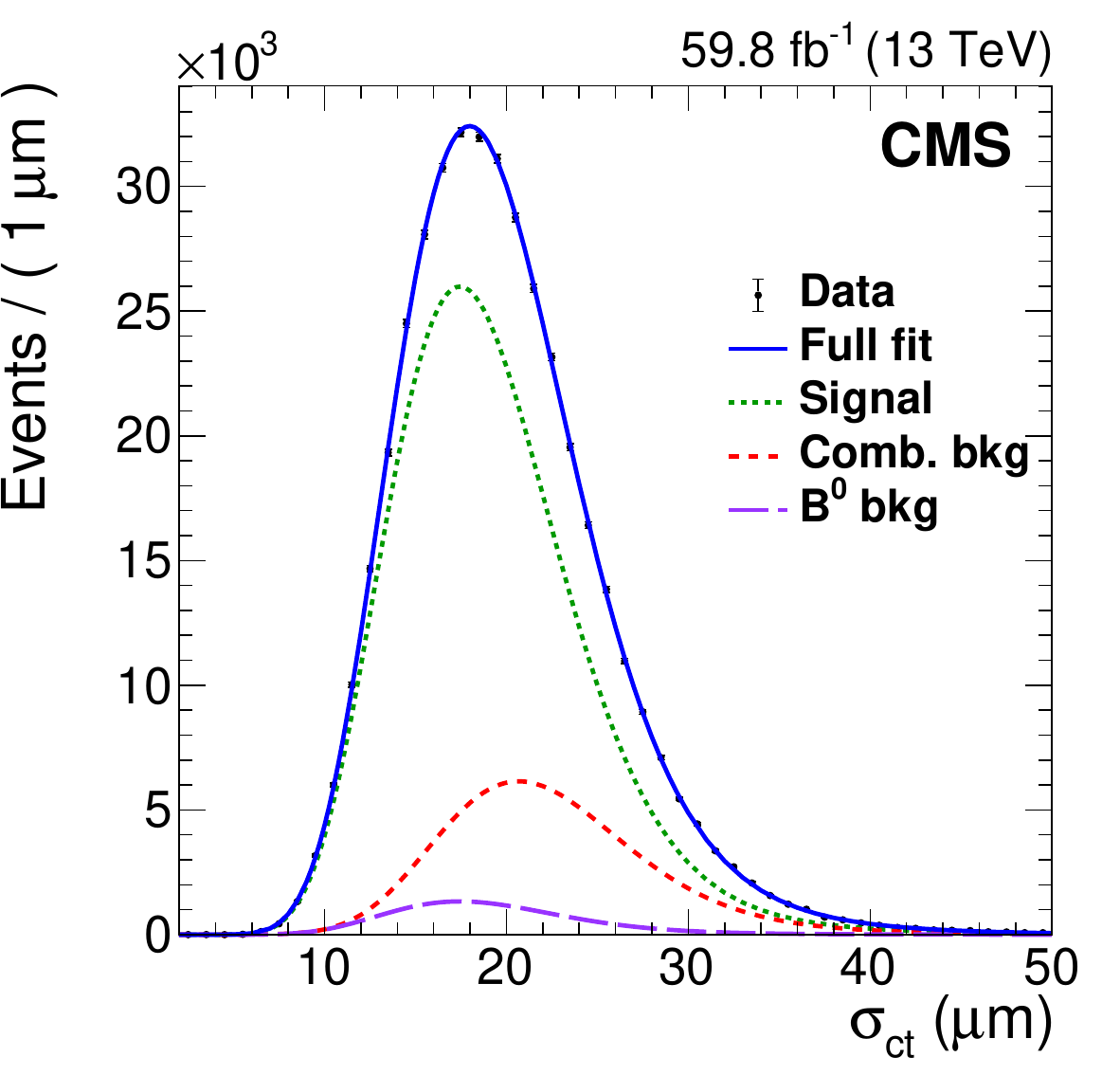} 
  \includegraphics[width=0.32\textwidth]{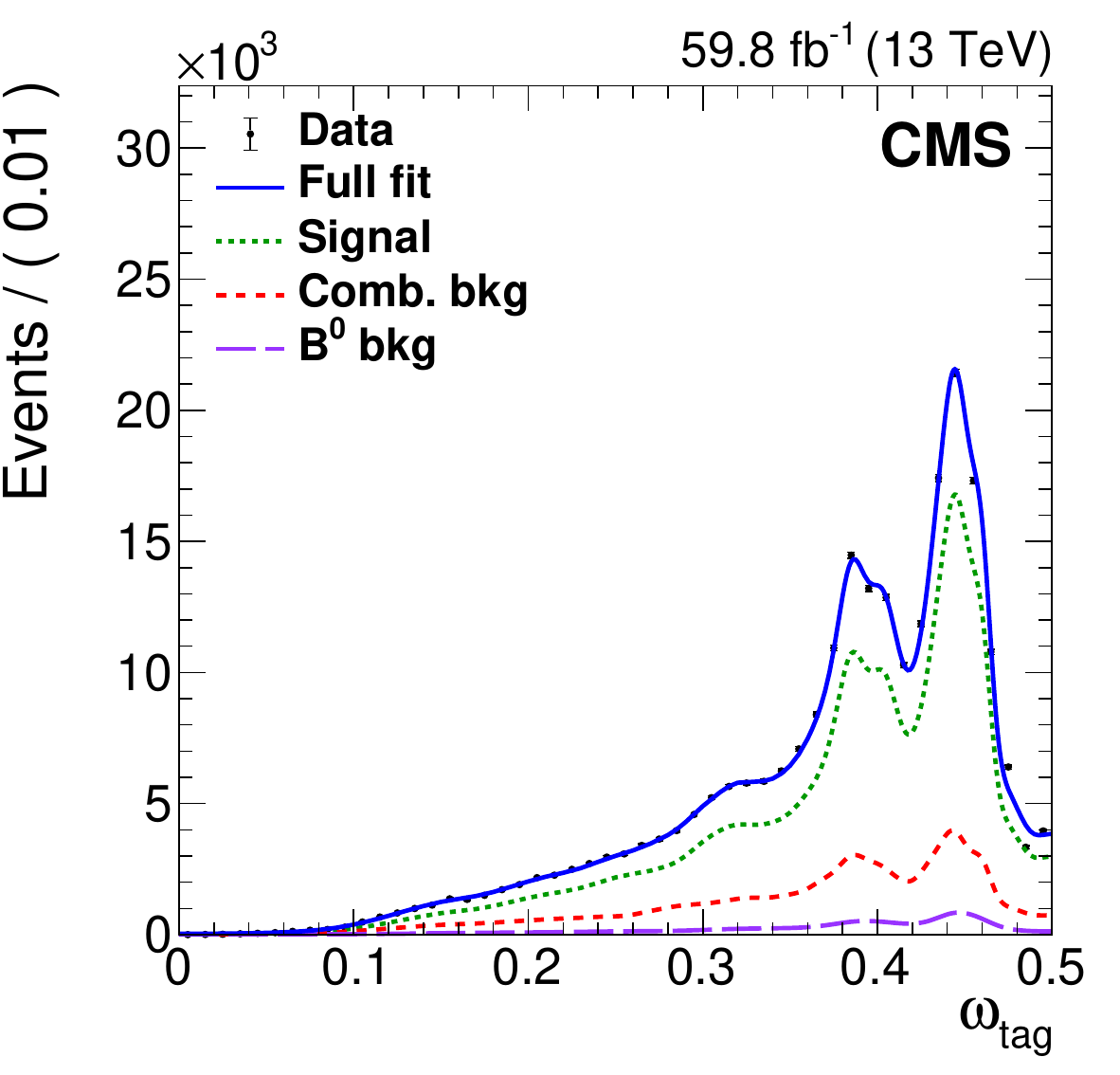}
  \caption{Distributions of the proper decay time \ct, its uncertainty \sct, and mistag probability \wtag of the selected candidates for the \setJpsiTrk trigger category (2018 data). The projections of the fitted model are also shown.}
  \label{fig:fig2}
\end{figure*}

The CMS apparatus is a multipurpose, nearly hermetic detector, designed to perform a wide range of particle physics measurements.
The \BsjpsiphiMMKK data sample (``\Bs data sample'') is collected with two independent triggers~\cite{CMS:2016ngn}.
The first trigger (``muon-tagging'', \setJpsiMu) requires a \jpsimumu candidate along with an additional muon, potentially usable for flavor tagging~\refCMSrunI.
In the second trigger, (``standard'', \setJpsiTrk), events with a displaced \jpsimumu candidate in conjunction with a \phiKK vertex are demanded.
To avoid double counting, events collected by the \setJpsiMu trigger are not considered in the \setJpsiTrk trigger.
To improve the signal purity, additional selection requirements are applied, including a requirement of $\ct > 60\,(100)\mum$ for the \setJpsiMu (\setJpsiTrk) triggered events.
A total of \nEvtHltMu and \nEvtHltTrk candidates are selected from the \setJpsiMu and \setJpsiTrk triggered events, respectively.
The distributions for the invariant mass \mBs, \ct, and its uncertainty \sct of the selected candidates in the \setJpsiTrk trigger category (2018 data) are shown in Figs.~\ref{fig:fig1} (right) and~\ref{fig:fig2} (left and center).
Appendices~\ifthenelse{\boolean{cms@external}}{A and B}{\ref{app:a} and \ref{app:b}} provide a more detailed description of the experimental setup and event reconstruction, while distributions for the other data samples can be found in \suppMaterial.

The flavor of the \Bs meson candidate, \ie, its particle or antiparticle state at production is determined using a novel flavor-tagging framework developed for this measurement.
It consists of four separate tagging algorithms (``taggers''): opposite-side (OS) muon, OS electron, OS jet, and same-side (SS).
The OS taggers infer the flavor of the \PQb quark contained in the signal \PQb meson by exploiting the decay products of the other \PQb quark in the event, relying on the fact that \PQb quarks are predominantly produced in \bbbar pairs.
The SS tagger takes advantage of the charge correlations between the \PQb quark in the signal \PQb meson and nearby particles produced in its hadronization, which reflect the underlying flavor content.
Each tagging algorithm produces a tagging inference $\xitag = 0, +1, -1$, indicating whether the event is untagged, tagged as \Bs or \Bsbar.
For tagged events, the algorithm also estimates the mistag probability \wtag.
The tagging efficiency (\etag) represents the fraction of tagged events, while the dilution factor $\Dtag \equiv 1 - 2\wtag$ quantifies the inference purity.
The tagging power $\Ptag \equiv \etag\sqDtag$ combines these parameters and reflects the effective number of correctly tagged signal events (\NBsPtag), contributing to flavor-sensitive measurements.

All four taggers are used for the \setJpsiTrk data sets, while only the OS muon algorithm is used in the \setJpsiMu ones.
The OS taggers are designed to be orthogonal to each other, meaning only one can provide an inference for each event.
To improve performance, \wtag is estimated on a per-event basis by dedicated deep neural networks (DNNs) developed using simulated events and calibrated in data.
For the OS-jet and SS taggers, the same DNNs are also used to infer \xitag.
These calibrated DNNs act as probability estimators for the tagging inference.
When an OS and the SS algorithm both provide tagging inferences, \xitag and \wtag are recalculated by combining the available information.
Appendix~\ifthenelse{\boolean{cms@external}}{B}{\ref{app:b}} provides more details about the tagging algorithms.

The combined tagging performance for the \Bs data sample is $\etag \approx 56\%$ and $\Ptag \approx 5.6\%$.
The performance breakdown in each mutually exclusive tagging category is reported in Table~\ref{tab:tagperf}.
The mistag probability distribution of the selected candidates for the \setJpsiTrk trigger category (2018 data) is shown in Fig.~\ref{fig:fig2} (right).

\tolerance=1800{
The flavor-tagging is validated with extensive studies on simulated data and \BdjpsikstarMMKP events (``\Bd control data sample'').
Several measurements are performed on this sample to ensure that the tagging inference is reliable and free from significant bias.
To test the robustness of the tagging framework in mixing scenarios, \Bd flavor oscillations are analyzed.
Using the tagging framework, the fitted value of the oscillation frequency is $\Dmd = 0.5081 \pm 0.0046\stat\ \hips$, which is in excellent agreement with the world-average value of $0.5069 \pm 0.0019\ \hips$~\refPDG.
The time-integrated mixing asymmetry in the \Bd control data sample is also examined, to investigate possible issues with the overall tagging calibration scale of the SS tagger. 
This is done by comparing the relationship between the measured mixing asymmetry and the estimated tagging dilution.
The observed experimental dilution is found to be consistent with the estimated value.
All validation studies are conducted using each tagger independently and in each tagging category, as well as using the tagging inference from the overall framework.
Moreover, the CPV measurement in the \Bsjpsiphi channel is repeated four additional times, each using only one of the four tagging algorithms, and the obtained results are compared.
No evidence of bias has been found in any of the studies.
}

The \Bs data are divided into four main data sets according to the two trigger categories and two data taking years, and further divided into a total of 12 data sets based on the three possible values of \xitag.
The POIs: \phis, \DGs, \Gs, \Dms, \abslambda, \sqAzero, \sqAperp, \sqAswav, \dpara, \dperp, and \dswpd, are extracted with a simultaneous unbinned multidimensional extended maximum likelihood fit to the \Bs data sets using seven observables as input: \mBs, \ct, \sct, \cthetaT, \cpsiT, \phiT, and \wtag.
The \Bsjpsiphi amplitudes are normalized to unity with the constraint $\sqApara \equiv 1 - \sqAperp - \sqAzero$.
The likelihood function is composed of probability density functions (pdfs) describing the \Bsjpsiphi signal, the combinatorial background, and the background from \BdjpsikstarMMKP, where the pion is misidentified as a kaon (``\Bd background'').
For each component, the pdf is the product of functions that model the distributions of the seven observables of the reconstructed candidates.
Correlations between the input observables have been studied in data and simulation and found to have a negligible effect.
The \Bsjpsiphi differential decay rate model utilized in the fit includes the flavor inference \xitag and the dilution term \Dtag.
Unless specified otherwise, all pdf parameters are left free in the final fit.
The fit model also includes the angular efficiency and time resolution functions.
These functions are computed separately for each main data set.

The event reconstruction and selection efficiency as a function of \ct, \ect, is determined from data utilizing the \Bd control data sample, which exhibits a similar topology to the \Bsjpsiphi signal.
The efficiency is parametrized using Bernstein polynomials to model the ratio between the observed \ct distribution for the signal and the expected distribution convolved with the time resolution.
The expected distribution is an exponential characterized by the \Bd lifetime $\smash[b]{\tauBd = 1.517\unit{ps}}$~\refPDG.
Residual efficiency differences between the control and measurement samples are corrected using simulation.
In the fit, each event in the \Bs data sample is weighted by $1/\ect$.
Since no dependence of the time resolution on \ct is observed in simulation, the \ct uncertainty is calibrated for each main data set in bins of \sct in a data sample of prompt events obtained by removing the displacement requirement from the \setJpsiMu trigger category.
Most of these events consist of prompt \jpsimumu decays accompanied by two tracks and are not anticipated to have a substantial lifetime.
The long-lived components are properly accounted for.
The negative portion of the \ct distribution in this sample arises from resolution effects and is used to calibrate \sct.
The \ct uncertainty matches the experimental time resolution well, with only a uniform correction of $+$1.5\mum being needed.
The calibrated \sct average value for signal candidates is $\approx$20\mum.
The angular efficiency functions $\varepsilon(\Theta)$ are derived from \Bsjpsiphi simulated events in which \DGs is set to zero to disentangle the angular and time components.
These simulated samples are corrected to better match the observed data, taking into account the kinematic properties of the final-state particles and the relevant physical parameters.
The angular efficiency is modeled using the Kernel Density Estimation (KDE) method~\cite{ref:rosenblatt1956, ref:parzen1962} from the reconstructed and generated angular distributions, and is incorporated in the signal pdf as a 3D histogram.
All parameters describing the efficiency and time resolution calibration are fixed in the final fit.

The mass shapes for the signal and the \Bd background are described by Johnson $S_{\mathrm{U}}$ distributions~\cite{Johnson:1949zj}, while an exponential distribution is used for the combinatorial background.
The \wtag distribution is modeled for each component using KDE distributions, with separate distributions used for events tagged as \Bs and \Bsbar.
For the \Bd and combinatorial backgrounds, the \ct component is modeled with one and two exponential distributions, respectively.
The angular variables are described with 3D Bernstein polynomials.
The pdfs describing \sct are modeled for each component by the sum of two Gamma distributions, with their parameters fixed in the final fit based on what is obtained from a series of pre-fits to the signal region $\mBs \in [5.28,\,5.45]$ and mass sideband region $\mBs \in [5.24,\,5.28] \cup [5.45,\,5.49]$~\GeV.
The shape of the \Bd background is obtained from simulated \Bdjpsikstar events (with the decay generated by \textsc{EvtGen} using the \textsc{EvtSVVHelAmp} model) reconstructed as \Bsjpsiphi, while its yield is determined from the \Bs data sample via a simultaneous fit to the invariant mass distributions in both the \Bsjpsiphi and \Bdjpsikstar reconstruction hypotheses.
The time resolution for the \Bd background is assumed to be the same as for the \Bs signal.
All parameters related to the \Bd background are fixed in the final fit.
The fit model is duplicated for the four main data sets, with the individual log-likelihoods summed to obtain the full log-likelihood.

The fit model is validated through studies on simulated samples and fits to individual data categories.
The statistical uncertainties on the POIs are determined using the profile-likelihood method, taking into account the presence of event weights.
The procedure is cross-checked with the parametric bootstrap method~\cite{EfroTibs93}, in which \nBootstrap pseudo-datasets are generated from the fitted model and refitted with the reference model.
The resulting distributions of the POIs are used to verify the coverage of the uncertainty estimates and to evaluate the  ``fit bias'', defined as the difference between the central values of the bootstrap distribution and the reference fit results.
This bias is treated as a systematic uncertainty.
As an additional cross-check, a set of 400 non-parametric bootstrap pseudo-datasets, obtained by sampling events with replacement directly from the data has also been generated.
Considering the associated uncertainties, the results obtained with the profile-likelihood, parametric bootstrap, and non-parametric bootstrap methods are consistent with each other.
The measured number of \Bsjpsiphi signal events from the fit is \nBs, corresponding to an effective number of tagged signal candidates of $\NBsPtag = \nBsEff \pm 100$.

The systematic uncertainties potentially impacting the POIs are categorized into two groups.
Type-I uncertainties stem from the finite number of events in the simulated samples and from uncertainties related to the parameters of the efficiency and calibration functions.
These uncertainties are evaluated by either sampling the function parameters based on their uncertainties or, in the case of nonparametric functions, by resampling the simulated samples.
The systematic uncertainties are then quantified as the root-mean-square of the resulting POI distributions.
Type-II uncertainties account for potential biases resulting from the assumptions made in the fit model and analysis methods.
For each tested assumption, an alternative model choice is selected, and the fit to the data is repeated.
A deviation from the reference results is significant if it exceeds 20\% of the statistical uncertainty associated with a parameter.
In such cases, a double-sided systematic uncertainty of size equal to the deviation is assigned.
The fit bias does not fall into either of these two categories.
It is evaluated, as discussed above, from the parametric bootstrap distributions and treated as a double-sided systematic uncertainty.
The systematic uncertainties are summarized in Table~\ref{tab:systematics}.

\begin{table*}[tb]
  \centering
  \topcaption{Summary of the systematic and statistical uncertainties.
  The symbol (\NA) indicates that no uncertainty is evaluated, as all tested Type-II effects are below the threshold to be considered significant.
  The combined systematic uncertainty is the sum in quadrature of the individual uncertainties.}
  \cmsTable{
  \begin{scotch}{lccccccccccc}
    & \phis            & \DGs            &  \Gs
    & \Dms             & \abslambda
    & \sqAzero         & \sqAperp         & \sqAswav
    & \dpara           & \dperp           & \dswpd
    \\
    & [mrad]           & [$\ips$]         & [$\ips$]
    & [$\hips$]        & 
    &                  &                  &
    & [rad]            & [rad]            & [rad]
    \\ \hline
    Statistical uncertainty
    & \phisErrSym     & \DGsErrSym        & \GsErrSym
    & \DmsErrSym      & \abslambdaErrSym
    & \sqAzeroErrSym  & \sqAperpErrSym    & \sqAswavErrSym
    & \dparaErrSym    & \dperpErrSym      & \dswpdErrSym
    \\ [\cmsTabSkip]
    Fit bias
    & \phisSystBias     & \DGsSystBias        & \GsSystBias
    & \DmsSystBias      & \abslambdaSystBias
    & \sqAzeroSystBias  & \sqAperpSystBias    & \sqAswavSystBias
    & \dparaSystBias    & \dperpSystBias      & \dswpdSystBias
    \\
    Flavor tagging
    & \phisSystTag     & \DGsSystTag        & \GsSystTag
    & \DmsSystTag      & \abslambdaSystTag
    & \sqAzeroSystTag  & \sqAperpSystTag    & \sqAswavSystTag
    & \dparaSystTag    & \dperpSystTag      & \dswpdSystTag
    \\
    Angular efficiency
    & \phisSystAng     & \DGsSystAng        & \GsSystAng
    & \DmsSystAng      & \abslambdaSystAng
    & \sqAzeroSystAng  & \sqAperpSystAng    & \sqAswavSystAng
    & \dparaSystAng    & \dperpSystAng      & \dswpdSystAng
    \\
    Time efficiency
    & \phisSystCt     & \DGsSystCt        & \GsSystCt
    & \DmsSystCt      & \abslambdaSystCt
    & \sqAzeroSystCt  & \sqAperpSystCt    & \sqAswavSystCt
    & \dparaSystCt    & \dperpSystCt      & \dswpdSystCt
    \\
    Time resolution
    & \phisSystReso     & \DGsSystReso        & \GsSystReso
    & \DmsSystReso      & \abslambdaSystReso
    & \sqAzeroSystReso  & \sqAperpSystReso    & \sqAswavSystReso
    & \dparaSystReso    & \dperpSystReso      & \dswpdSystReso
    \\
    Model assumptions
    & \phisSystModel     & \DGsSystModel       & \GsSystModel
    & \DmsSystModel      & \abslambdaSystModel
    & \sqAzeroSystModel  & \sqAperpSystModel    & \sqAswavSystModel
    & \dparaSystModel    & \dperpSystModel      & \dswpdSystModel
    \\
    \Bd background
    & \phisSystBd     & \DGsSystBd        & \GsSystBd
    & \DmsSystBd      & \abslambdaSystBd
    & \sqAzeroSystBd  & \sqAperpSystBd    & \sqAswavSystBd
    & \dparaSystBd    & \dperpSystBd      & \dswpdSystBd
    \\
    \Lambdab background
    & \phisSystLb     & \DGsSystLb        & \GsSystLb
    & \DmsSystLb      & \abslambdaSystLb
    & \sqAzeroSystLb  & \sqAperpSystLb    & \sqAswavSystLb
    & \dparaSystLb    & \dperpSystLb      & \dswpdSystLb
    \\
    $S$-$P$ wave interference
    & \phisSystKsp     & \DGsSystKsp        & \GsSystKsp
    & \DmsSystKsp      & \abslambdaSystKsp
    & \sqAzeroSystKsp  & \sqAperpSystKsp    & \sqAswavSystKsp
    & \dparaSystKsp    & \dperpSystKsp      & \dswpdSystKsp
    \\
    $P(\sct)$ uncertainty
    & \phisSystFix     & \DGsSystFix        & \GsSystFix
    & \DmsSystFix      & \abslambdaSystFix
    & \sqAzeroSystFix  & \sqAperpSystFix    & \sqAswavSystFix
    & \dparaSystFix    & \dperpSystFix      & \dswpdSystFix
    \\ [\cmsTabSkip]
    Total systematic uncertainty
    & \phisSyst     & \DGsSyst        & \GsSyst
    & \DmsSyst      & \abslambdaSyst
    & \sqAzeroSyst  & \sqAperpSyst    & \sqAswavSyst
    & \dparaSyst    & \dperpSyst      & \dswpdSyst
    \\
  \end{scotch}
  }
  \label{tab:systematics}
\end{table*}

Type-I uncertainties are evaluated for the \Bd background yield, the values of \ksp and \tauBd, and the parameters of the time resolution, time efficiency and flavor tagging calibration functions.
Similarly, uncertainties arising from the fixed \sct pdfs in the final fit are also evaluated.
Statistical uncertainties associated with the angular efficiency function are estimated using resampling techniques on the \Bsjpsiphi simulated events.

Type-II uncertainties are assigned by conducting extensive investigations and cross-verification processes to validate the model assumptions.
We explore alternative hypotheses regarding the pdf shapes of the mass, \ct, \sct, \wtag, and angular variable distributions for both the signal and background components.
For the pre-fits, different signal and sideband regions were tried and the \splot method~\cite{Pivk:2004ty} was used in place of simple sideband subtraction.
Additionally, we examine alternative models for the efficiency functions to identify potential sources of bias.
The impact of neglecting background from \LambdabjpsiKpMMKp is assessed by incorporating a dedicated component into the fit model, following an approach similar to the one for the \Bd background, and repeating the fit to the \Bs data sample.
Various validation studies are carried out on the correction procedure applied to the simulated samples used to evaluate the angular efficiency.
The calibration procedure for the OS tagging algorithms is evaluated using results from simulation to correct for differences between \Bsjpsiphi and \BujpsiK decays.
The calibration for the SS tagger already accounts for these differences.
We investigate its calibration strategy by removing these corrections, which introduce a significant perturbation of approximately 10\% to the calibration parameters, and testing the stability of the physics results in response to these changes.
Additionally, possible differences in the calibration between \Bs/\Bu and \Bsbar/\Bm decays are tested with flavor-specific calibrations, and no significant impact on the POIs is observed.

\begin{figure*}[tb]
  \centering
  \includegraphics[width=0.32\textwidth]{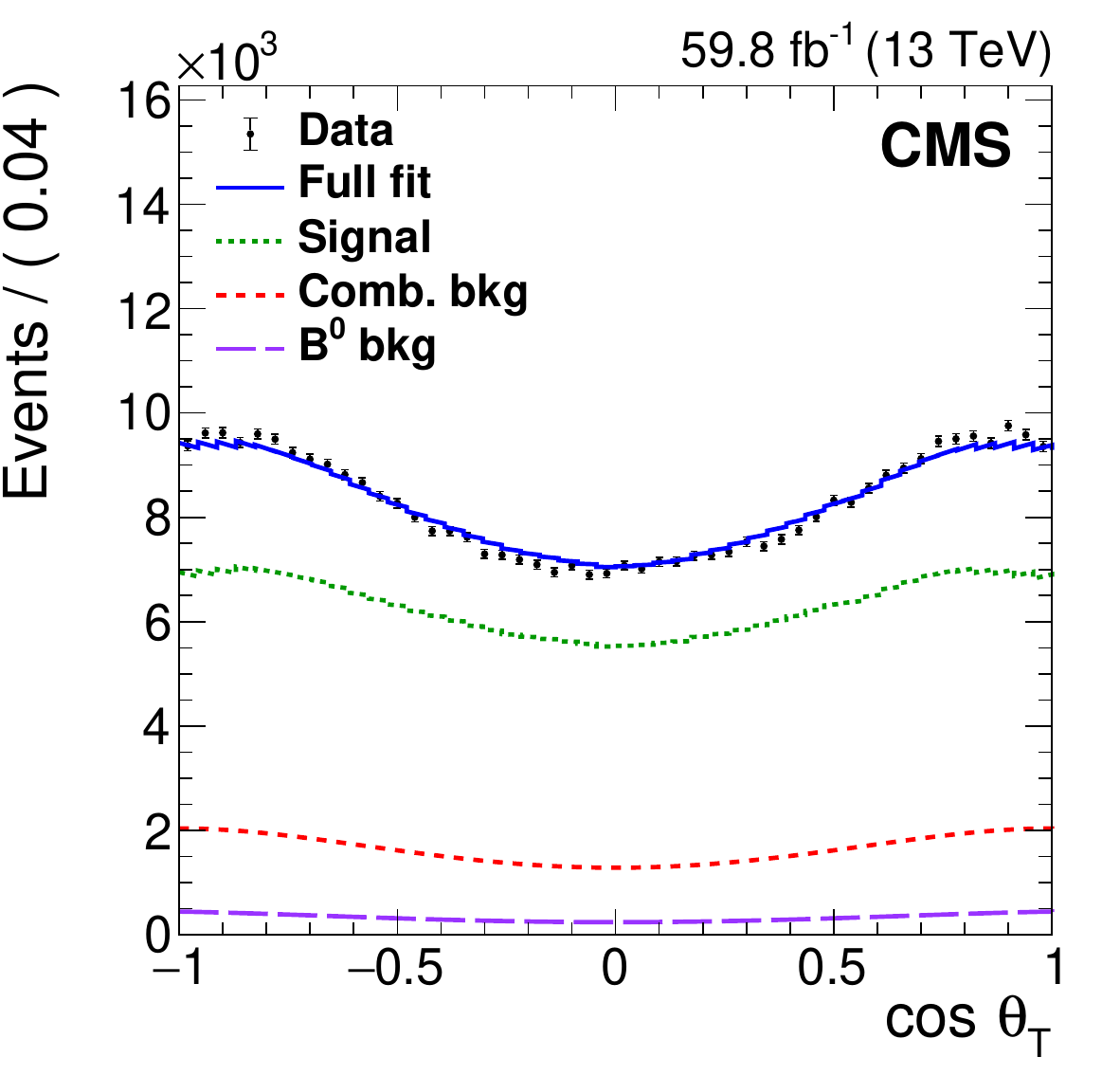} 
  \includegraphics[width=0.32\textwidth]{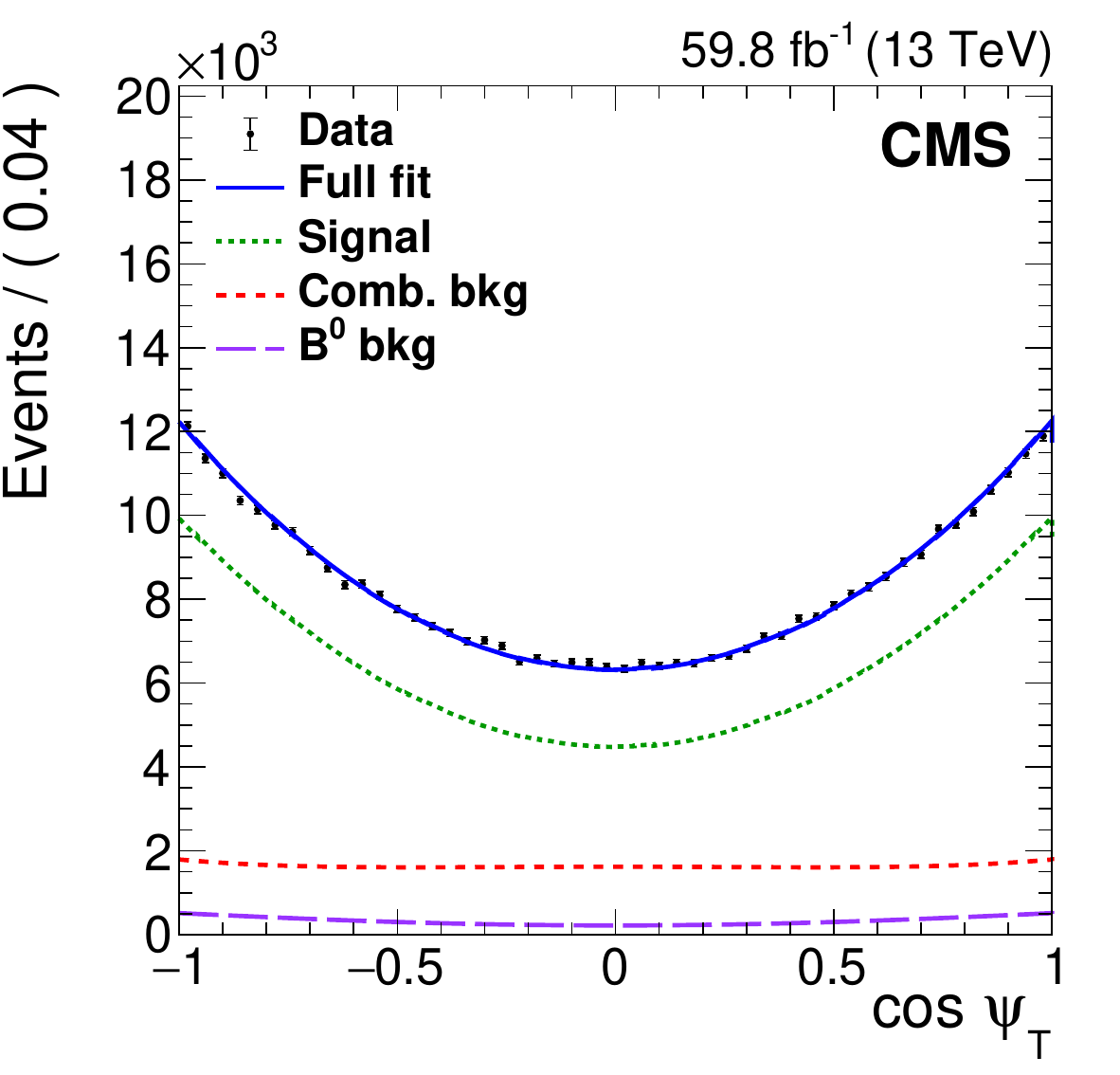} 
  \includegraphics[width=0.32\textwidth]{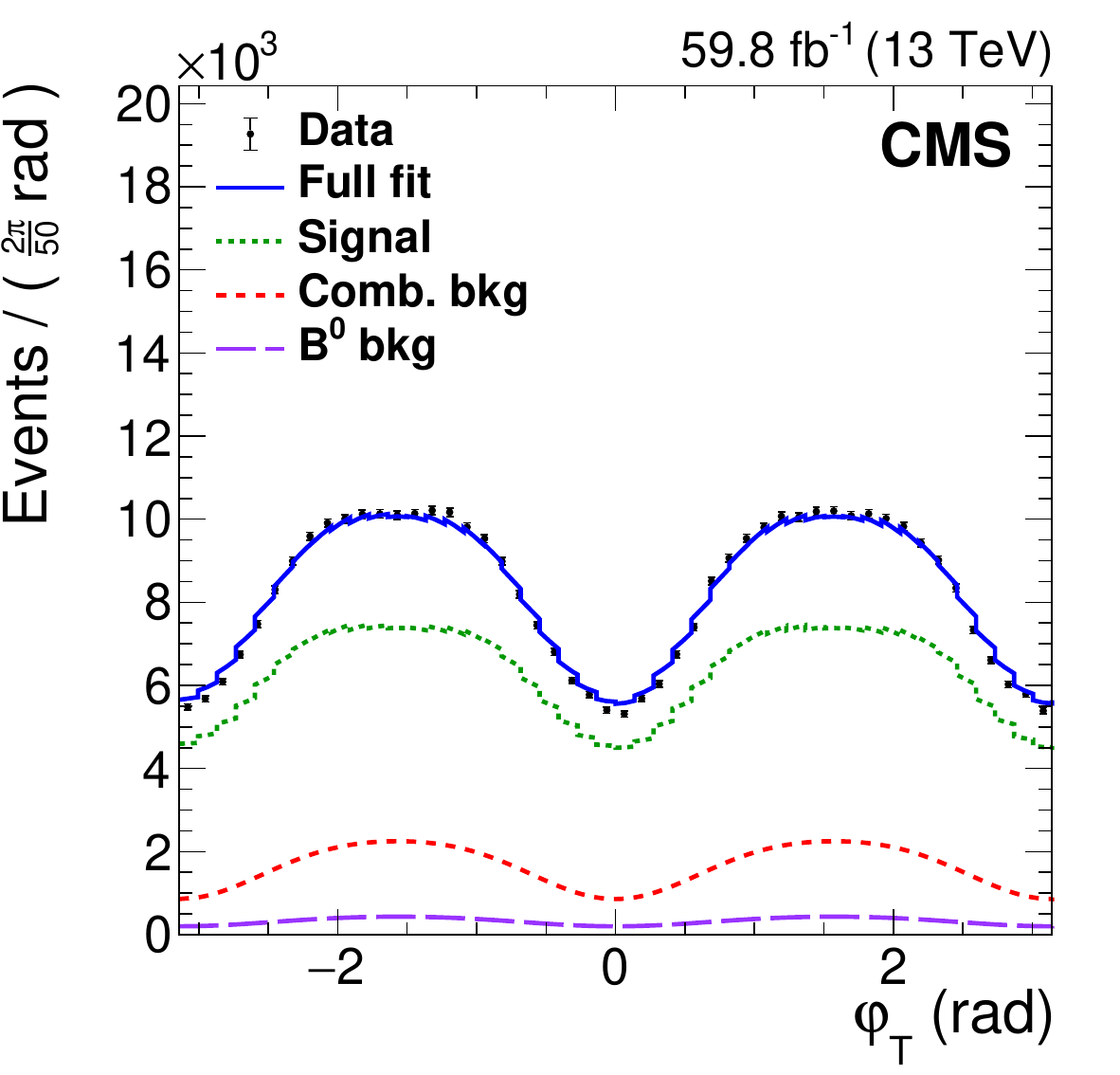}
  \caption{Angular observables distributions of the selected candidates and fit projection for the \setJpsiTrk trigger category (2018 data). The projections of the fitted model are also shown.}
  \label{fig:fig3}
\end{figure*}

\begin{figure}[ht!]
  \centering
  \includegraphics[width=0.49\textwidth]{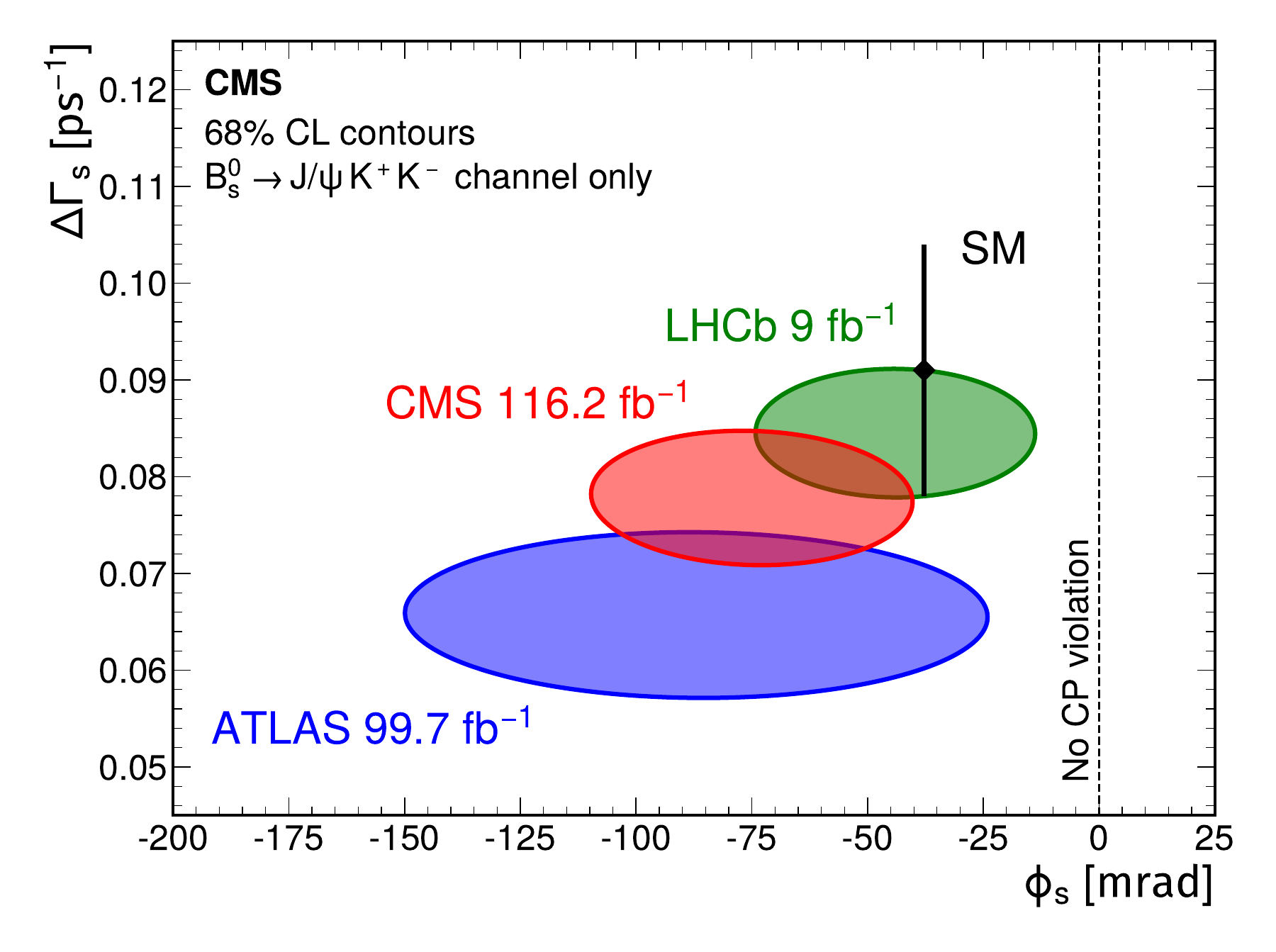}
  \caption{
    The two-dimensional 68\% \CL contours in the $\phis$-$\DGs$ plane for the combined CMS (red), ATLAS (blue)~\cite{Aad2021}, and LHCb (green)~\cite{PhysRevLett.132.051802} results.
    Results refer only to \BsjpsiKK measurements in the $\DGs > 0$ region.
    The contours account for both statistical and systematic uncertainties.
    The SM prediction neglects possible contributions from higher-order penguin diagrams and is represented by the thin black band, with the central value indicated by the black diamond~\cite{ref:Charles11prd, ref:UTfit22, ref:Lenz2019lvd}.
  }
  \label{fig:fig4}
\end{figure}

The fit results, along with the associated statistical and systematic uncertainties, are provided in Table~\ref{tab:fitresult}.
\begin{table}[tb]
    \centering
    \topcaption{Measured values and uncertainties of the parameters of interest in the $\DGs > 0$ region.}
    \begin{scotch}{l r l l l}
        Parameter & Fit value & Stat. unc. & Syst. unc.\\
        \hline
        \phis [mrad]    & \phisRes      & $ \phisErr $      & $\pm\ \phisSyst $ \\
        \DGs [$\ips$]   & \DGsRes       & $ \DGsErr $       & $\pm\ \DGsSyst $ \\
        \Gs [$\ips$]    & \GsRes        & $ \GsErr $        & $\pm\ \GsSyst $ \\
        \Dms [$\hips$]  & \DmsRes       & $ \DmsErr $       & $\pm\ \DmsSyst $ \\
        \abslambda      & \abslambdaRes & $ \abslambdaErr $ & $\pm\ \abslambdaSyst $ \\
        \sqAzero        & \sqAzeroRes   & $ \sqAzeroErr $   & $\pm\ \sqAzeroSyst $ \\
        \sqAperp        & \sqAperpRes   & $ \sqAperpErr $   & $\pm\ \sqAperpSyst $ \\
        \sqAswav        & \sqAswavRes   & $ \sqAswavErr $   & $\pm\ \sqAswavSyst $ \\
        \dpara [rad]    & \dparaRes     & $ \dparaErr $     & $\pm\ \dparaSyst $ \\
        \dperp [rad]    & \dperpRes     & $ \dperpErr $     & $\pm\ \dperpSyst $ \\
        \dswpd [rad]    & \dswpdRes     & $ \dswpdErr $     & $\pm\ \dswpdSyst $ \\
    \end{scotch}
    \label{tab:fitresult}
\end{table}
The projection of the fit onto each of the seven fitted distributions for the \setJpsiTrk trigger category (2018 data) is shown in Figs.~\ref{fig:fig1}, \ref{fig:fig2}, and \ref{fig:fig3}.
The results are found to be consistent with SM-based predictions~\cite{ref:Charles11prd, ref:UTfit22, ref:Lenz2019lvd} and the latest world-average values~\refPDG.
The weak CPV phase is measured to be $\phis = \phisRes \pm \phisErrSym\stat \pm \phisSyst\syst$\mrad, which is within 1.5 standard deviations of the SM-based prediction.
The measured value of \abslambda is consistent with no direct CPV ($\abslambda = 1$).
These results are comparable in precision to the world's most precise single measurement~\cite{PhysRevLett.132.051802} and supersede the previous results in Ref.~\refCMSrunIIpartial.
The results are combined with those obtained by CMS at \cmeRunI~\refCMSrunI using the BLUE method~\cite{ref:Lyons1988rp, ref:Valassi2003mu} to obtain $\phis = \phisResComb \pm \phisErrComb$\mrad and $\DGs = \DGsResComb \pm \DGsErrComb \ips$, which are both compatible with SM-based predictions.
The uncertainties include both statistical and systematic components.
The full set of combined results is reported in \suppMaterial.
Figure~\ref{fig:fig4} shows the two-dimensional \phis vs. \DGs contour at 68\% \CL for the combined results, alongside the SM-based prediction and the latest results from other LHC experiments.
Tabulated results are provided in the HEPData record for this analysis~\cite{hepdata}.
The combined \phis value exhibits a deviation from zero of 3.2 standard deviations, providing the first evidence for mixing-induced CPV in \Bsjpsiphi decays.

\begin{acknowledgments}
We congratulate our colleagues in the CERN accelerator departments for the excellent performance of the LHC and thank the technical and administrative staffs at CERN and at other CMS institutes for their contributions to the success of the CMS effort. In addition, we gratefully acknowledge the computing centres and personnel of the Worldwide LHC Computing Grid and other centres for delivering so effectively the computing infrastructure essential to our analyses. Finally, we acknowledge the enduring support for the construction and operation of the LHC, the CMS detector, and the supporting computing infrastructure provided by the following funding agencies: SC (Armenia), BMBWF and FWF (Austria); FNRS and FWO (Belgium); CNPq, CAPES, FAPERJ, FAPERGS, and FAPESP (Brazil); MES and BNSF (Bulgaria); CERN; CAS, MoST, and NSFC (China); MINCIENCIAS (Colombia); MSES and CSF (Croatia); RIF (Cyprus); SENESCYT (Ecuador); ERC PRG, RVTT3 and MoER TK202 (Estonia); Academy of Finland, MEC, and HIP (Finland); CEA and CNRS/IN2P3 (France); SRNSF (Georgia); BMBF, DFG, and HGF (Germany); GSRI (Greece); NKFIH (Hungary); DAE and DST (India); IPM (Iran); SFI (Ireland); INFN (Italy); MSIP and NRF (Republic of Korea); MES (Latvia); LMTLT (Lithuania); MOE and UM (Malaysia); BUAP, CINVESTAV, CONACYT, LNS, SEP, and UASLP-FAI (Mexico); MOS (Montenegro); MBIE (New Zealand); PAEC (Pakistan); MES and NSC (Poland); FCT (Portugal); MESTD (Serbia); MCIN/AEI and PCTI (Spain); MOSTR (Sri Lanka); Swiss Funding Agencies (Switzerland); MST (Taipei); MHESI and NSTDA (Thailand); TUBITAK and TENMAK (Turkey); NASU (Ukraine); STFC (United Kingdom); DOE and NSF (USA).
\end{acknowledgments}
\bibliography{auto_generated}

\clearpage
\appendix

\section{The CMS detector, simulated samples, and event reconstruction}\label{app:a}

The CMS apparatus is a multipurpose, nearly hermetic detector, designed to trigger on~\cite{CMS:2020cmk, CMS:2016ngn} and identify electrons, muons, photons, and hadrons~\cite{CMS:2020uim, CMS:2018rym,CMS:2014pgm}. A global event description algorithm~\cite{CMS:2017yfk} aims to reconstruct all individual particles in an event, combining information provided by the all-silicon inner tracker and by the crystal electromagnetic, and a brass and scintillator hadron calorimeters, operating inside a 3.8\unit{T} superconducting solenoid, with data from the gas-ionization muon detectors embedded in the flux-return yoke outside the solenoid.
Simulated events are employed to determine the angular efficiency and to train the various machine-learning models.
These samples are generated with the software described in Refs.~\cite{Sjostrand:2014zea, Sirunyan:2019dfx, Ball:2017nwa, ref:Lange01nim, ref:Barberio94cpc, ref:Agostinelli03nim} and validated via comparison with background-subtracted data in several control distributions.

To improve the purity of the signal, additional requirements are imposed on events selected by the trigger system.
Muons and kaons must have pseudorapidity $\abs{\eta} < 2.4$ and $\abs{\eta} < 2.5$, respectively.
For the \setJpsiMu (\setJpsiTrk) trigger category, muons are required to have transverse momentum $\pt > 3.5\,(4.0)\GeV$, while tracks are required to have $\pt > 1.15\,(0.90)\GeV$.
The opposite-charge muons forming the \jpsi candidate are then constrained to a common vertex, obtained via a Kalman fit~\cite{ref:Fruhwirth87nim}, with an invariant mass within 150\MeV of the world-average mass of the \jpsi meson~\refPDG.
The \phimes meson candidates are retained only if their invariant mass, assuming both oppositely charged particles to be kaons, lies within 10\MeV of the world-average mass of the \phimes meson~\refPDG.
Finally, the \jpsi and \phimes mesons candidates are combined to create \Bs meson candidates through a kinematic fitting process involving the four tracks, where the dimuon mass is constrained to the world-average \jpsi meson mass~\refPDG.
The \Bs meson candidates are accepted if their $\pt > 9.5$\GeV, their invariant mass is in the 5.24--5.49\GeV range, and the vertex fit $\chi^2$ probability exceeds 2\%.
The primary vertex (PV) is chosen as the vertex closest to the line passing through the \Bs decay vertex and parallel to its momentum.
The \ct is measured as $\ct = c\,m^\text{PDG}_\Bs \, \vec{L}_\xy \cdot \vec{p}_\text{T} / p^2_\text{T}$, where $m^\text{PDG}_\Bs$ is the world-average \Bs mass~\refPDG and $\vec{L}_\xy$ is the vector from the PV to the \Bs vertex in the transverse plane.
The \ct uncertainty \sct is obtained by propagating the uncertainties in $L_\xy$ and \pt of the \Bs meson candidate to \ct.
It is required that  $\ct \geq 60\,(100)\mum$ for the \setJpsiMu (\setJpsiTrk) trigger category, with $\sct < 50\mum$.
Additionally, events in the \setJpsiTrk trigger category need to satisfy $\ct/\sct>3$.
In approximately 2\% of the events, multiple \Bs meson candidates are reconstructed, and in these cases the candidate with the highest vertex-fit probability is chosen.

\section{Flavor tagging}\label{app:b}

The OS muon and OS electron taggers follow the same strategy.
They search for semileptonic $\PQb \to \ell^- + \PX$ decays of the OS \PQb quark, using the lepton charge as the tagging variable and subsequently employing DNNs to estimate \wtag.
For each event, we search for a candidate OS lepton consistent with being produced by the OS \PQb hadron.
To reduce contamination from background sources, loose kinematic requirements are imposed on the candidates, and DNN discriminators are used to reduce contributions from hadrons and photons misreconstructed as muons and electrons, respectively.
To avoid overlaps, an OS electron is searched only if no OS muon is found.
In both algorithms, conditions are introduced to separate the OS lepton from the \Bs meson candidate.
An ensemble of DNNs is used to discriminate the right- and wrong-tag leptons, as well as to estimate the event mistag probability.
The DNNs have been developed using simulated \Bsjpsiphi events and calibrated with self-tagging \BujpsiK data samples, as described below.
The input features of the lepton tagging DNNs consist of variables extracted from the kinematic properties of the lepton candidate and the surrounding activity.
The structure of all DNN models has been optimized to maximize the tagging power.
The architecture is purposefully designed so that its output score ``$s_\text{DNN}$'' corresponds directly to the probability of correctly tagging an event.

The OS jet tagger is based on DNNs, referred to collectively as ``DeepJetCharge'' (DJC), derived from the \textsc{DeepSets} architecture~\cite{zaheer2018deep} and exploits charge asymmetries in OS \PQb jets to predict the probability of being produced by \PQb versus \PAQb quarks.
The resulting probability serves as the basis for determining both \xitag and \wtag.
An OS jet candidate is selected if it passes loose kinematic requirements, does not contain tracks from the signal \Bs, the probability $P_{\PQb}$ of originating from a bottom or anti-bottom quark, as estimated by \textsc{DeepJet}~\cite{BTV-16-002} exceeds 20\%, and contains at least two tracks from the PV.
To ensure the OS algorithms are orthogonal to each other, an OS jet is searched for only if no OS muon or OS electron is found.
Features used by DJC include \pt and $\eta$ of the \Bs meson, \pt, $\eta$, and $P_{\PQb}$ of the jet, and the angular distances $\dR \equiv \sqrt{\smash[b]{(\Delta\eta)^2 + (\Delta\phi)^2}}$ and $\Delta\phi$ between the direction of the jet and that of the \Bs.
Features from at most 15 charged particles (ordered by the longitudinal impact parameter relative to the PV, \IPz) are also included: charge, \pt, $\eta$, \dR and $\Delta\phi$ relative to the \Bs momentum, impact parameters with respect to the PV, and their uncertainties.
Besides the jet constituents, other tracks in a $\dR < 0.5$ cone around the jet axis are used to recover tracks not found by the jet reconstruction.
The DJC is trained on simulated \Bsjpsiphi events of the \setJpsiTrk trigger category and its structure is composed of four blocks.
First, three dense layers act on every input feature except track charge, making this block explicitly flavor invariant.
Then, the charge information is added and two more dense layers are executed.
Subsequently, all the hidden features are summed across all tracks, flattening the track vector to a single dimension.
Finally, four interlaced dense and dropout layers convert the feature vector to a probability, which is the model output.
All dense layers use an exponential linear unit activation function~\cite{clevert2016fast}.
To preserve symmetry between the two classes, that is \PQb and \PAQb jets, the algorithm is constructed as flavor invariant, that is the probabilities for the \PQb hypothesis are identical to those of the \PAQb hypothesis when inverting all the charges.
This is achieved by explicitly symmetrizing the output as $s_\text{DJC}(x) = [s_\text{DJC}(x) + (1-s_\text{DJC}(\overline{x}))]/2$, where $x$ is the model input and $\overline{x}$ is the input with all the track charges reversed.
The output of DJC coincides with the probability of the OS jet being produced by a bottom quark, which can be straightforwardly used to perform the tagging inference and \wtag estimation.

The SS tagger, referred to as ``DeepSameSide'' (DSS), exploits the correlation between the quark content of the \Bs meson at production time and the charges of the fragmentation particles surrounding it.
Consequently, it is independent of the OS tag, is not diluted by mixing in the other \PQb quark, and does not depend on the presence of its decay products from the other \PQb quark, such as leptons or jets, within the acceptance of the CMS detector.
The architecture of DSS is modeled on that of DJC (including output symmetrization), exploiting the features from the \Bs meson candidate and all the tracks, other than its constituents, in a $\dR = 0.8$ cone around its direction (care is taken to avoid geometrical overlap with the OS methods).
Up to 20 tracks, ordered by \IPz, are considered; in contrast to DJC, no jet features are used.
The DSS is trained on a sample of \Bs and \Bu simulated events to compute the probability of having a bottom quark at production time.
The \Bu events are used to render the algorithm invariant for $\Bs \leftrightarrow \Bm$ and $\Bsbar \leftrightarrow \Bu$ transformation and allow the calibration in data with \BujpsiK decays.
The DSS exploits the charge of the particles produced in the hadronization for the discrimination, which depends on the light quark in the meson.
The DSS is trained to produce the probability of an initial signal \Bs meson, from which the tagging inference and \wtag estimation can be then performed.

The event mistag probability, as estimated by the tagging algorithms, undergoes calibration utilizing a self-tagging data sample of \BujpsiK events, where the kaon's charge corresponds to the \PBp meson flavor.
The calibration procedure is executed separately for each data set, by comparing the measured mistag fraction (\wmeas) with the \wtag predicted by the taggers.
The \Bu events are divided into bins in \wtag, and, within each bin, the count of right- and wrong-tag events is computed independently on the $\jpsi\,\PKp$ invariant mass distribution to derive the corresponding \wmeas value.
The $(\wtag,\, \wmeas)$ pairs are then used to refine the DNNs' predictions through the Platt scaling method~\cite{platt1999probabilistic}.
This method fits a linear function to the model's output before the final sigmoid layer, effectively calibrating the predicted probabilities to the measured outcomes.
In the case of the OS lepton taggers, both parameters of the linear function are left free to float, while for the OS jet and SS tagger, the constant term is set to zero to preserve symmetry between the two flavor hypotheses.
The calibration of DSS is further corrected to account for potential differences between \Bs and \Bu mesons by using the calibrations obtained in simulated \Bsjpsiphi and \BujpsiK events.
The size of this correction is found to be of the order of 10\%.
For the OS-jet and SS tagger, events with  $\wtag > 0.48$ and $\wtag > 0.46$, respectively, are classified as untagged.
These events contribute minimally to the performance, and their removal enhances the calibration reliability.

All tagging DNNs are developed in simulated events and employ the cross-entropy loss function combined with the Adam optimizer~\cite{kingma2014adam}.
Input features are preprocessed so that the mean and variance for each of the input variables are, respectively, zero and one.
The four tracks forming the \Bs meson candidate are explicitly excluded from consideration by the taggers and precautions are taken to reduce the effect of multiple collisions in the same or neighboring bunch crossings.

The combined performance of the tagging framework yields a tagging power of $\Ptag = (5.59 \pm 0.02)\%$.
The tagging efficiency, tagging power, and effective dilution for each exclusive tagging category are presented in Table~\ref{tab:tagperf}.

\begin{table}[ht!]
    \centering
    \topcaption{Calibrated flavor tagging performance as measured in the \Bs data sample in the various mutually exclusive categories.
    The effective dilution $\mathcal{D}_\text{tag,eff}^2$ is obtained from the measured tagging efficiency \etag and tagging power \Ptag.
    The uncertainties are statistical only.}
    \begin{scotch}{ l c c c }
        Category           & $\etag\ [\%]$      & $\mathcal{D}_\text{tag,eff}^2$  & $\Ptag\ [\%]$ \\
        \hline
        Only OS muon       & $\hphantom{3}6.07 \pm 0.05$    & $0.212$    & $1.29 \pm 0.07$   \\
        Only OS electron   & $\hphantom{3}2.72 \pm 0.02$    & $0.079$    & $0.214 \pm 0.004$ \\
        Only OS jet        & $\hphantom{3}5.16 \pm 0.03$    & $0.045$    & $0.235 \pm 0.003$ \\
        Only SS            & $33.12 \pm 0.07$               & $0.080$    & $2.64 \pm 0.01$   \\
        SS + OS muon       & $\hphantom{3}0.62 \pm 0.01$    & $0.202$    & $0.125 \pm 0.003$ \\
        SS + OS electron   & $\hphantom{3}2.77 \pm 0.02$    & $0.150$    & $0.416 \pm 0.005$ \\
        SS + OS jet        & $\hphantom{3}5.40 \pm 0.03$    & $0.124$    & $0.671 \pm 0.006$ \\ [\cmsTabSkip]
        Combined           & $55.9 \pm 0.1$                 & $0.100$    & $5.59 \pm 0.02$   \\
    \end{scotch}
    \label{tab:tagperf}
\end{table}

\ifthenelse{\boolean{cms@external}}{}{
\clearpage
\numberwithin{table}{section}
\numberwithin{figure}{section}
\section{Additional tables and figures\label{app:suppMat}}

\input{supplemental_material}

}
\cleardoublepage \section{The CMS Collaboration \label{app:collab}}\begin{sloppypar}\hyphenpenalty=5000\widowpenalty=500\clubpenalty=5000\input{BPH-23-004-public-authorlist.tex}\end{sloppypar}
\end{document}

%% file: supplemental_material.tex
The matrices of the statistical and systematic uncertainty correlations between pairs of physics parameters, as obtained from the analysis of the 13\TeV data, are reported in Tables~\ref{tab:corrStat} and~\ref{tab:corrSyst}.

The values and uncertainties of the physics parameters obtained from the combination of the CMS 8\TeV and 13\TeV results using the BLUE method are provided in Table~\ref{tab:combRes}, while the matrix of the correlations between the physics parameters is reported in Table~\ref{tab:corrComb}.

\begin{figure*}[!ht]
  \centering
  \includegraphics[width=0.31\textwidth]{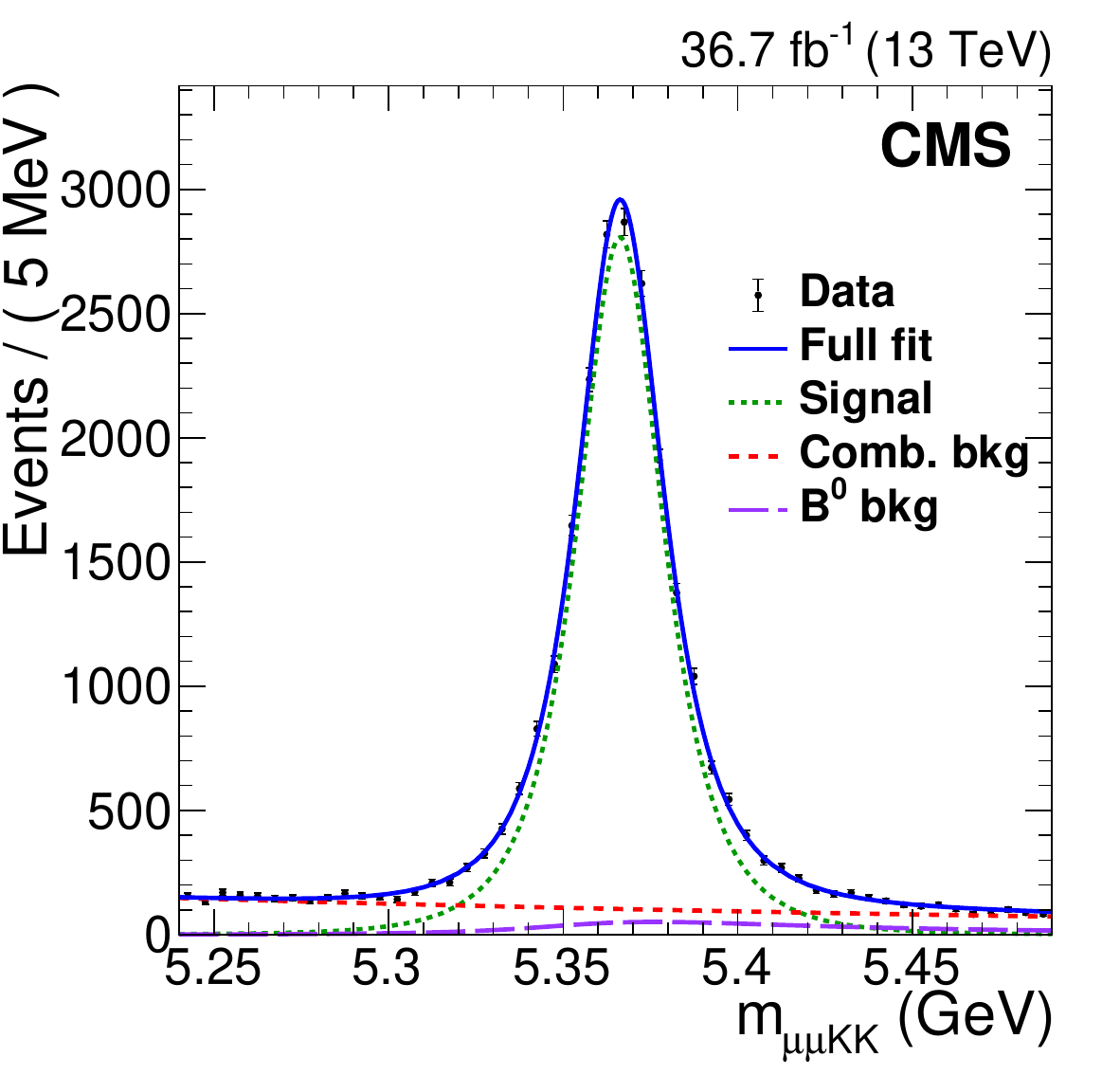} \hfil
  \includegraphics[width=0.31\textwidth]{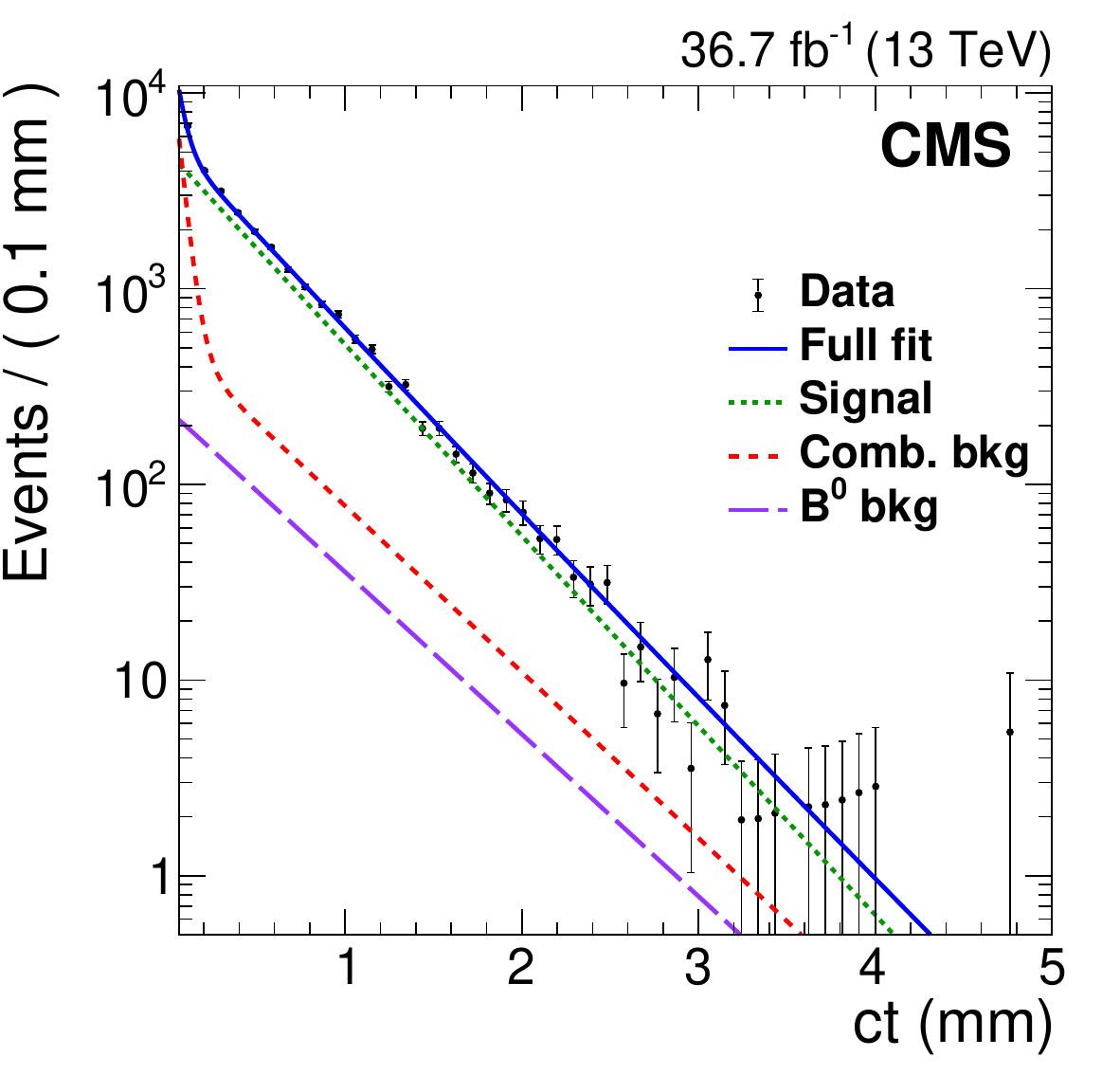} \hfil
  \includegraphics[width=0.31\textwidth]{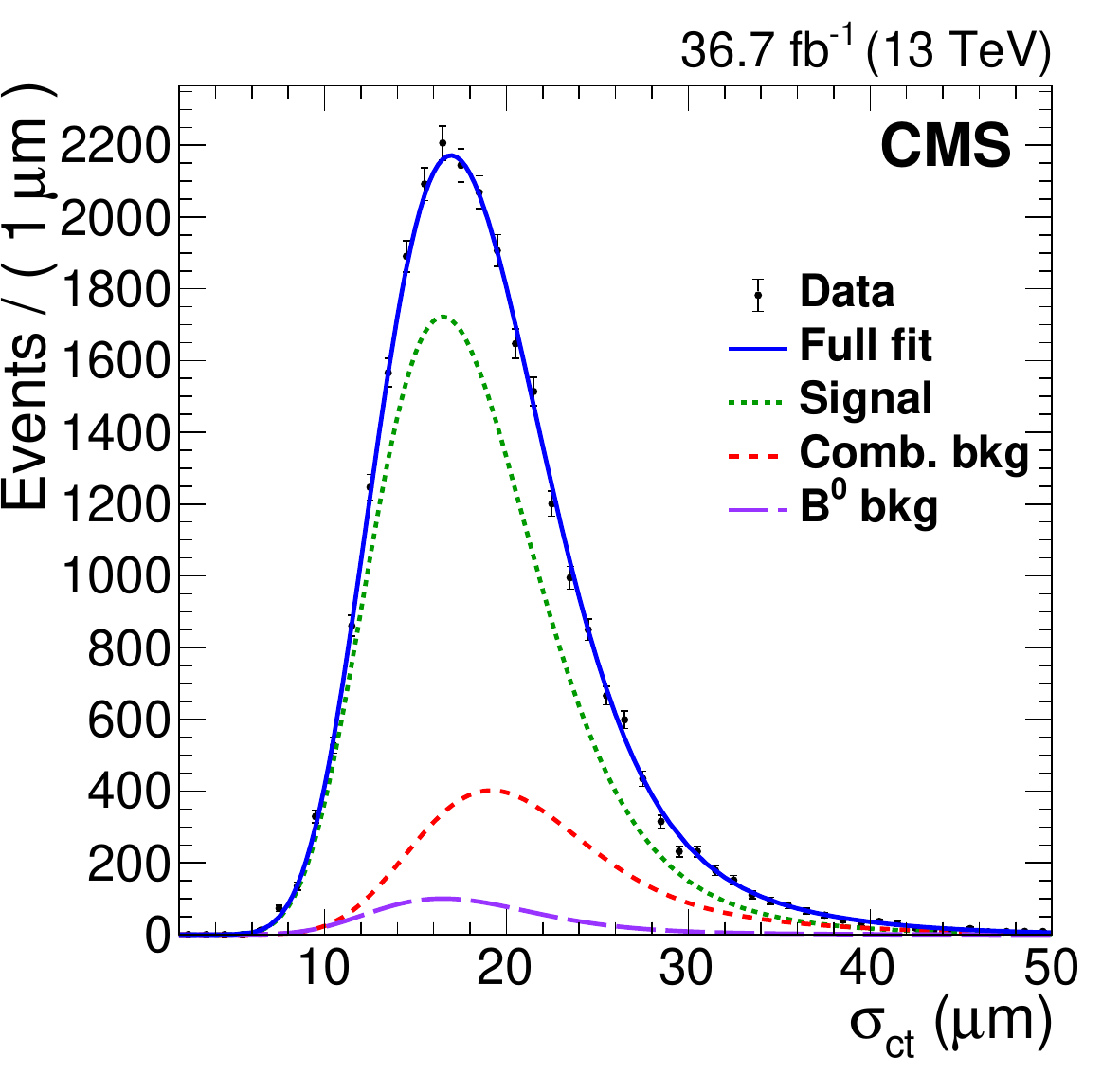} \hfil \\
  \includegraphics[width=0.31\textwidth]{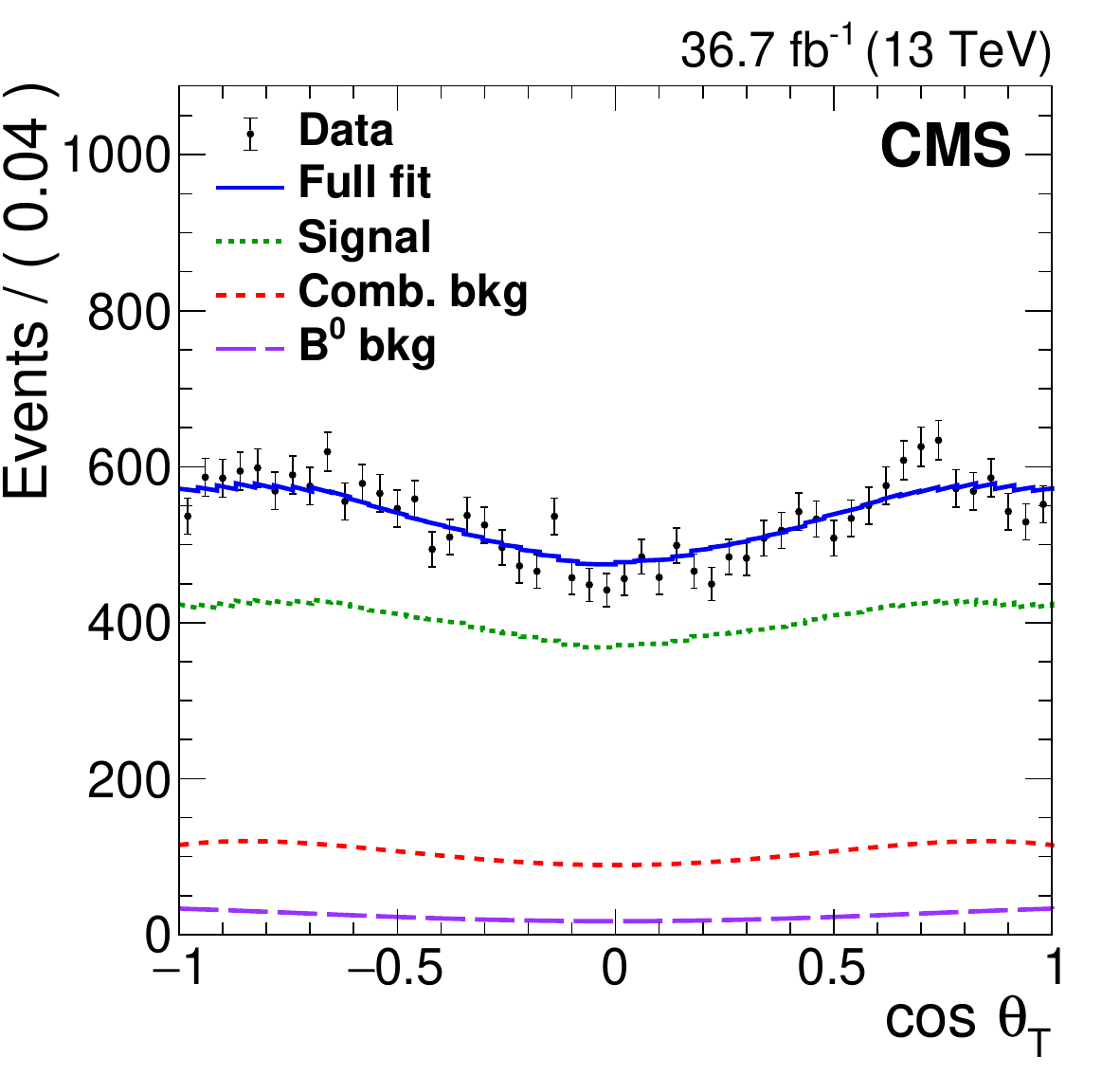} \hfil
  \includegraphics[width=0.31\textwidth]{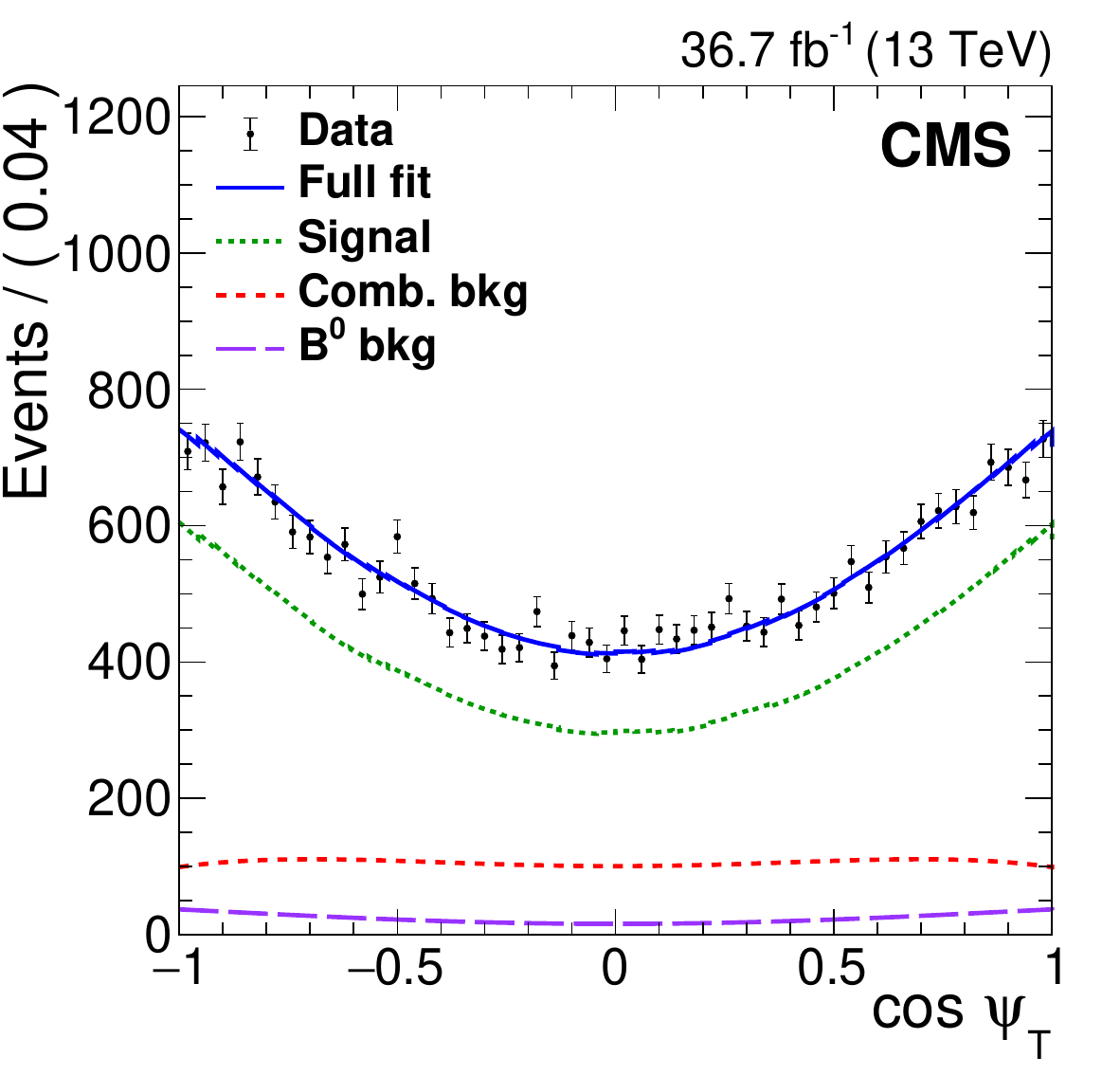} \hfil
  \includegraphics[width=0.31\textwidth]{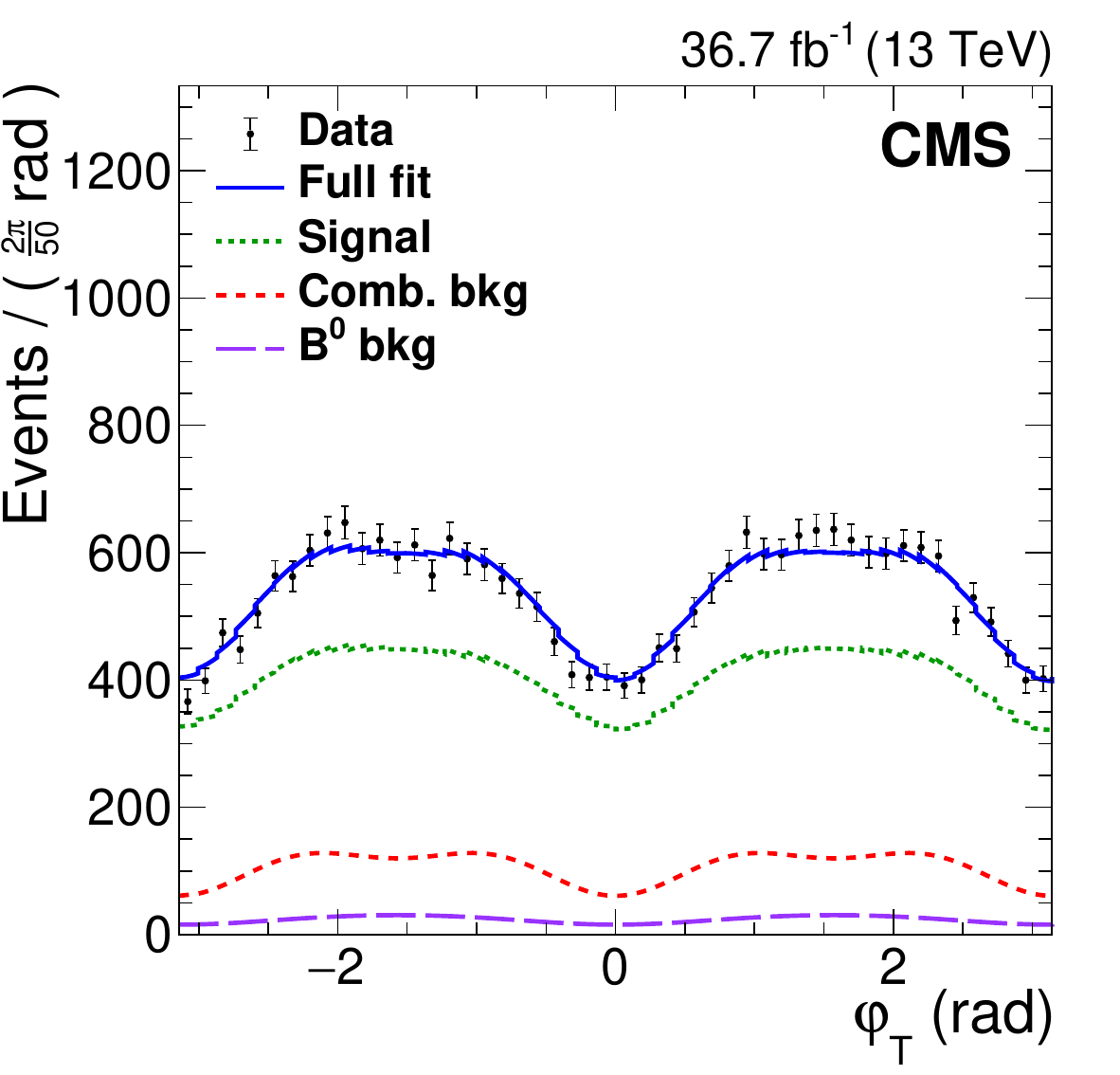} \\
  \includegraphics[width=0.31\textwidth]{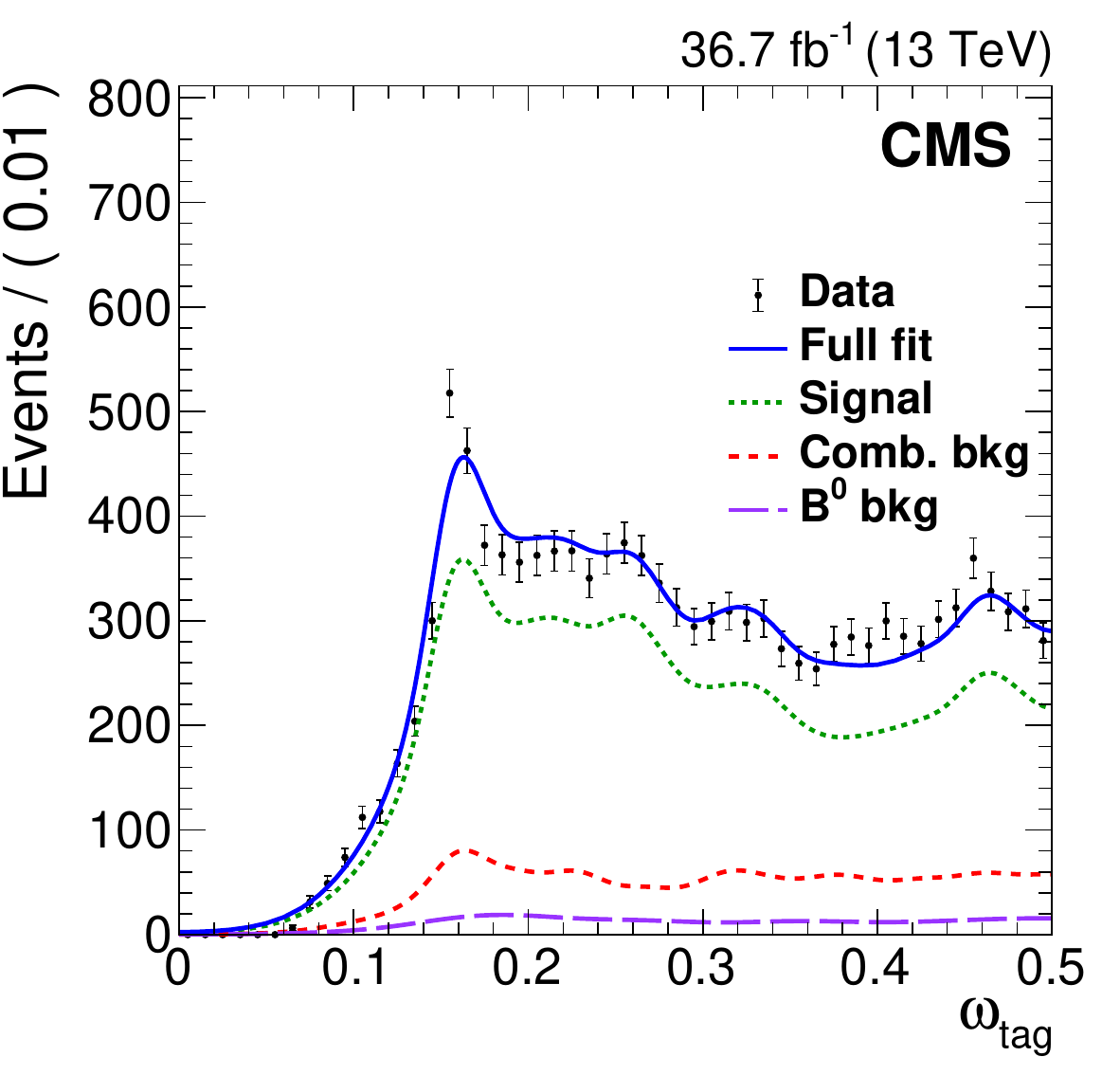}
  \caption{The distributions for the input observables of the selected candidates and the projections from the fit for the \setJpsiMu trigger category (2017 data).}
  \label{fig:projJpsiMuon17}
\end{figure*}

\begin{figure*}[!ht]
  \centering
  \includegraphics[width=0.31\textwidth]{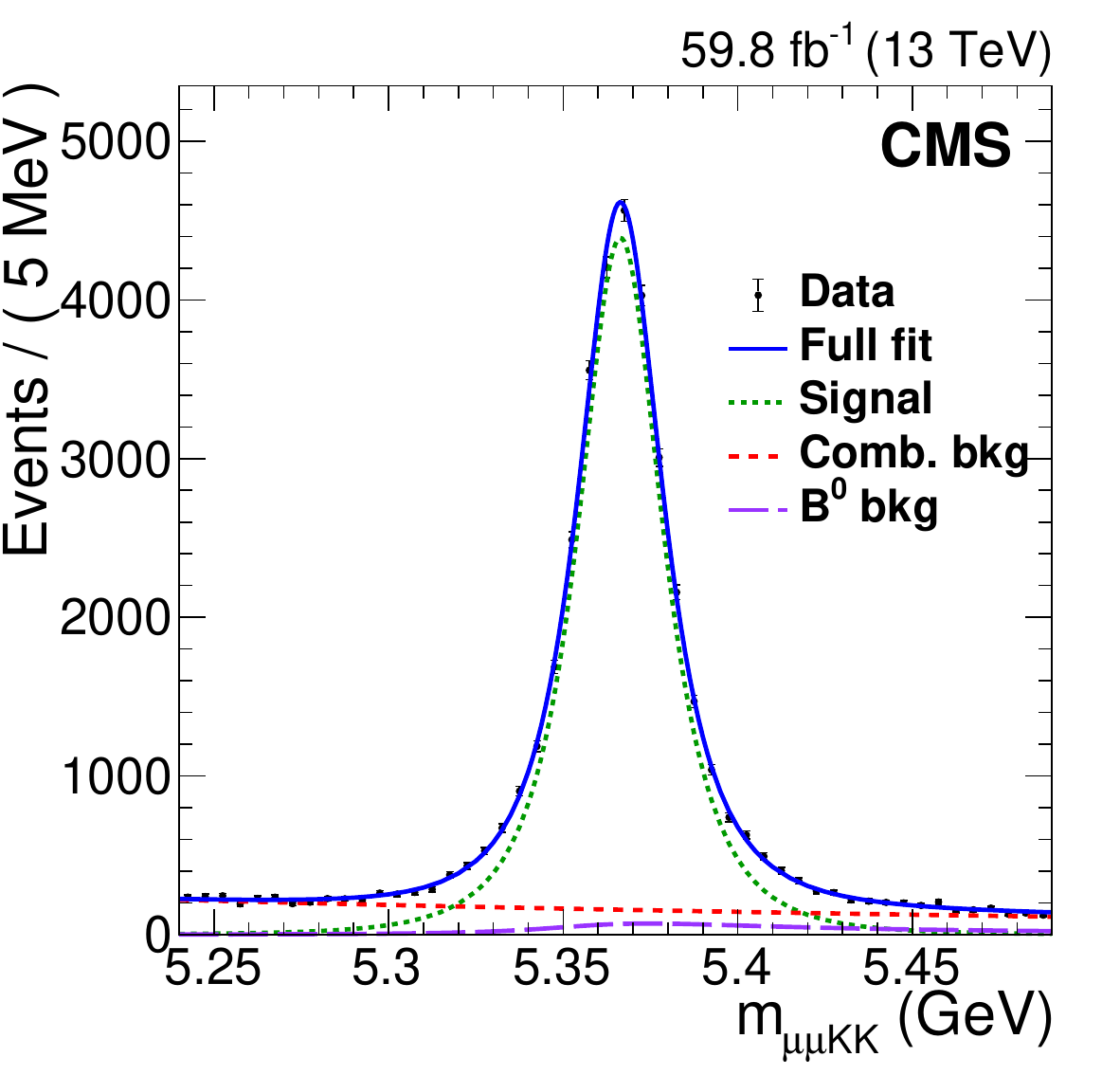} \hfil
  \includegraphics[width=0.31\textwidth]{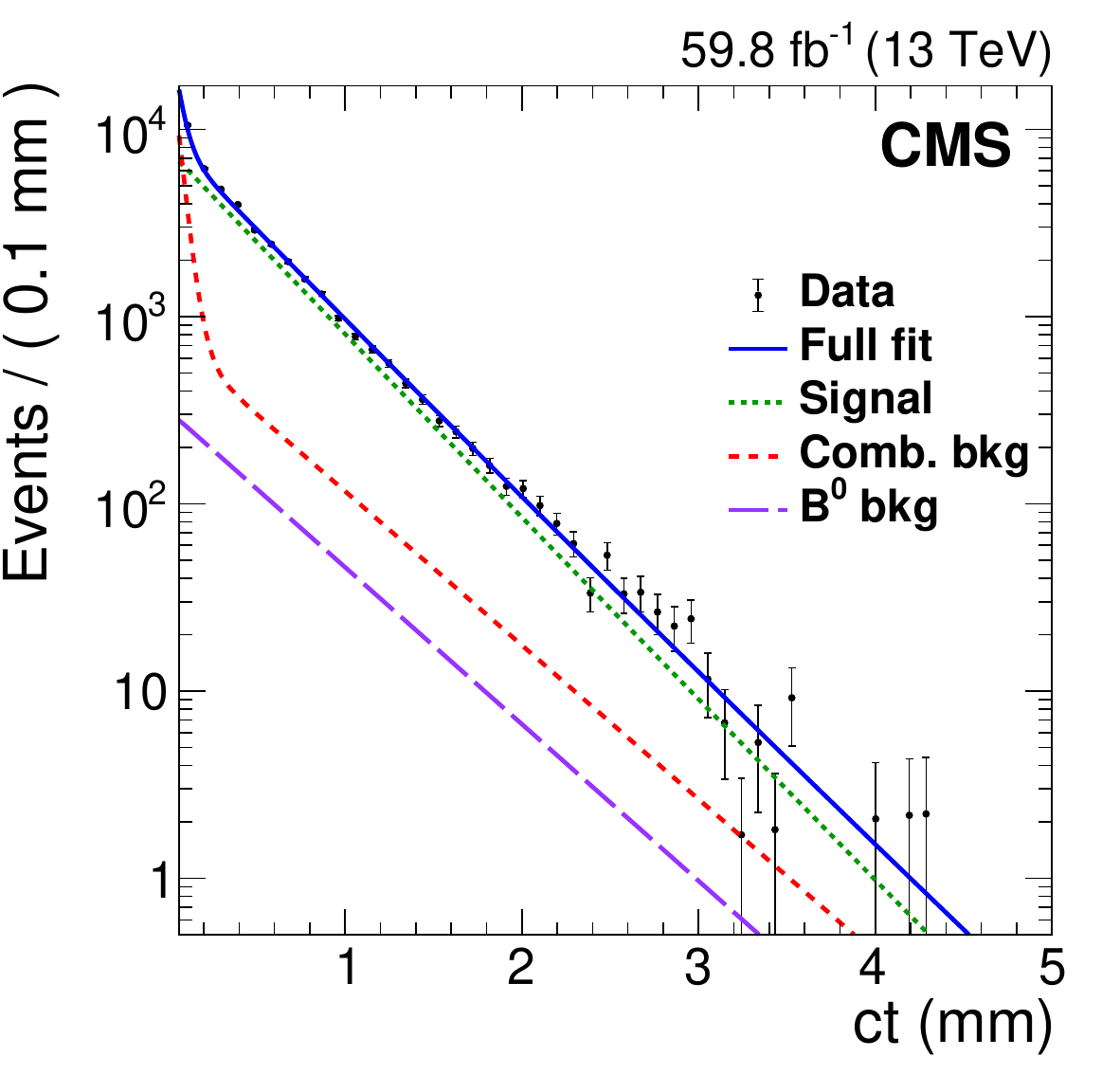} \hfil
  \includegraphics[width=0.31\textwidth]{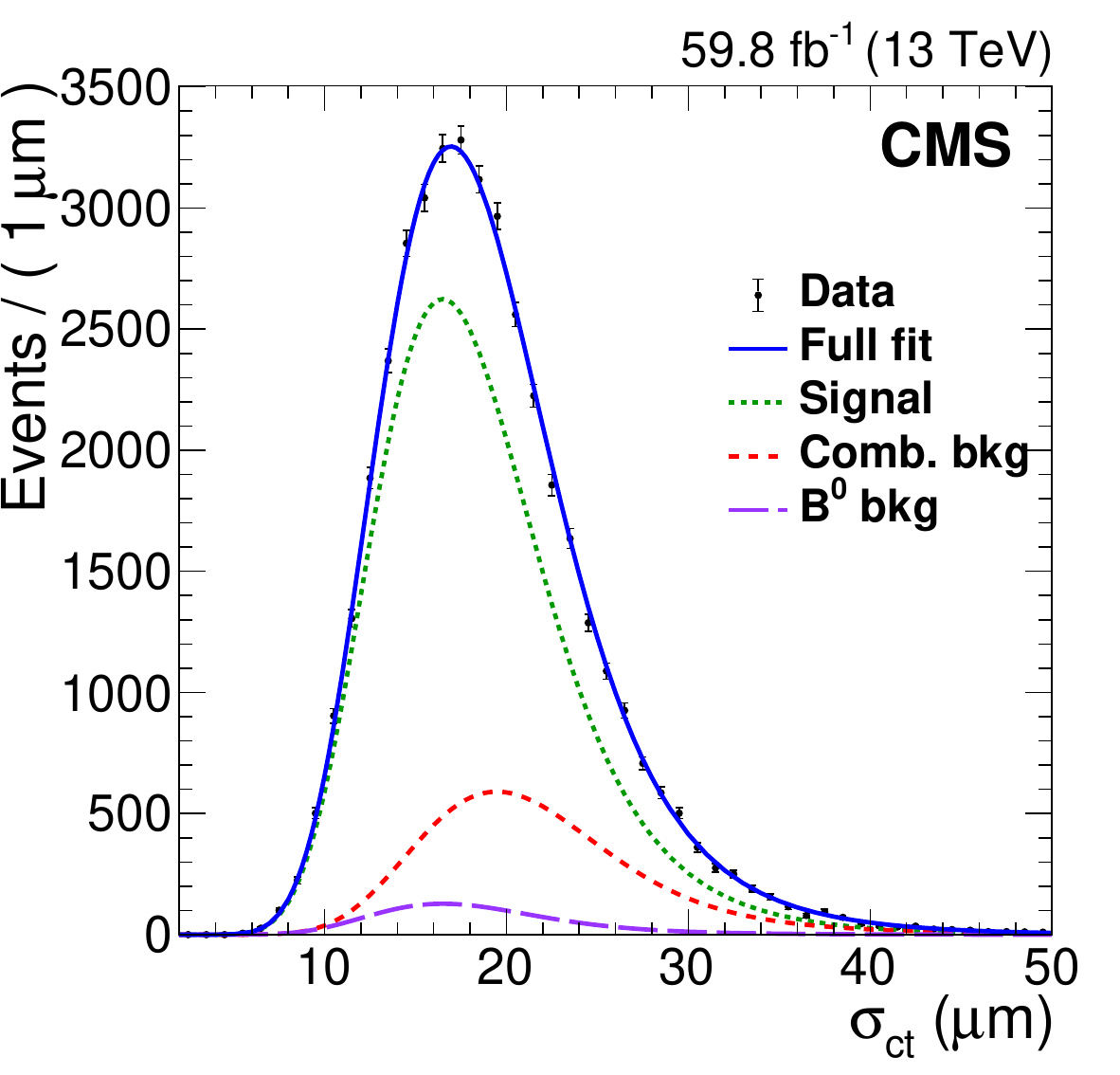} \\
  \includegraphics[width=0.31\textwidth]{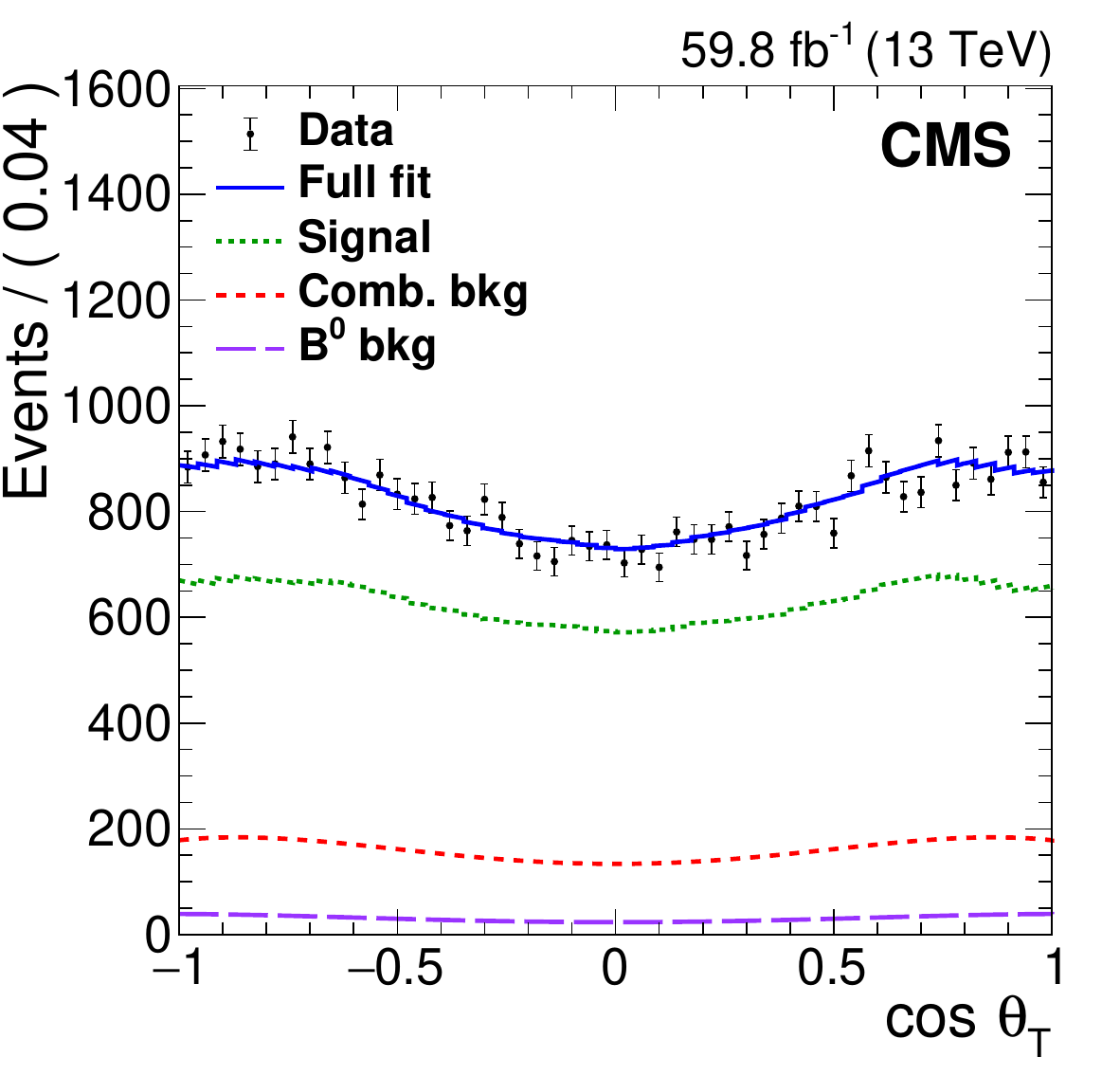} \hfil
  \includegraphics[width=0.31\textwidth]{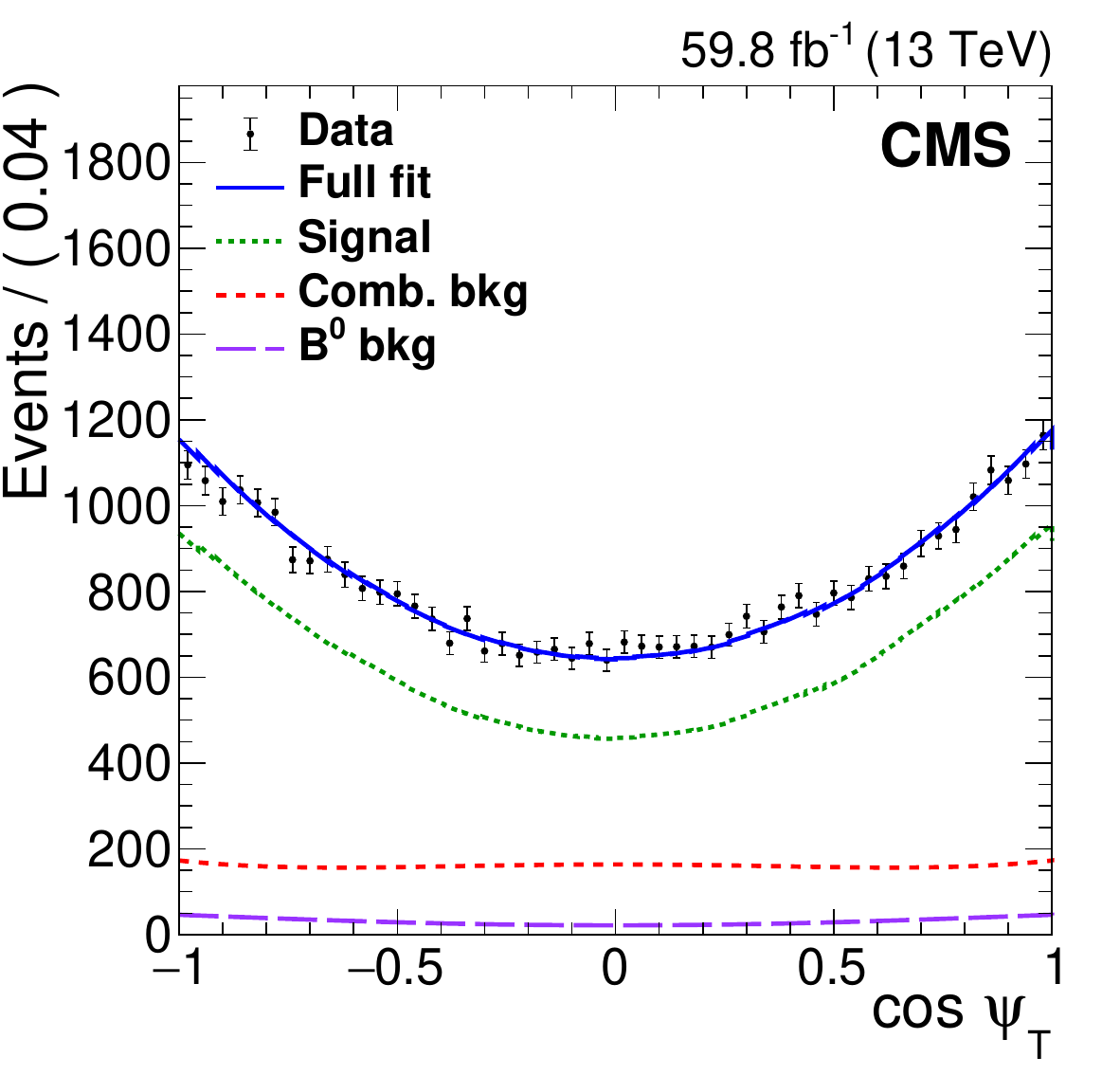} \hfil
  \includegraphics[width=0.31\textwidth]{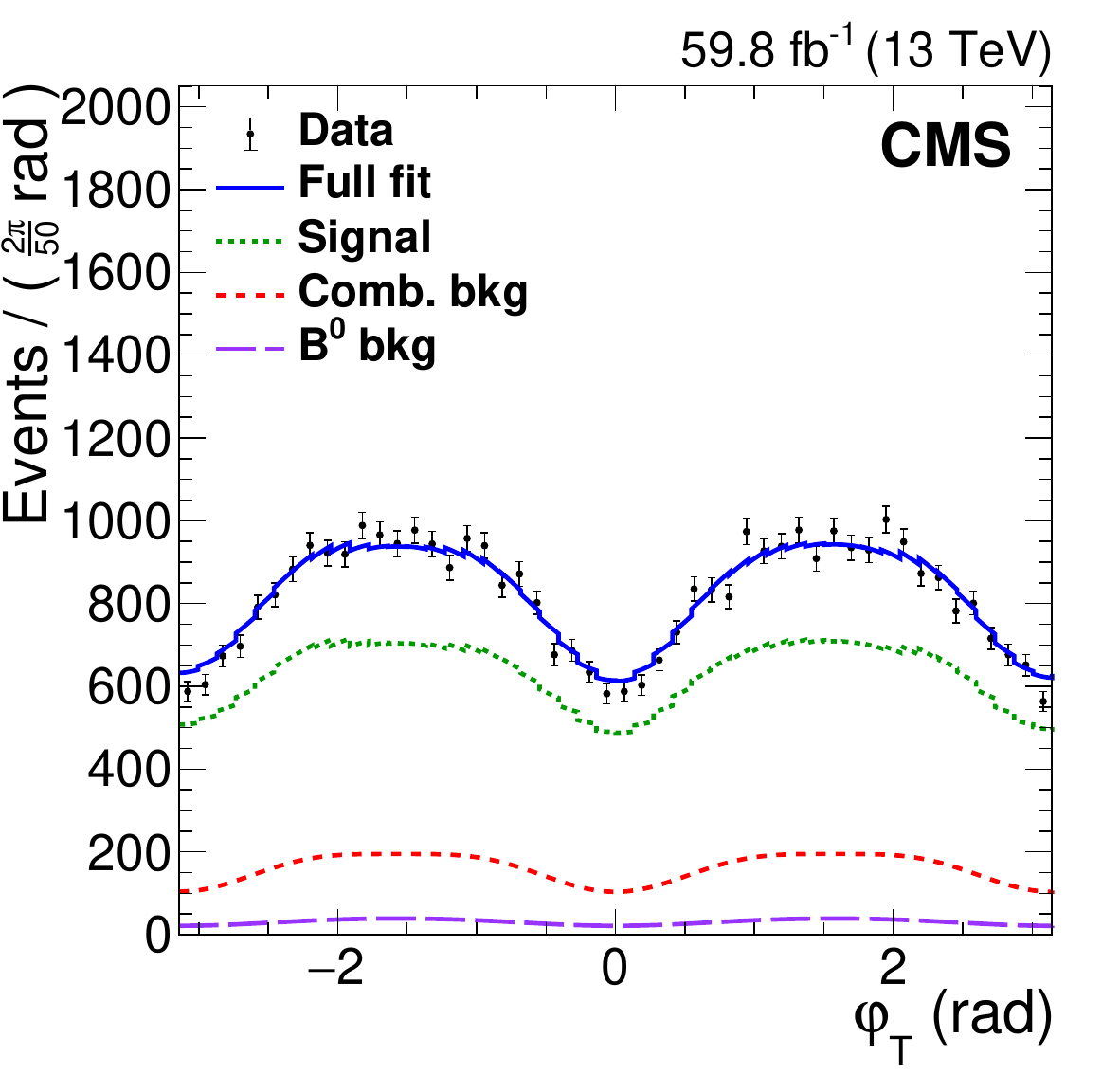} \\
  \includegraphics[width=0.31\textwidth]{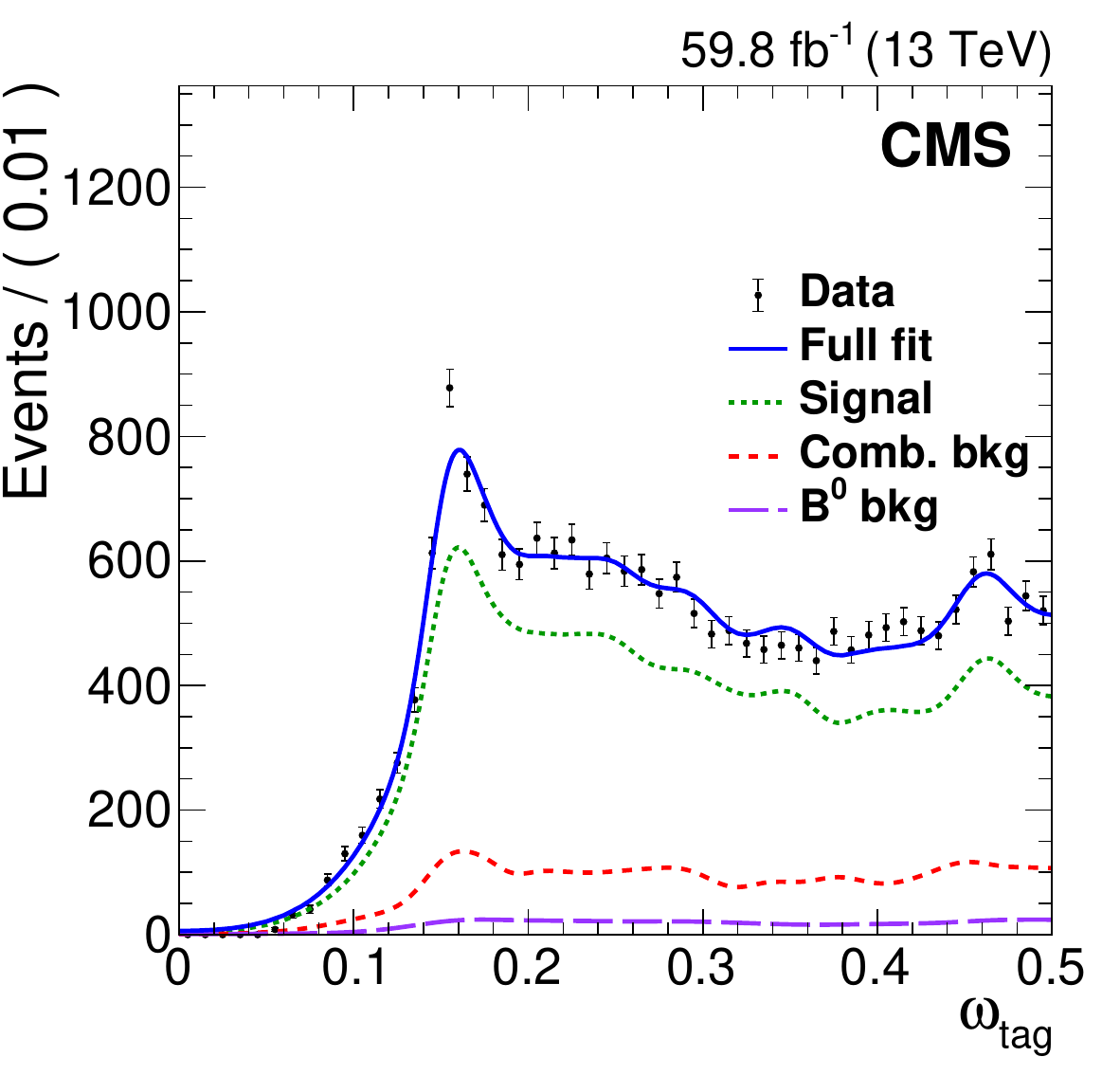}
  \caption{The distributions for the input observables of the selected candidates and the projections from the fit for the \setJpsiMu trigger category (2018 data).}
  \label{fig:projJpsiMuon18}
\end{figure*}

\begin{figure*}[!ht]
  \centering
  \includegraphics[width=0.31\textwidth]{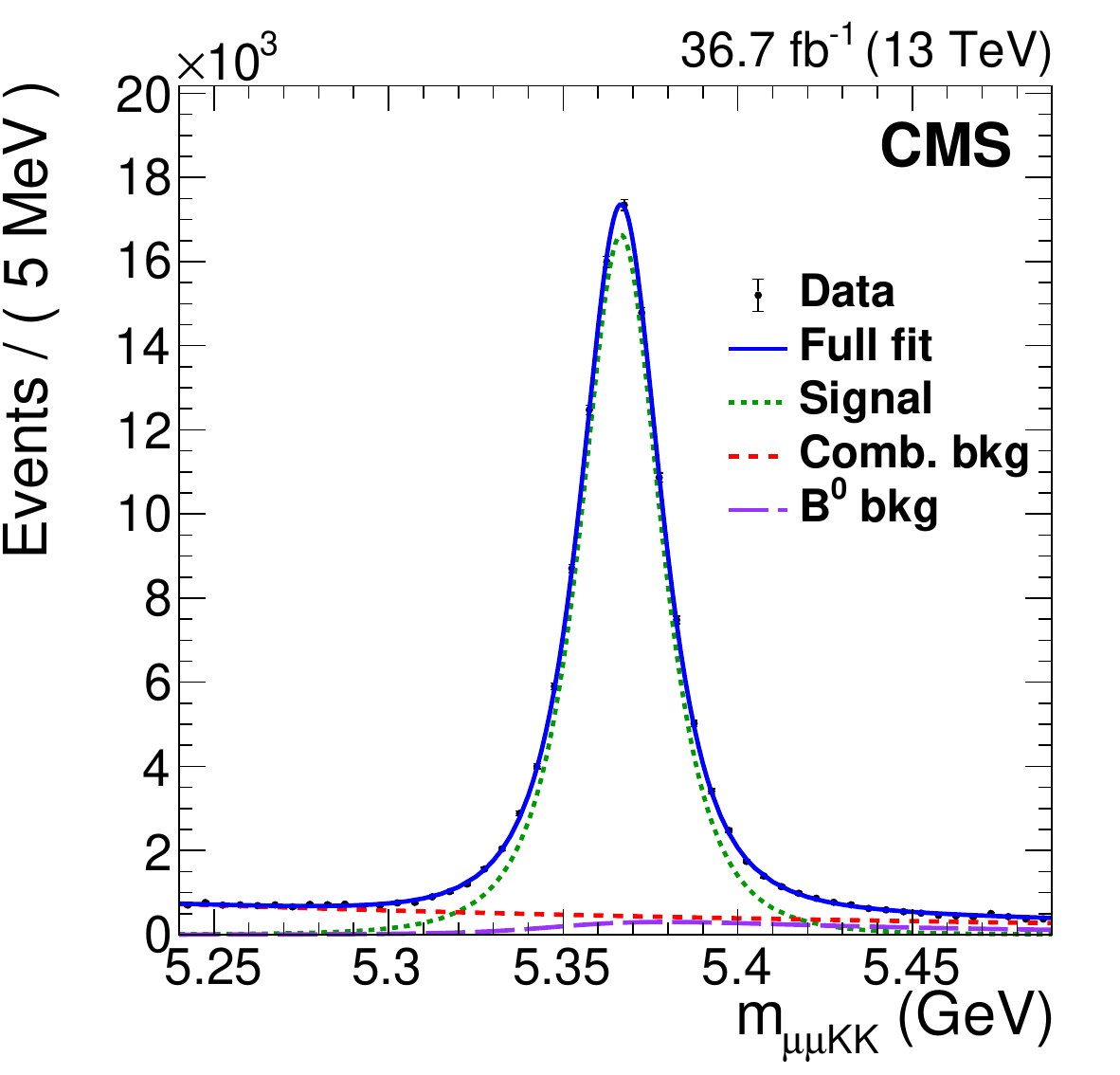} \hfil
  \includegraphics[width=0.31\textwidth]{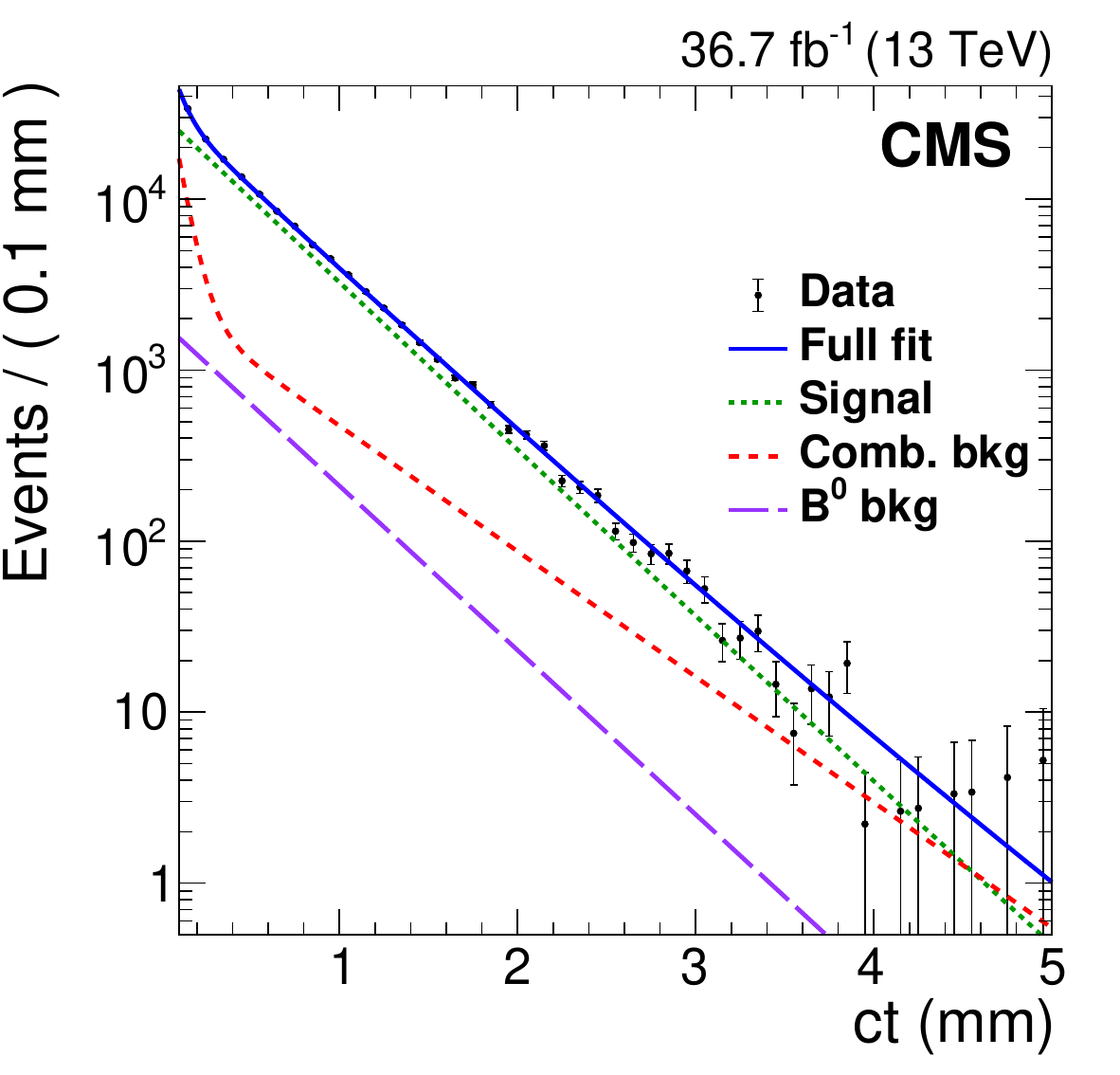} \hfil
  \includegraphics[width=0.31\textwidth]{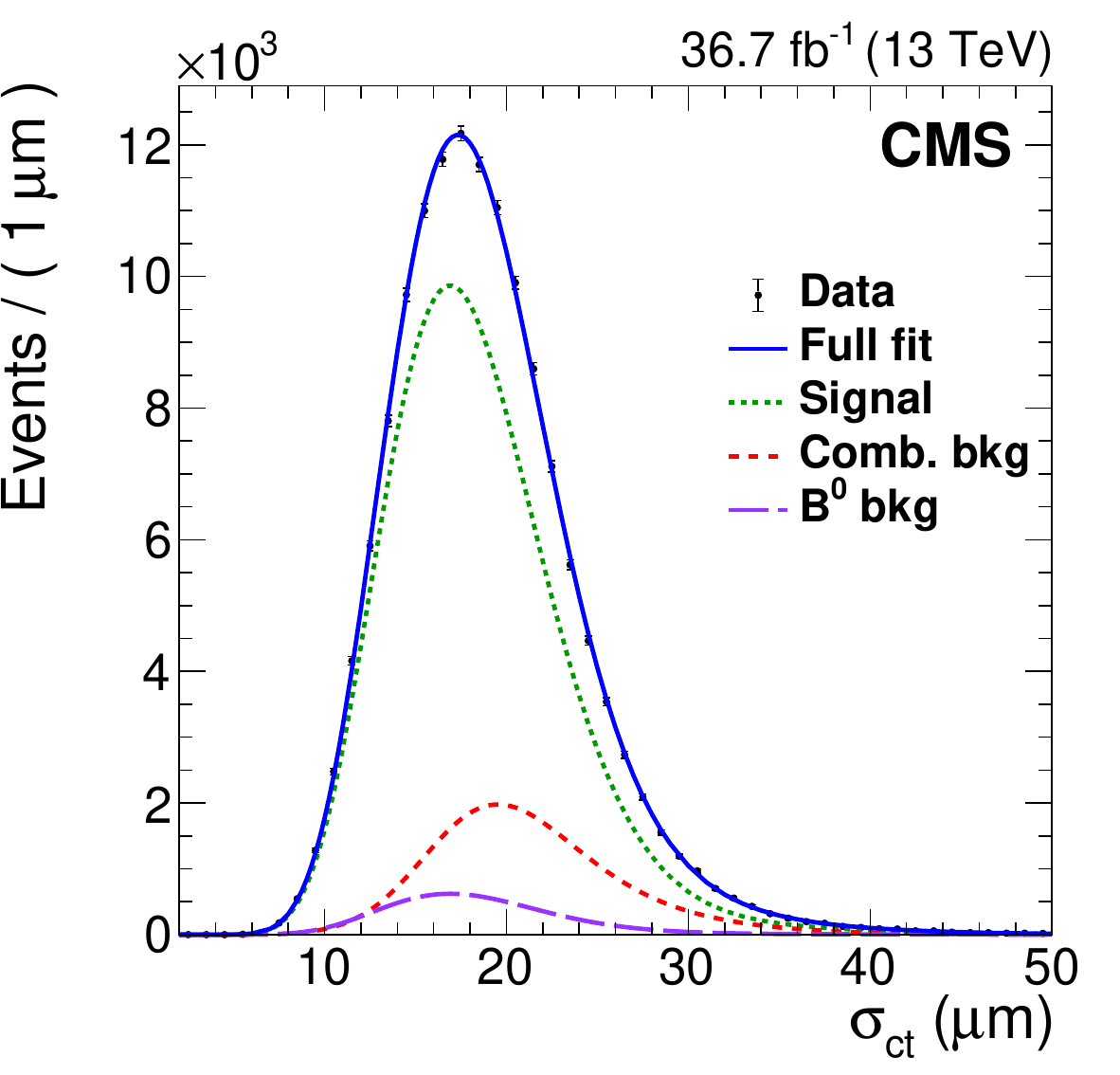} \\
  \includegraphics[width=0.31\textwidth]{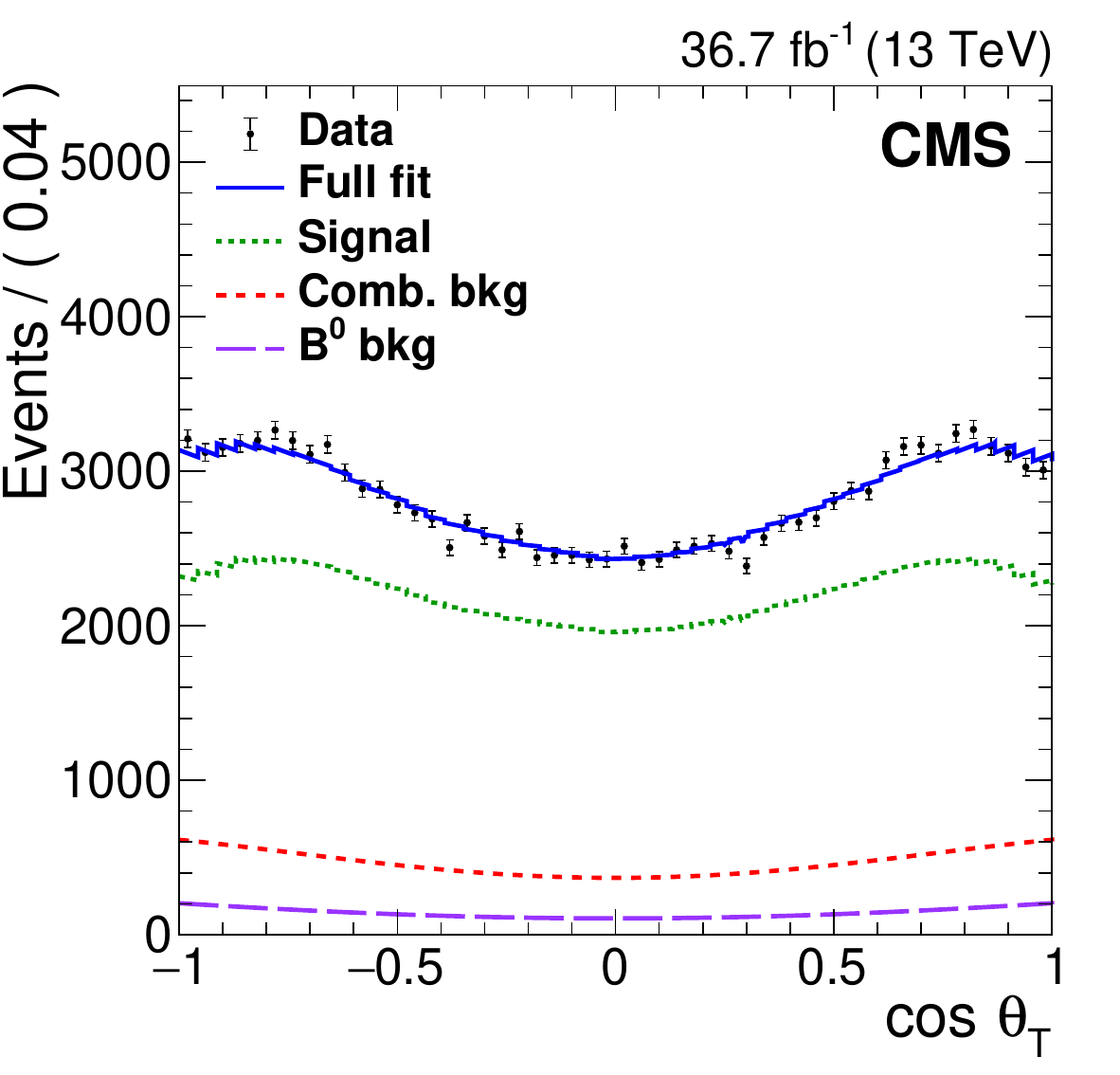} \hfil
  \includegraphics[width=0.31\textwidth]{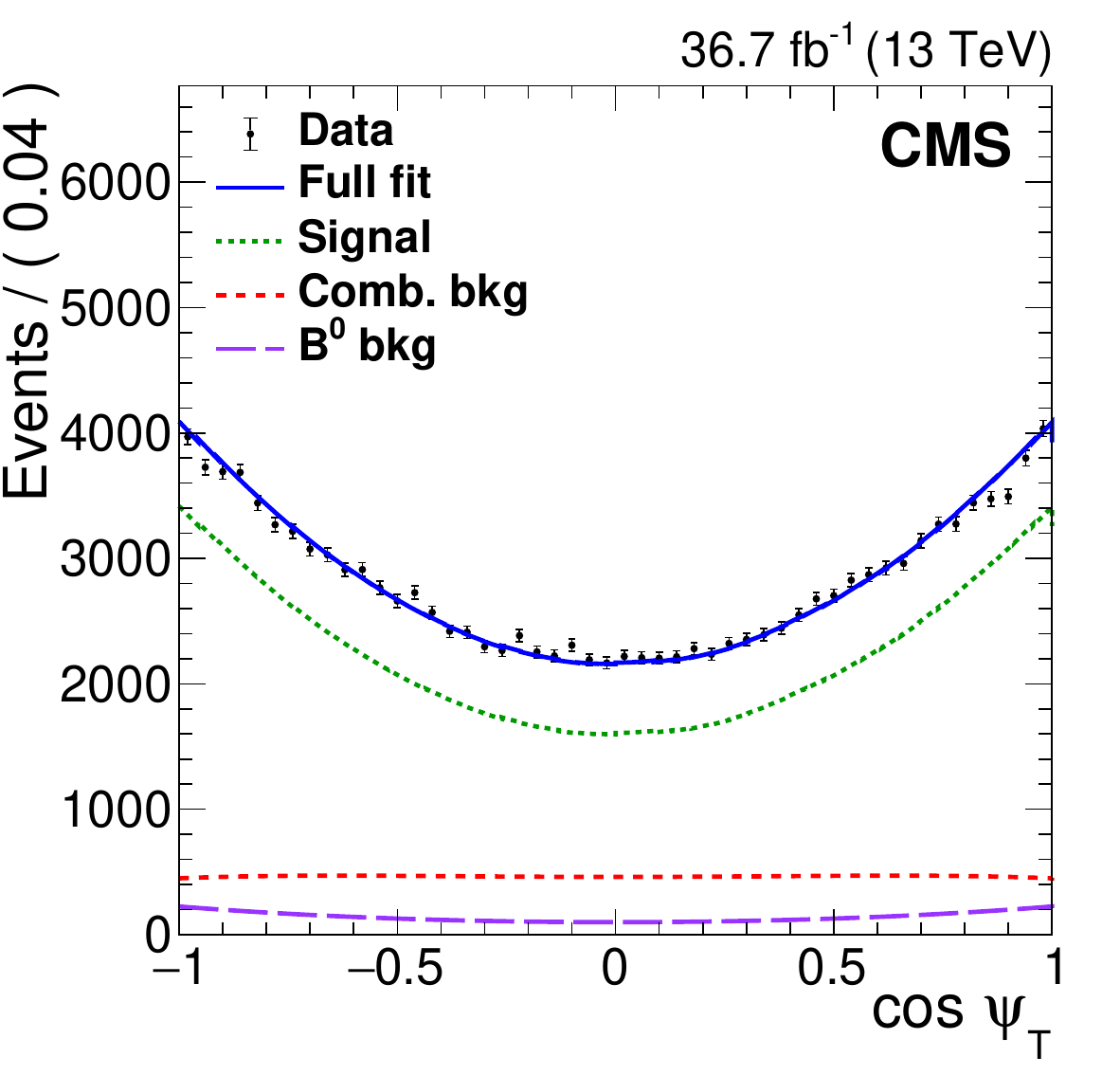} \hfil
  \includegraphics[width=0.31\textwidth]{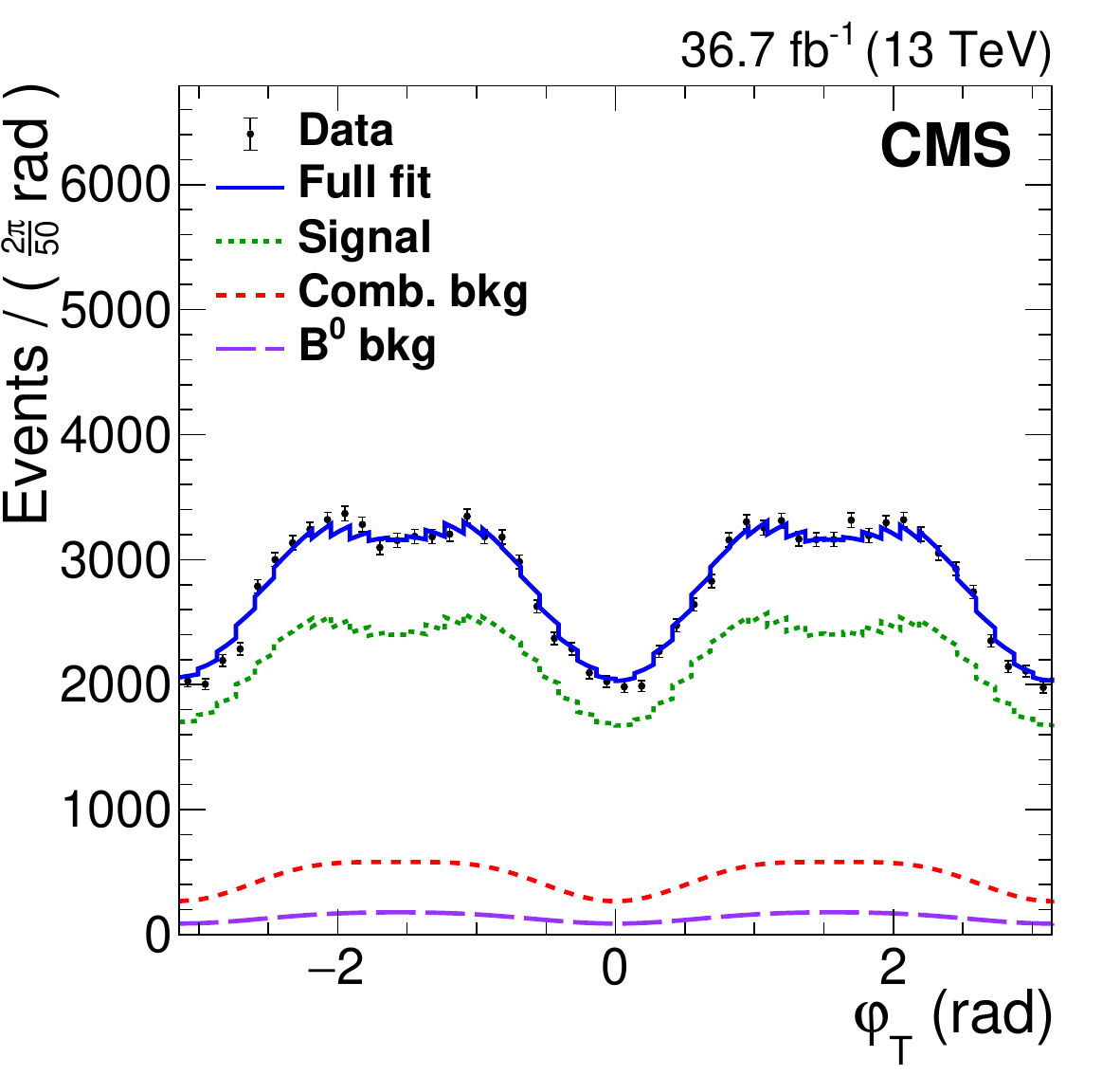} \\
  \includegraphics[width=0.31\textwidth]{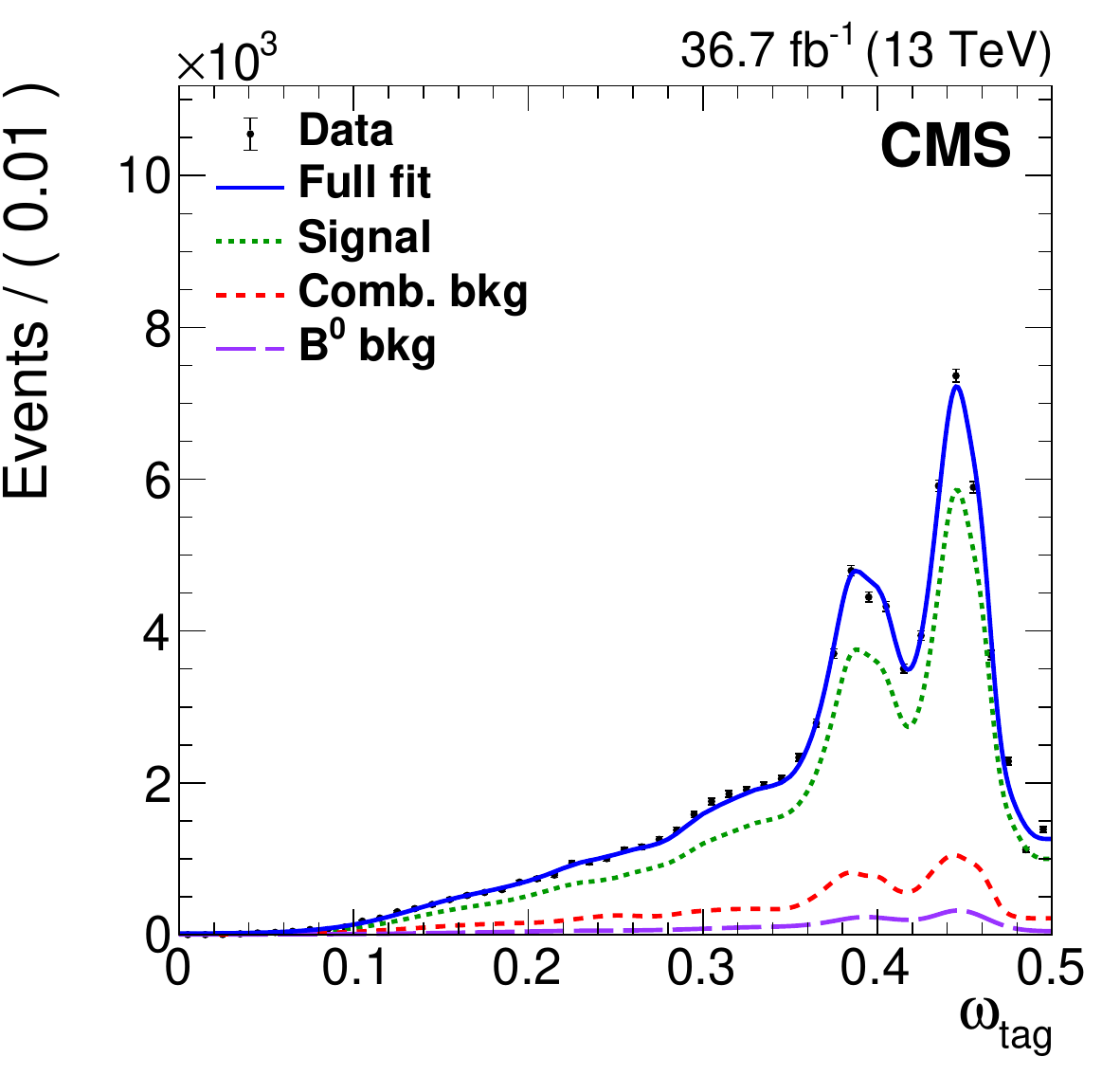}
  \caption{The distributions for the input observables of the selected candidates and the projections from the fit for the \setJpsiTrk trigger category (2017 data).}
  \label{fig:projJpsiTrk17}
\end{figure*}

The distributions for the input observables of the selected candidates and the projections from the fit for the \setJpsiMu categories and the \setJpsiTrk trigger category (2017 data) are shown in Figs.~\ref{fig:projJpsiMuon17}, \ref{fig:projJpsiMuon18}, and~\ref{fig:projJpsiTrk17}.

\begin{figure*}[!ht]
  \centering
  \includegraphics[width=0.45\textwidth]{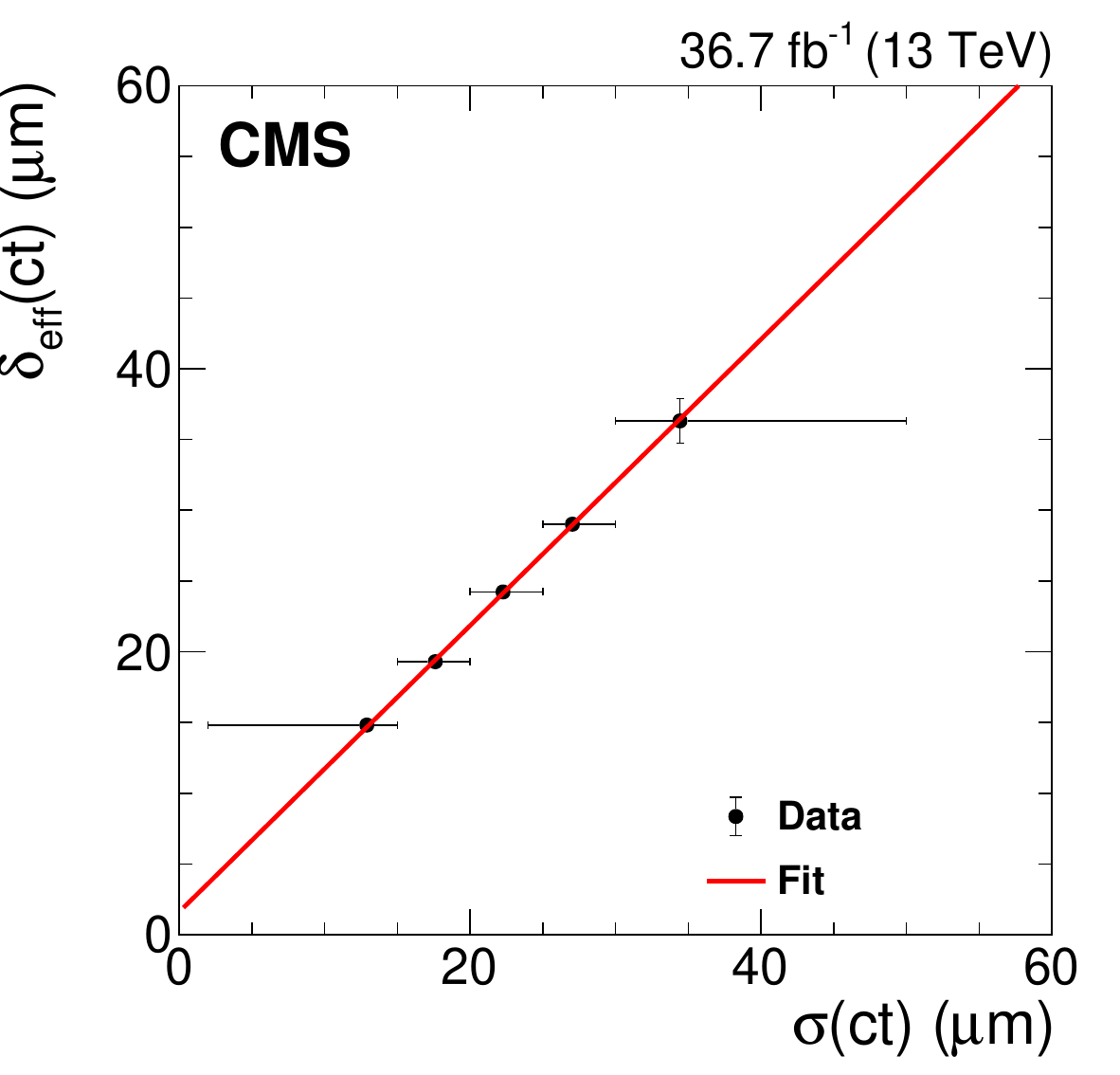} \hfil
  \includegraphics[width=0.45\textwidth]{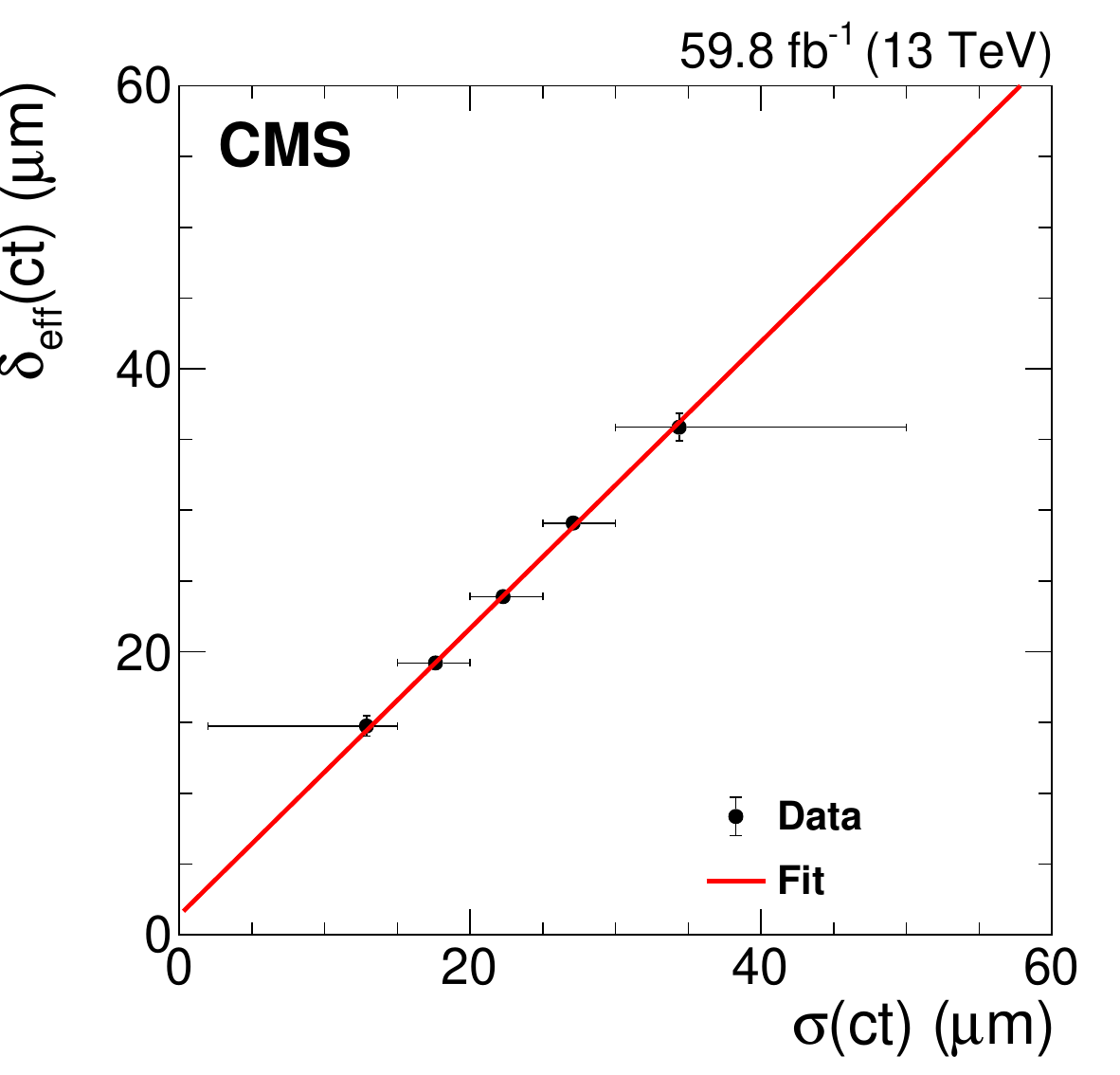}
  \caption{The \ct uncertainty calibration fits for the 2017 (left) and 2018 (right) data samples.
  The measured effective \ct resolution is plotted versus the average (calculated) value of \sct in bins of \sct.
  The results of a fit to a line are shown with a solid red line.
  The horizontal error bars indicate the bins'edges and are not used in the fit.}
  \label{fig:ctrescal}
\end{figure*}

\begin{figure*}[!ht]
  \centering
  \includegraphics[width=0.45\textwidth]{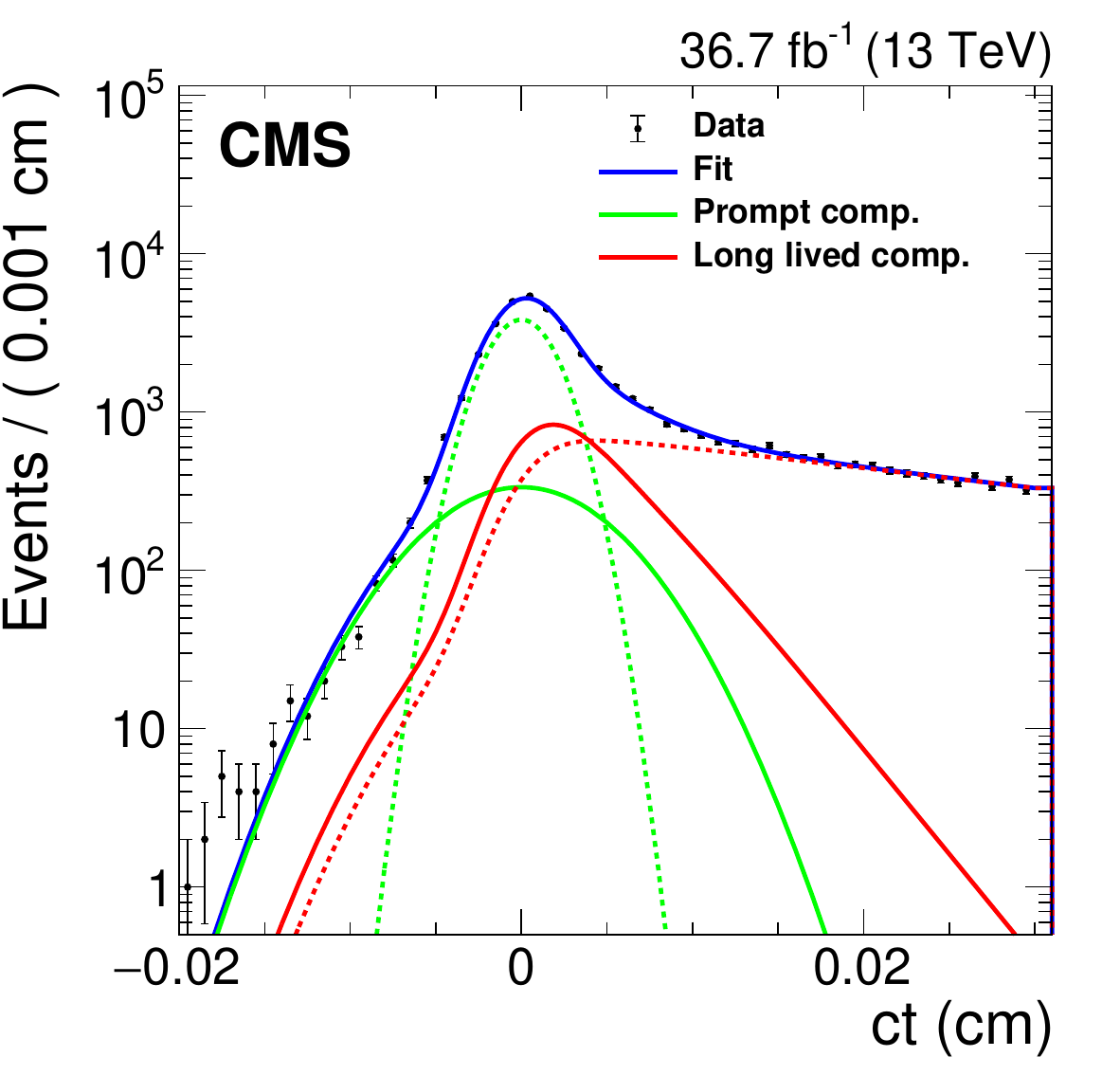} \hfil
  \includegraphics[width=0.45\textwidth]{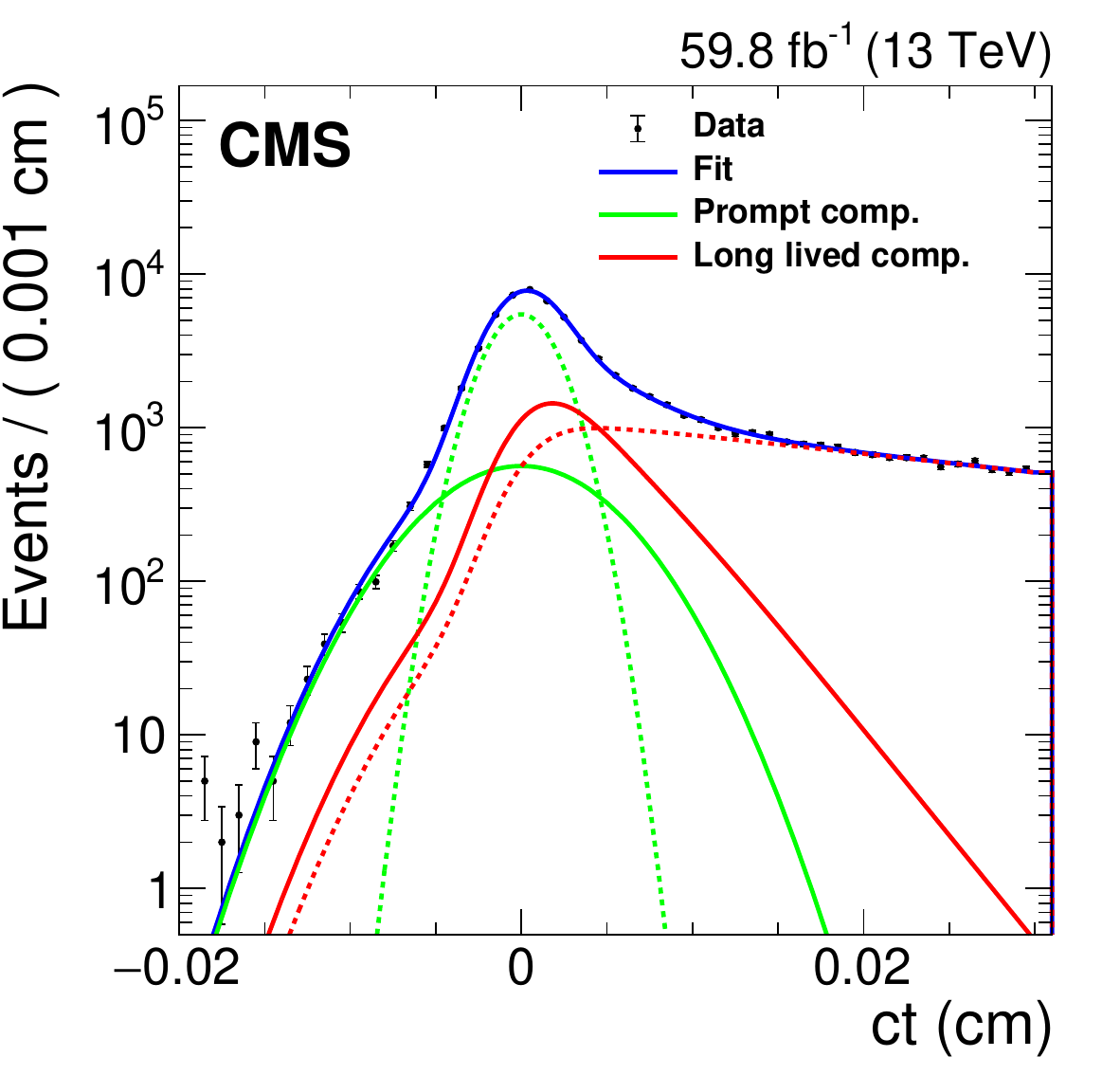}
  \caption{Proper decay time distribution of prompt events in the \setJpsiMu trigger category for 2017 (left) and 2018 (right) data.
  The two components of the resolution function are modeled with Gaussian distributions and shown in solid and dashed green. The resolution function is the model for the prompt component, and the long-lived components, shown in solid and dashed red, are modeled as two exponential distributions convolved with the resolution function.}
  \label{fig:ctresfit}
\end{figure*}

Figures~\ref{fig:ctrescal} and~\ref{fig:ctresfit} show the \ct uncertainty calibration fits and the \ct distribution of prompt events used for calibration, respectively.

\begin{figure*}[!ht]
  \centering
  \includegraphics[width=0.45\textwidth]{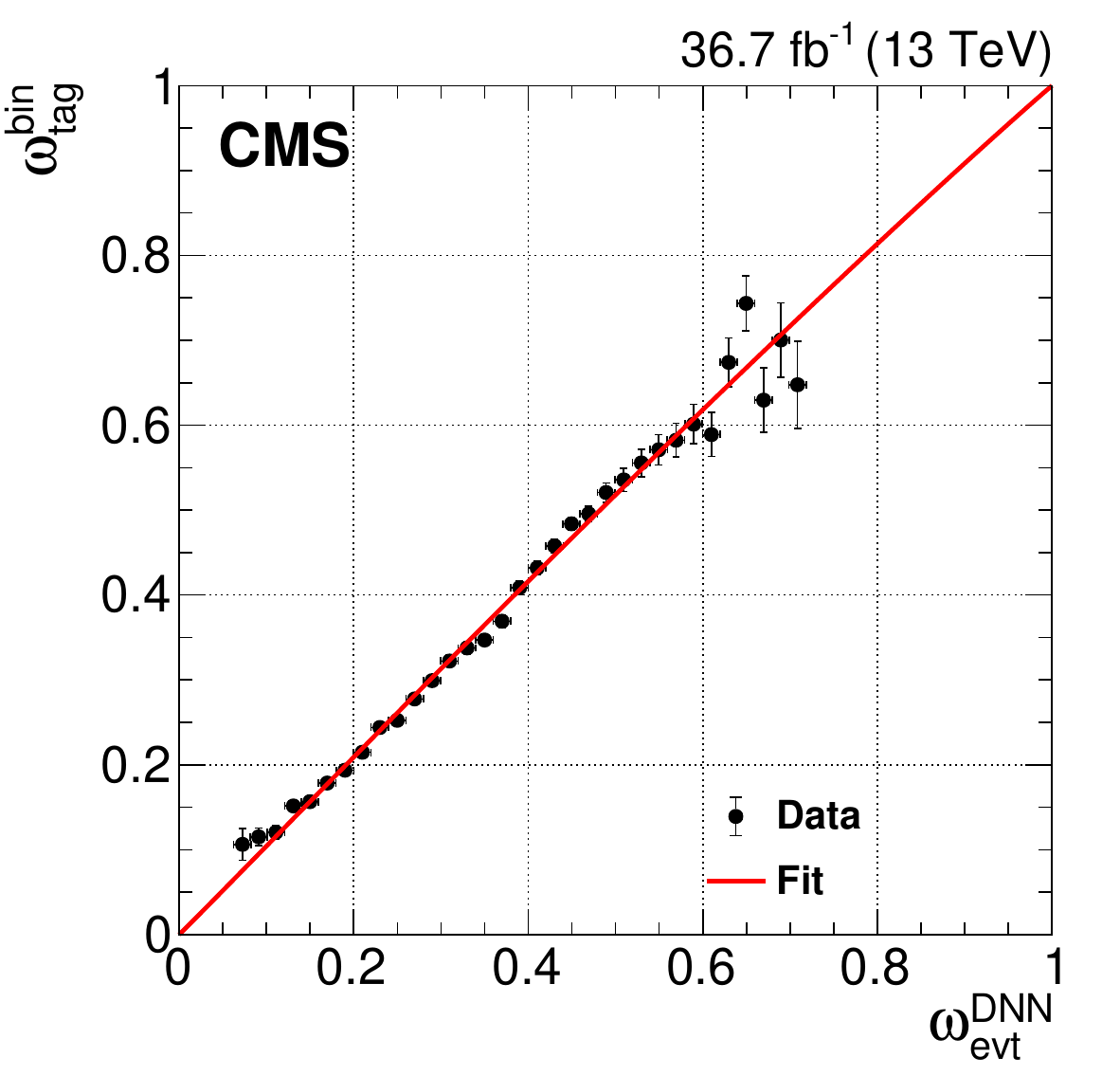} \hfil
  \includegraphics[width=0.45\textwidth]{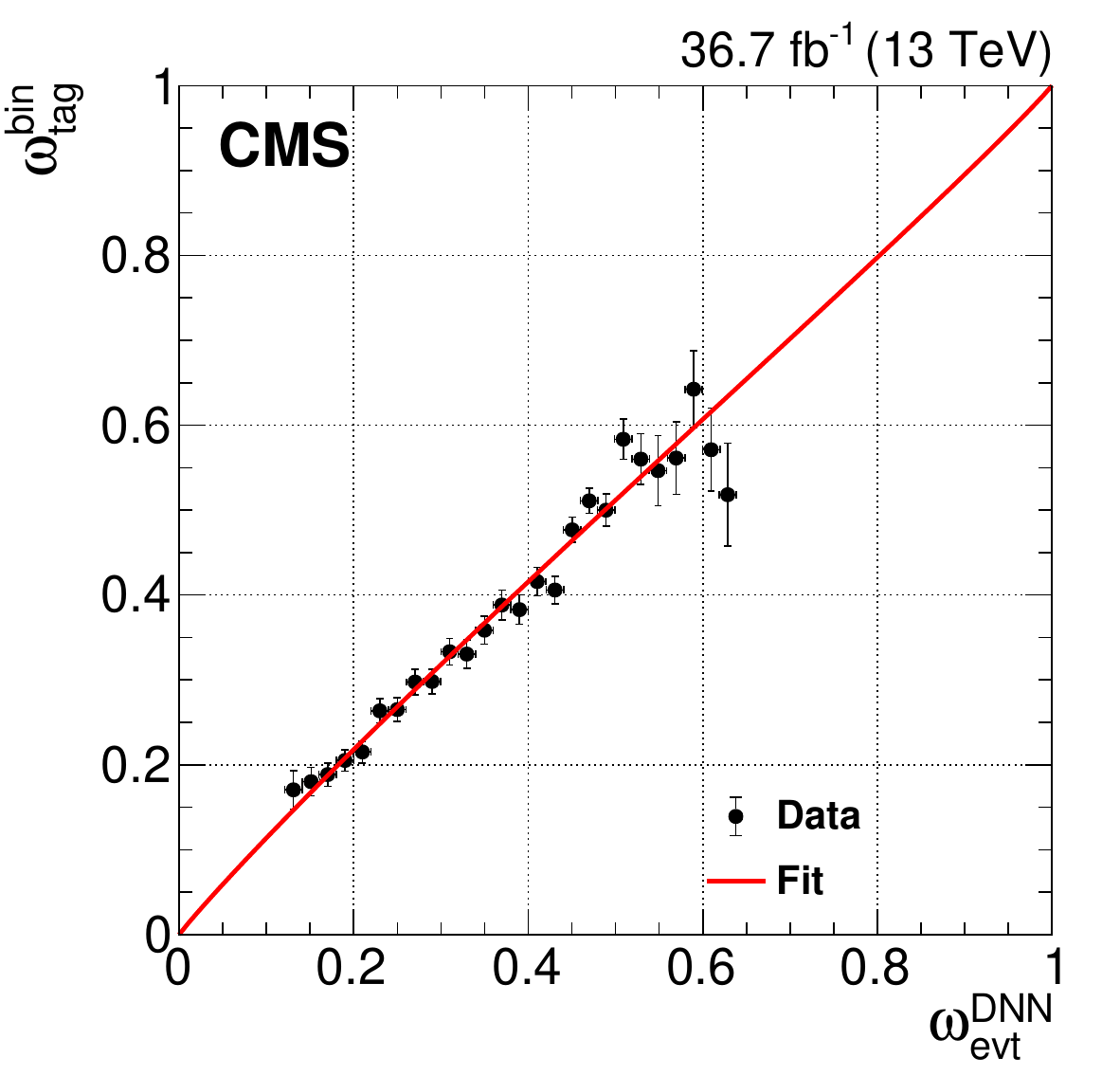} \\
  \includegraphics[width=0.45\textwidth]{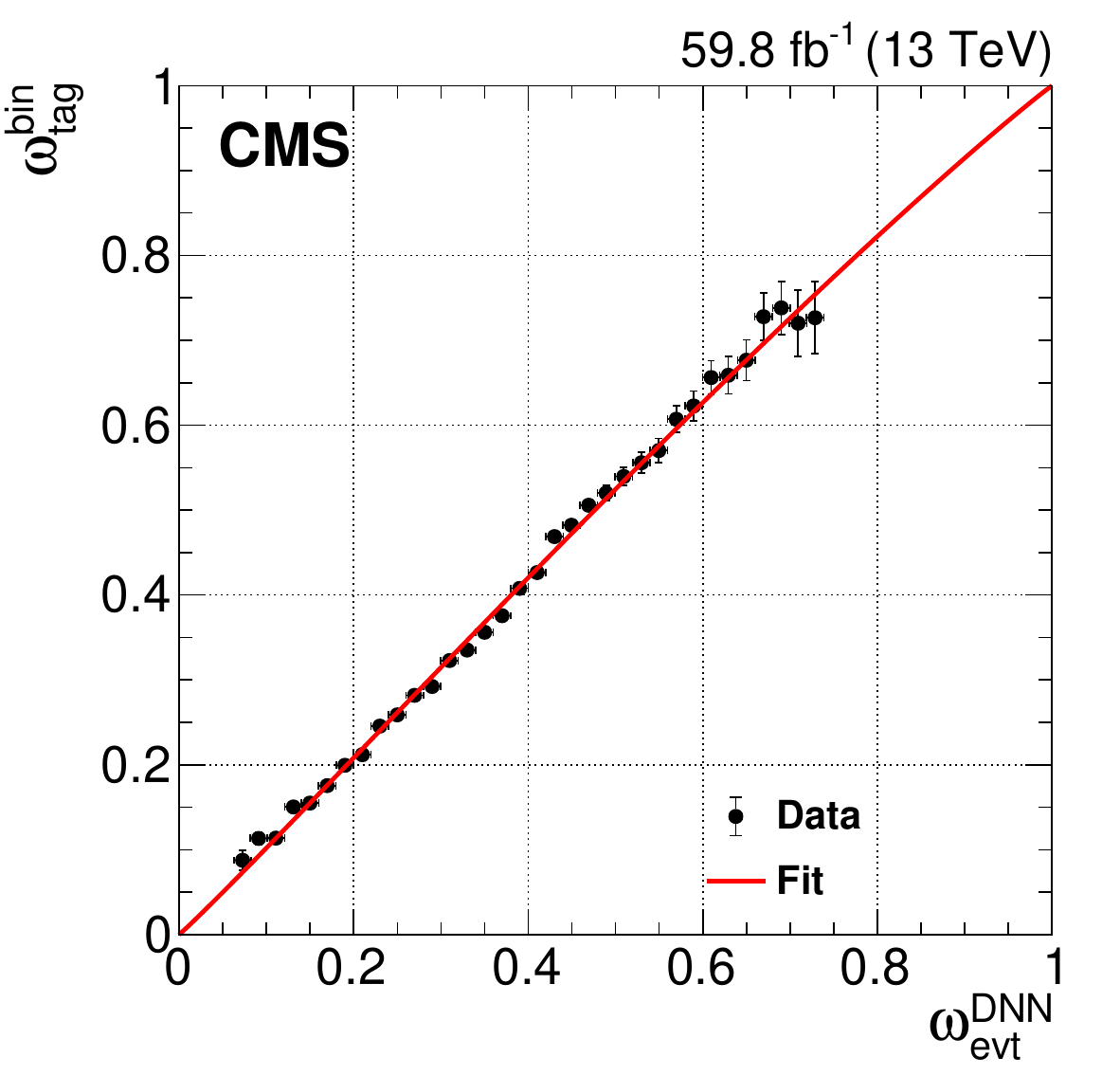} \hfil
  \includegraphics[width=0.45\textwidth]{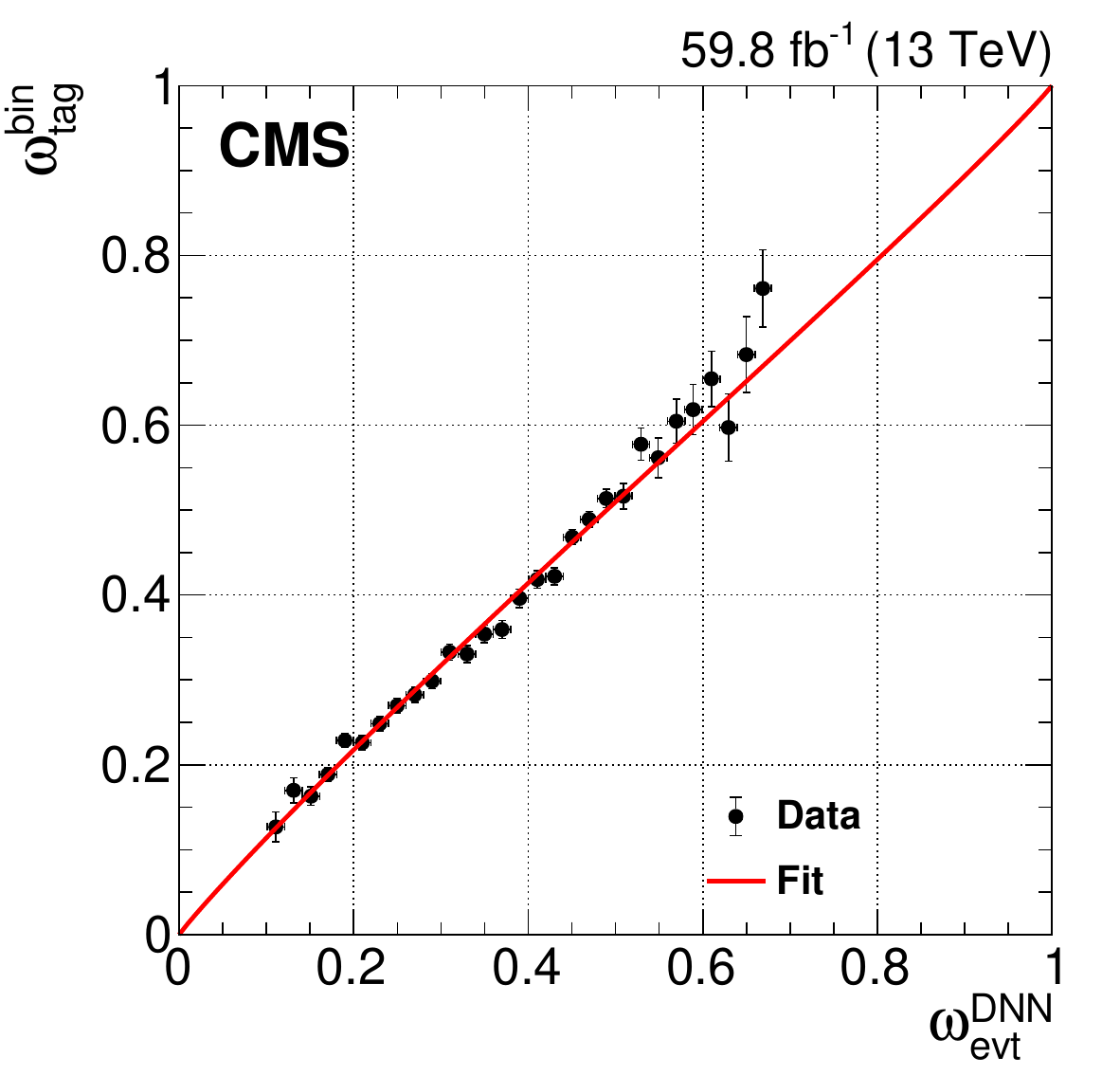} \\
  \caption{Results of the mistag probability calibration fit for the OS-muon tagger on \BujpsiK decays for 2017 (top row) and 2018 (bottom row) data.
  The measured mistag probability is plotted versus the value predicted by the tagging algorithm.
  The solid red line shows the results of the Platt scaling calibration fit to data.
  The calibration model is a linear function applied to the tagger DNN's output before the final sigmoid layer.
  The plots in the left column refer to the \setJpsiMu trigger category, while the ones in the right column to the \setJpsiTrk trigger category.}
  \label{fig:calosmuon}
\end{figure*}

\begin{figure*}[!ht]
  \centering
  \includegraphics[width=0.45\textwidth]{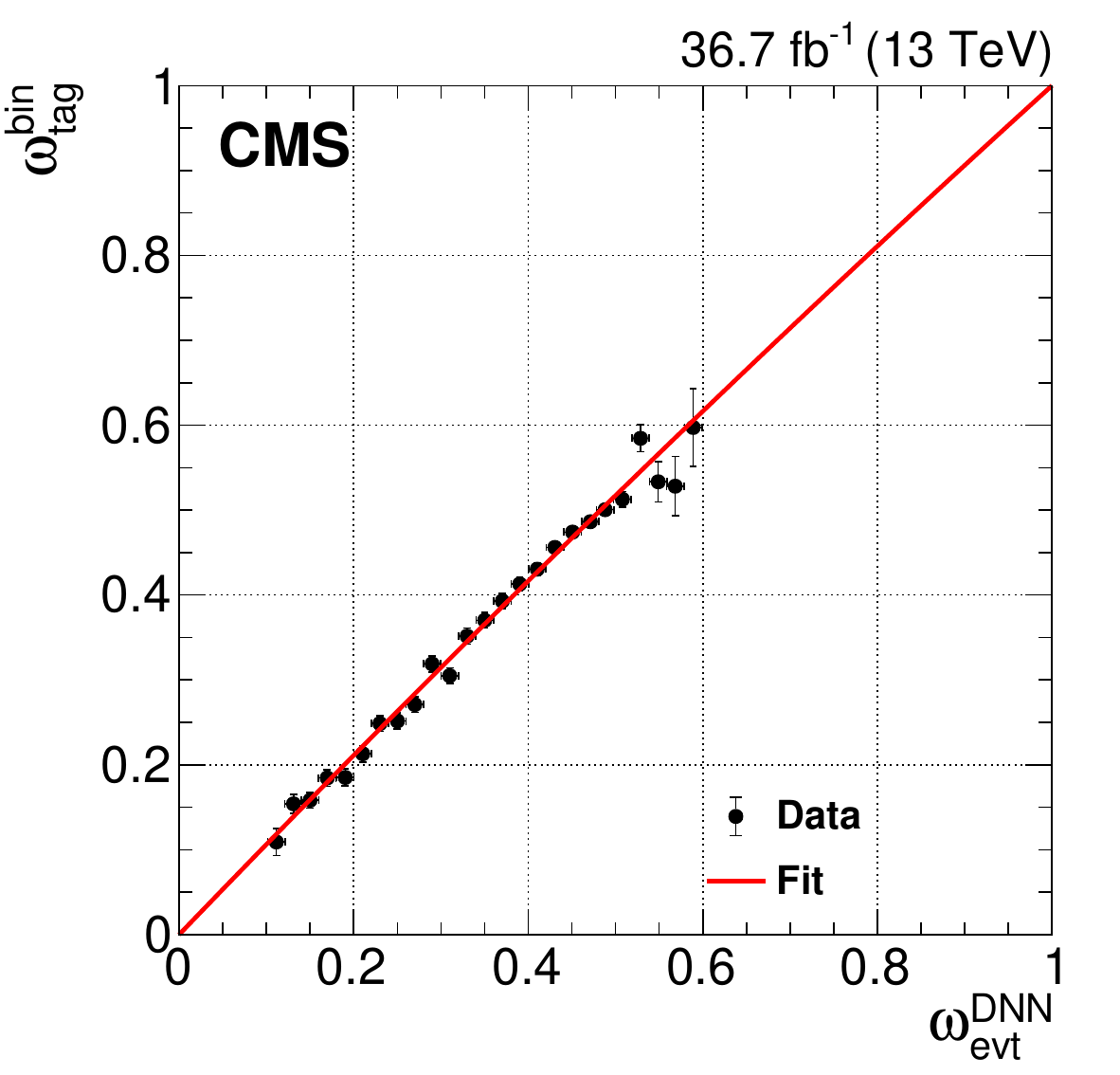} \hfil
  \includegraphics[width=0.45\textwidth]{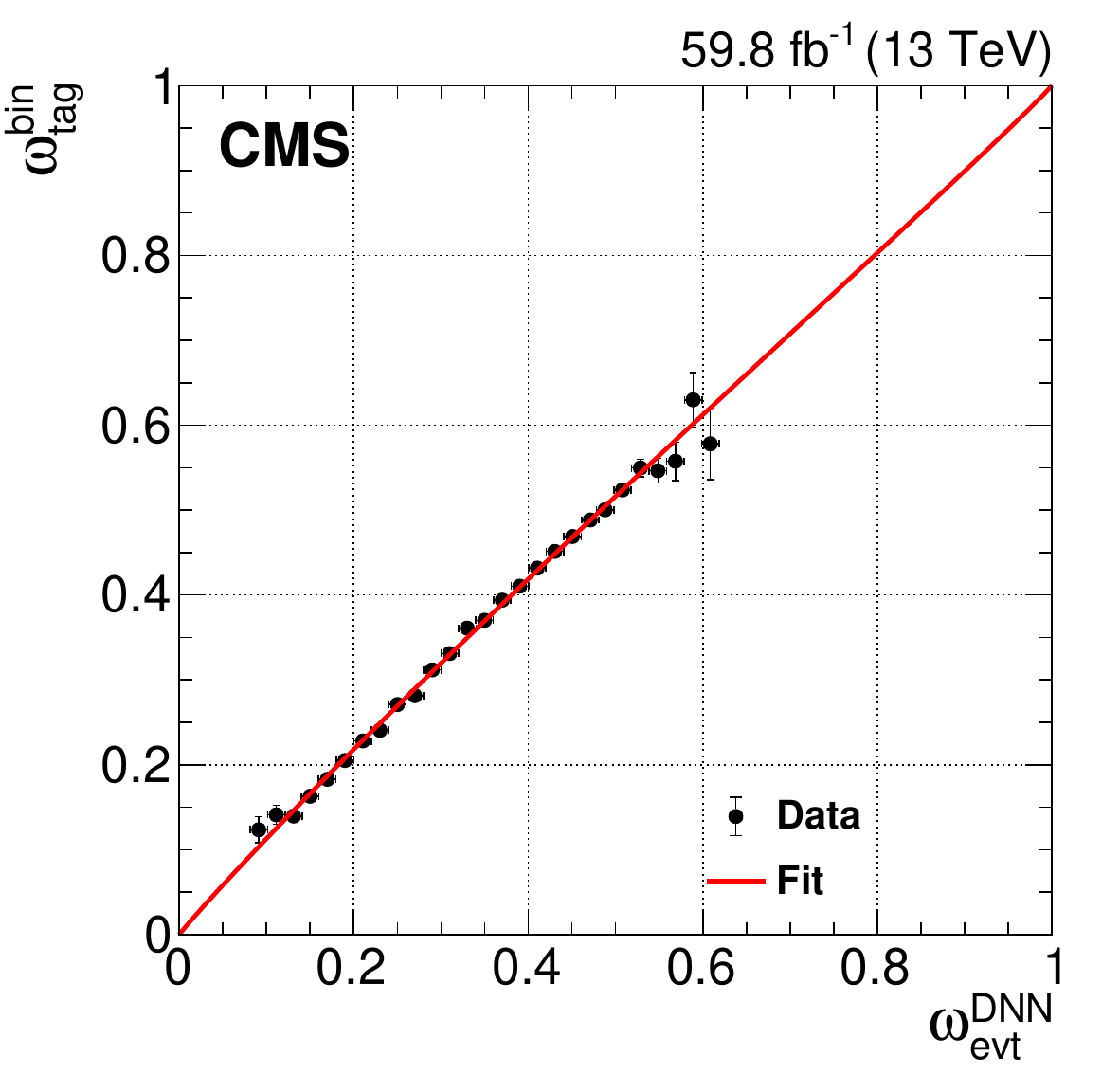}
  \caption{Results of the mistag probability calibration fit for the OS-electron tagger on \BujpsiK decays for 2017 (left) and 2018 (right) data.
  The measured mistag probability is plotted versus the value predicted by the tagging algorithm.
  The solid red line shows the results of the Platt scaling calibration fit to data.
  The calibration model is a linear function applied to the tagger DNN's output before the final sigmoid layer.}
  \label{fig:calosele}
\end{figure*}

\begin{figure*}[!ht]
  \centering
  \includegraphics[width=0.45\textwidth]{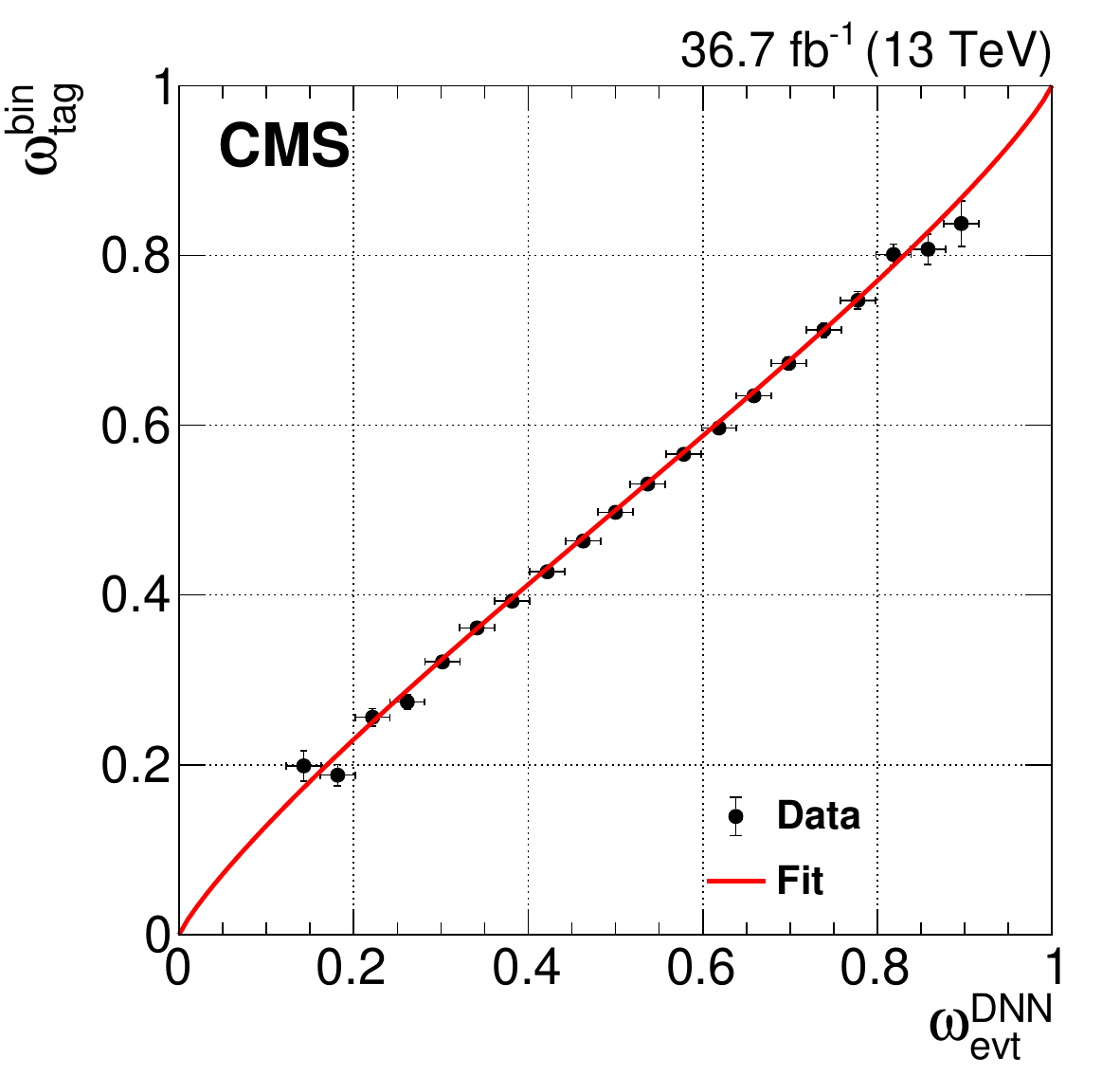} \hfil
  \includegraphics[width=0.45\textwidth]{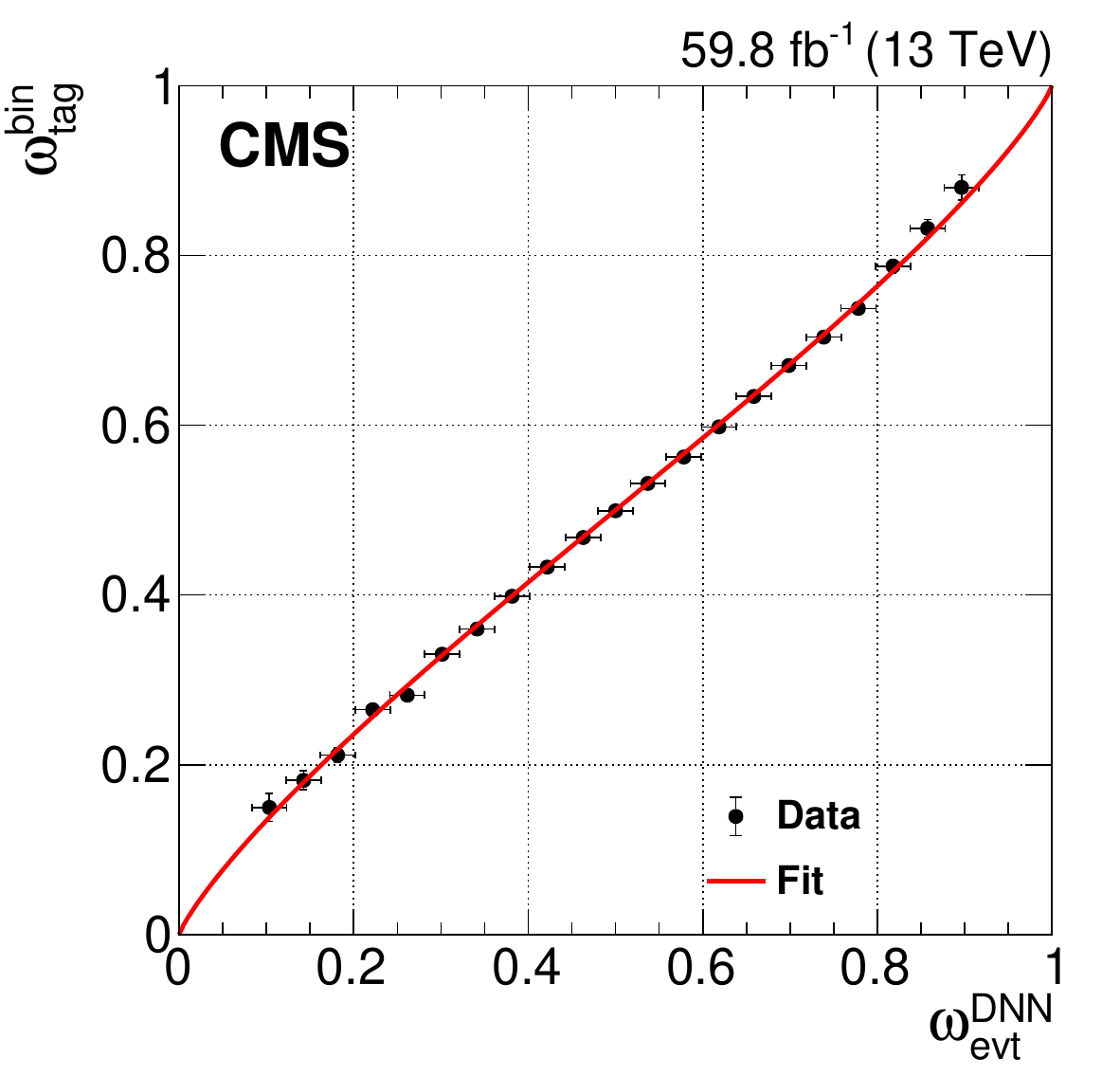}
  \caption{Results of the mistag probability calibration fit for the OS-jet tagger on \BujpsiK decays for 2017 (left) and 2018 (right) data.
  The measured mistag probability is plotted versus the value predicted by the tagging algorithm.
  The solid red line shows the results of the Platt scaling calibration fit to data.
  The calibration model is a linear function with the intercept fixed to 0 applied to the tagger DNN's output before the final sigmoid layer.}
  \label{fig:calosjet}
\end{figure*}

\begin{figure*}[!ht]
  \centering
  \includegraphics[width=0.45\textwidth]{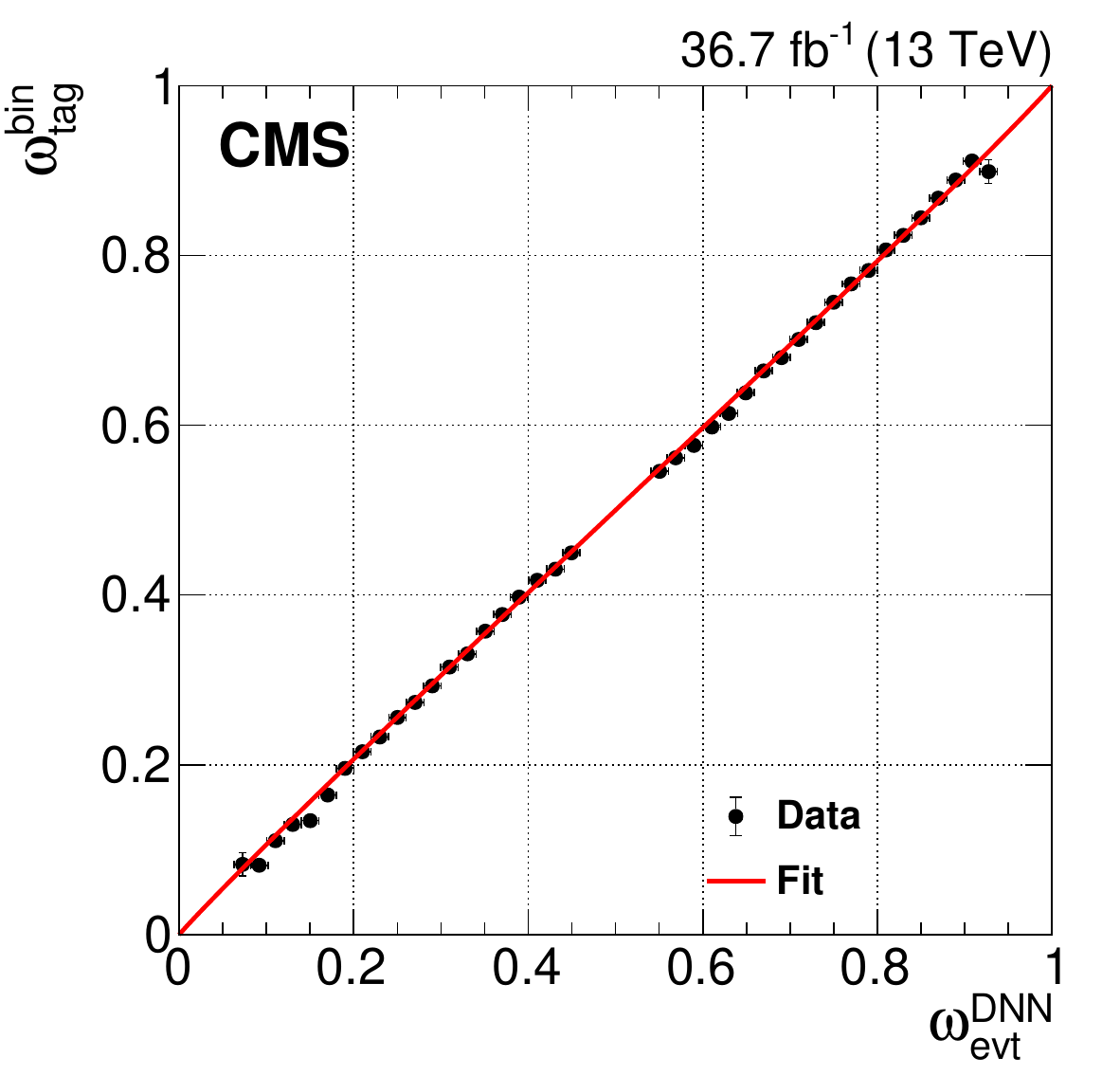} \hfil
  \includegraphics[width=0.45\textwidth]{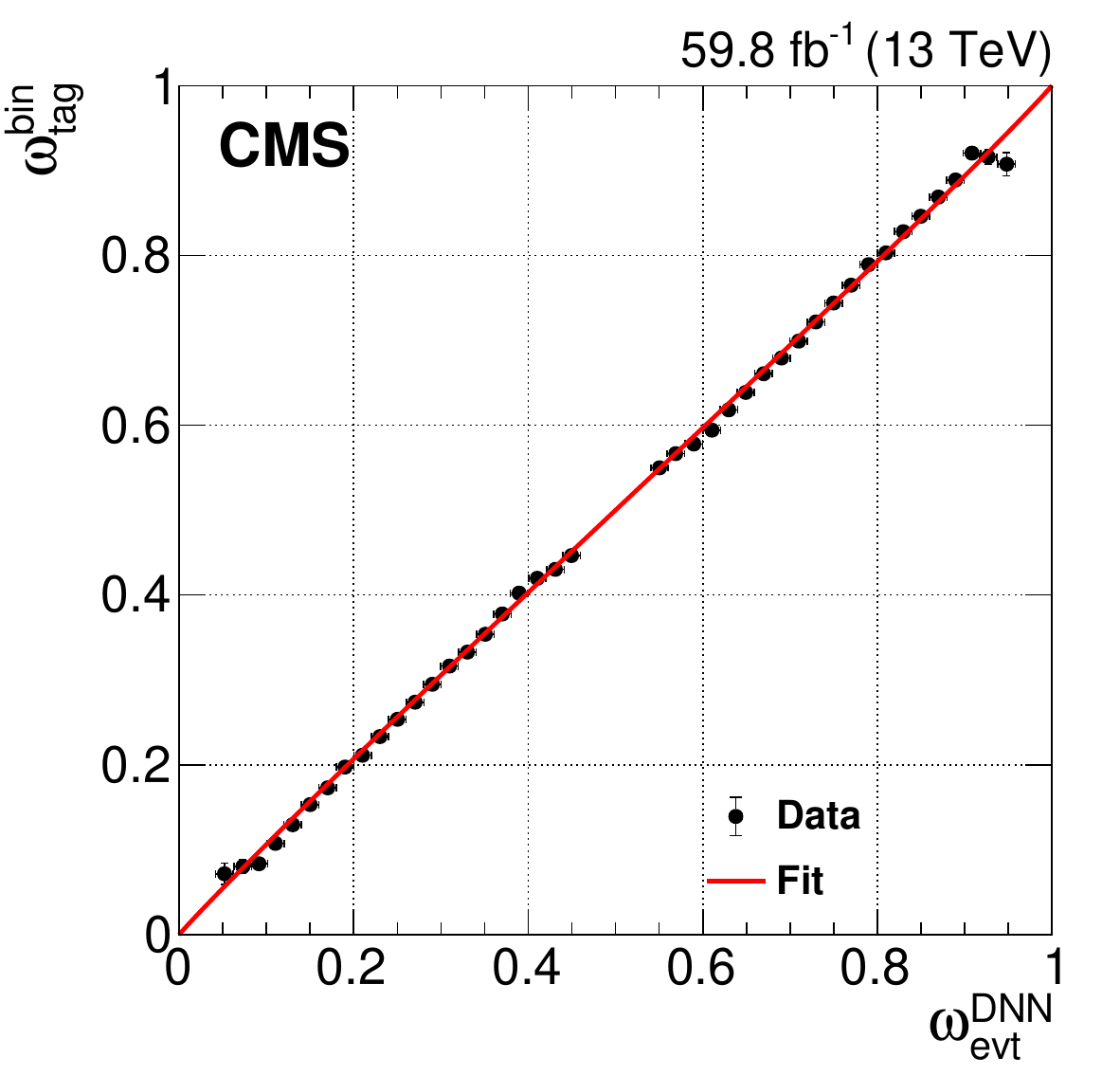}
  \caption{Results of the mistag probability calibration fit for the SS tagger on \BujpsiK decays for 2017 (left) and 2018 (right) data.
  The measured mistag probability is plotted versus the value predicted by the tagging algorithm.
  The solid red line shows the results of the Platt scaling calibration fit to data.
  The calibration model is a linear function with the intercept fixed to 0 applied to the tagger DNN's output before the final sigmoid layer.}
  \label{fig:calss}
\end{figure*}

\begin{figure*}[!ht]
  \centering
  \includegraphics[width=0.45\textwidth]{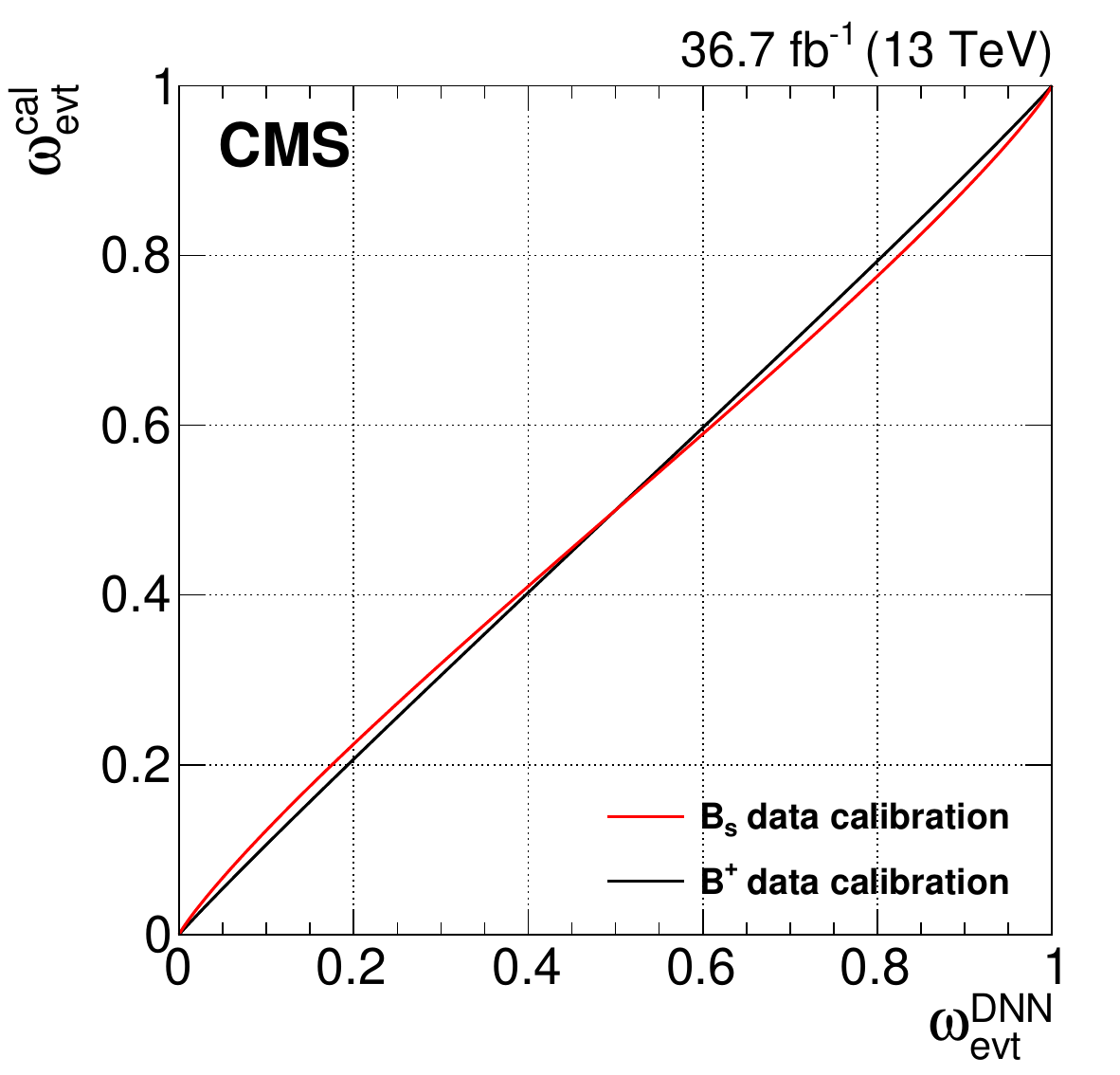} \hfil
  \includegraphics[width=0.45\textwidth]{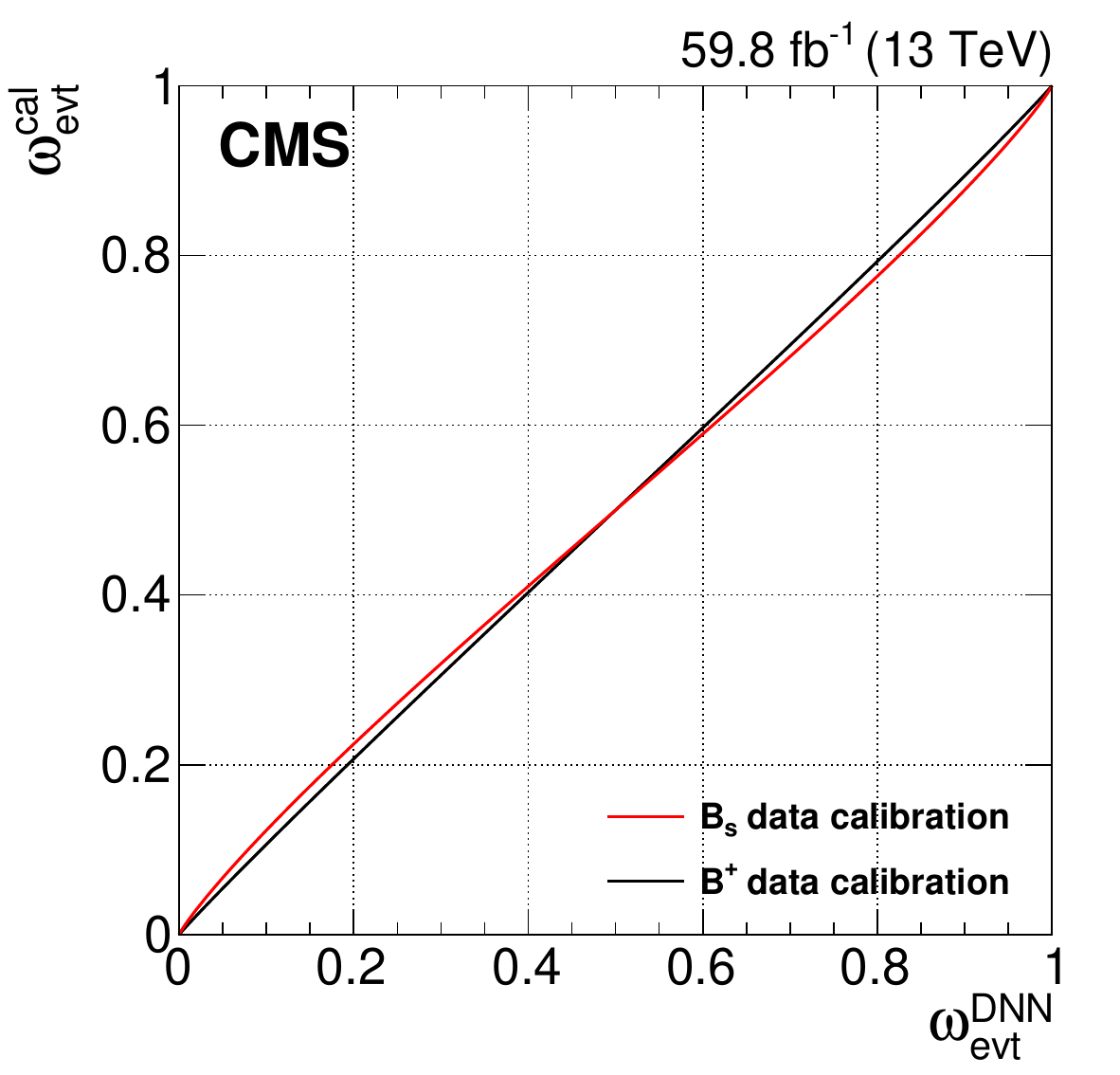}
  \caption{Comparison of the calibration curves before (black, \Bu) and after (red, \Bs) the application of the corrections from simulation to the calibration obtained in \BujpsiK decays.
  The left plot refers to 2017 data, while the right one to 2018 data.
  The red function is the one used as calibration for the CPV measurement.}
  \label{fig:bubsss}
\end{figure*}

\begin{figure*}[!ht]
  \centering
  \includegraphics[width=0.5\textwidth]{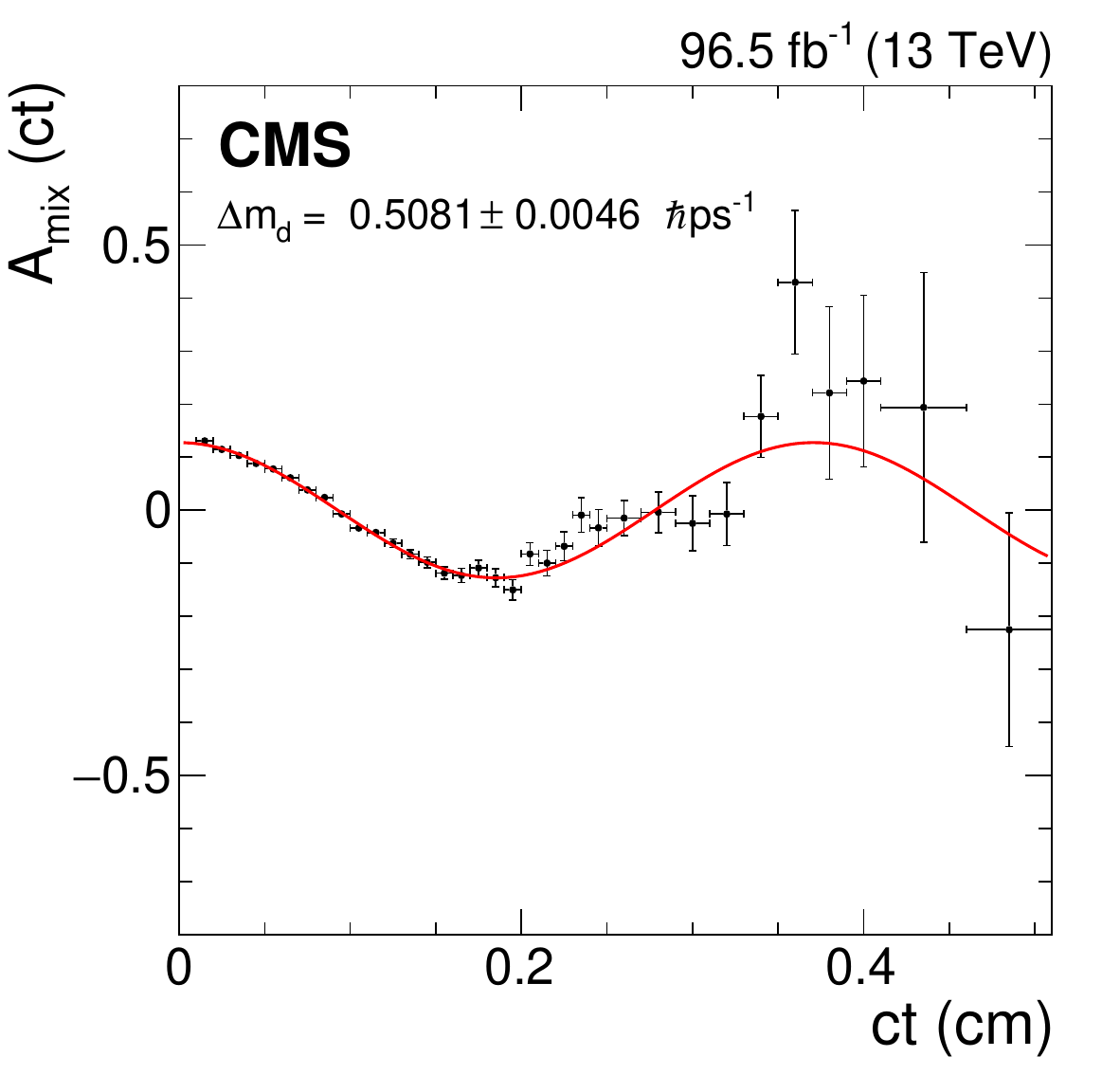}
  \caption{The \Bd mixing asymmetry as a function of the proper decay time as measured in the \Bdjpsikstar control data sample as a validation to the tagging framework. The measured oscillation frequency is consistent with the world-average value $\Dmd^\text{PDG} = 0.5069 \pm 0.0019\ \hips$~\refPDG.}
  \label{fig:dmd}
\end{figure*}

\begin{figure*}[!ht]
  \centering
  \includegraphics[width=0.24\textwidth]{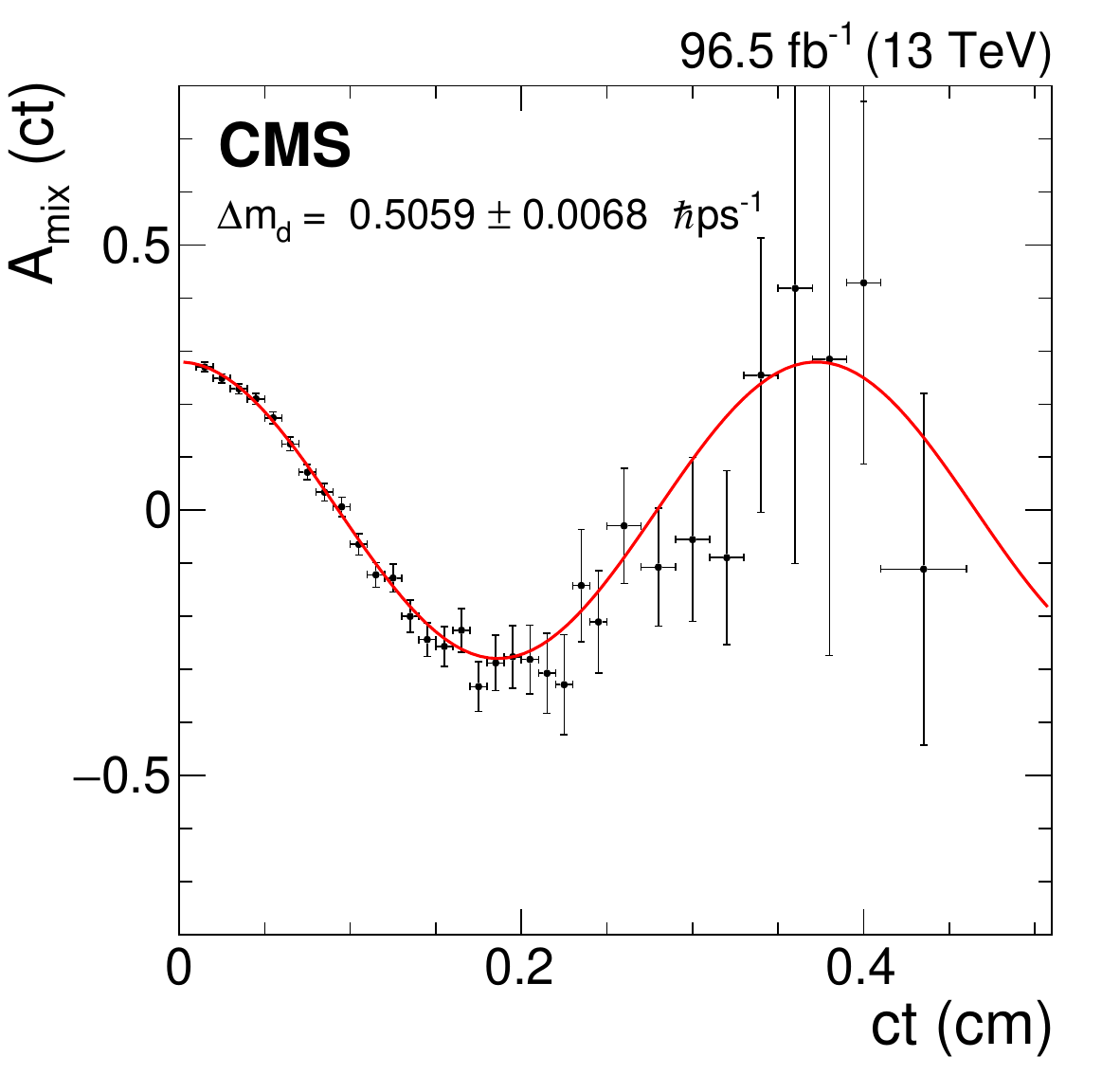} \hfil
  \includegraphics[width=0.24\textwidth]{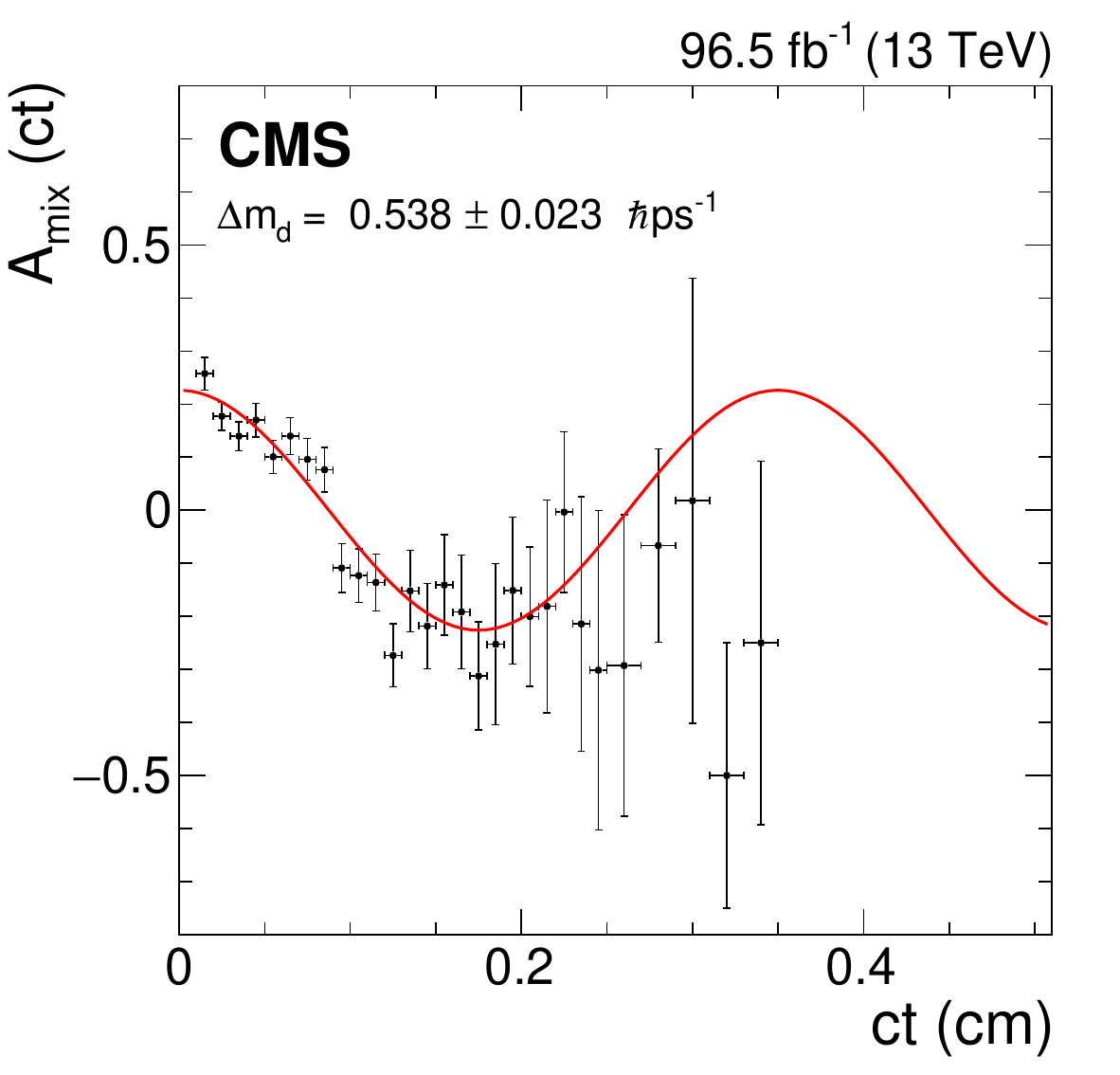} \hfil
  \includegraphics[width=0.24\textwidth]{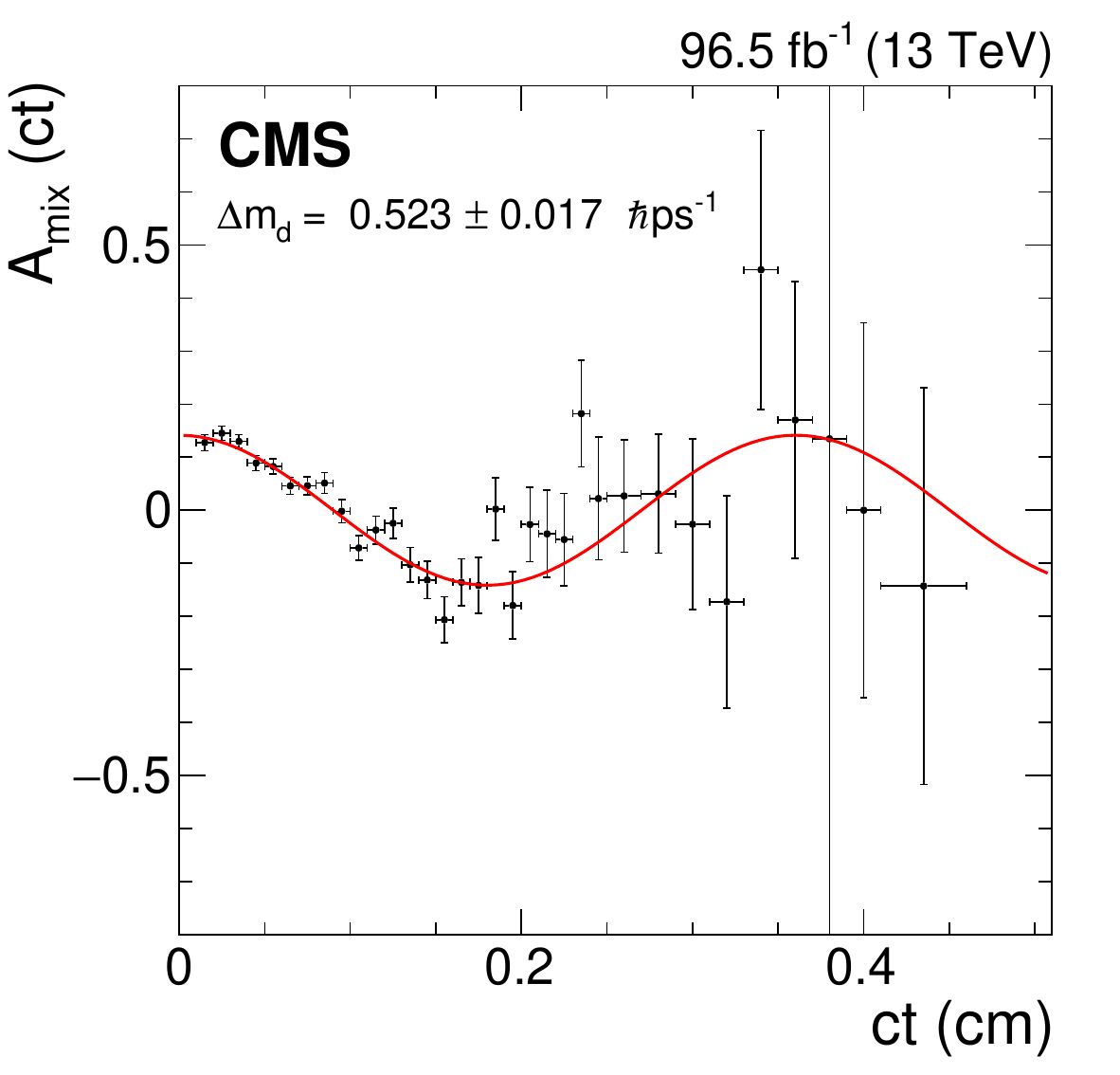} \hfil
  \includegraphics[width=0.24\textwidth]{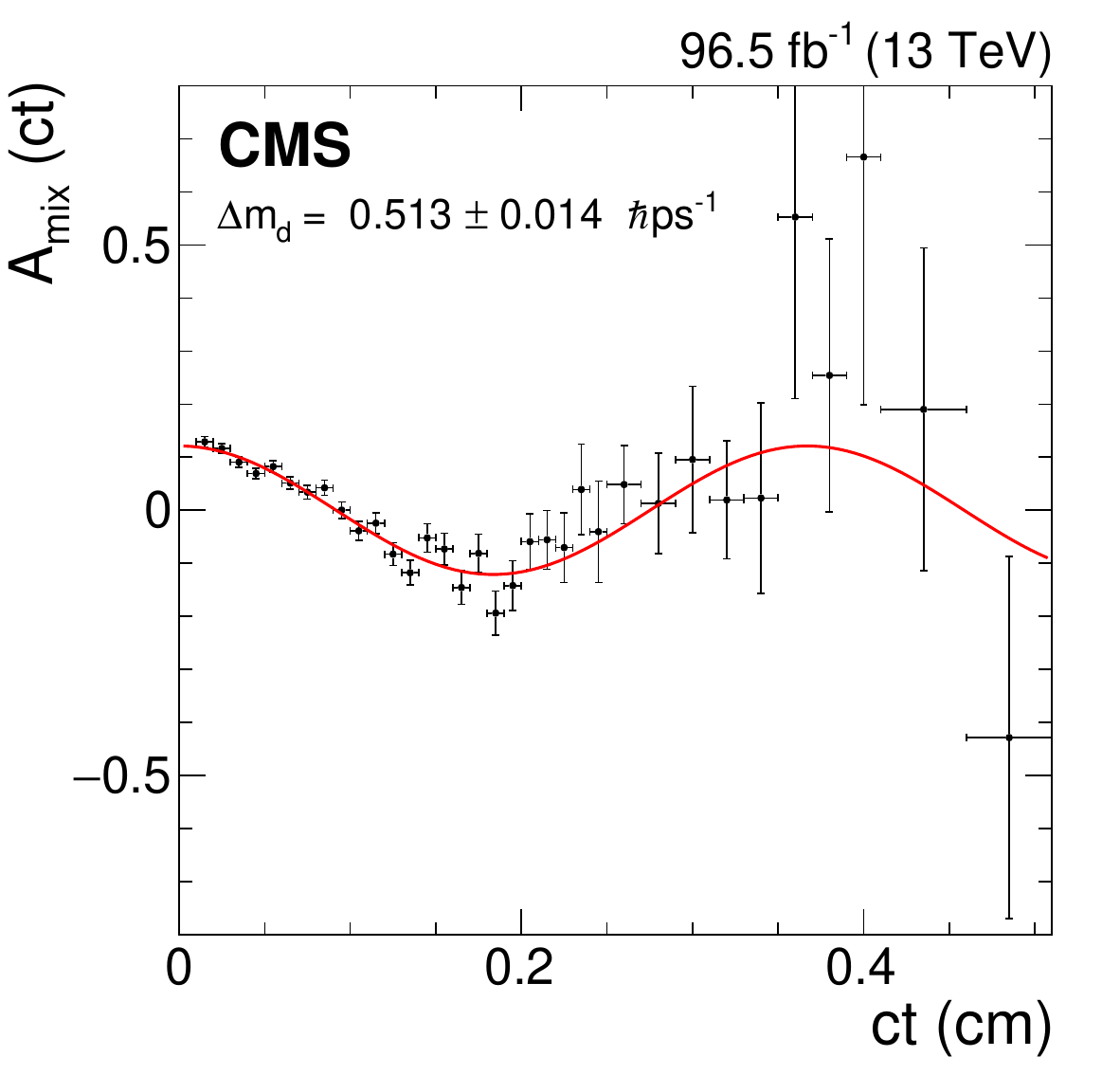} \\
  \includegraphics[width=0.24\textwidth]{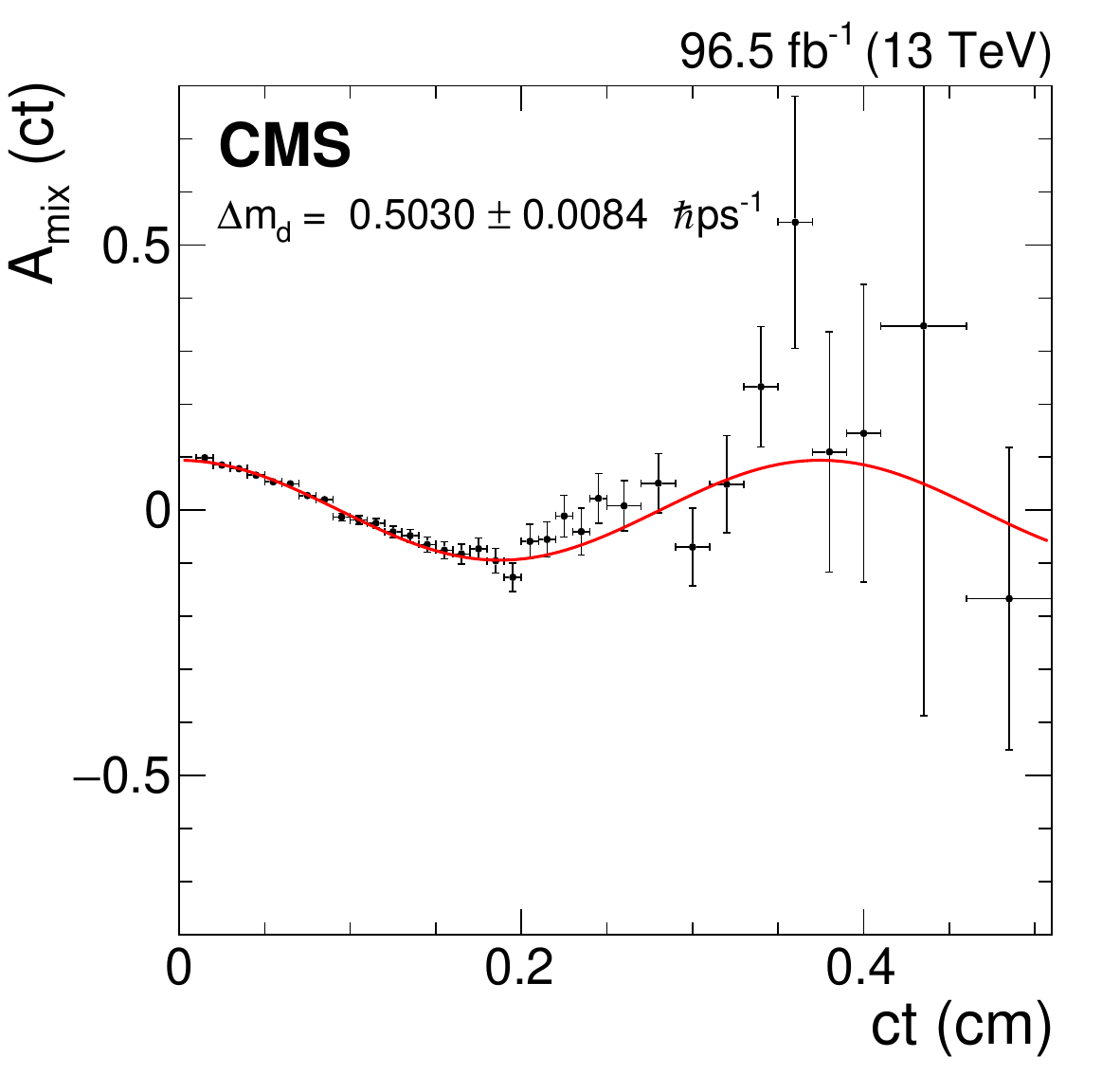} \hfil
  \includegraphics[width=0.24\textwidth]{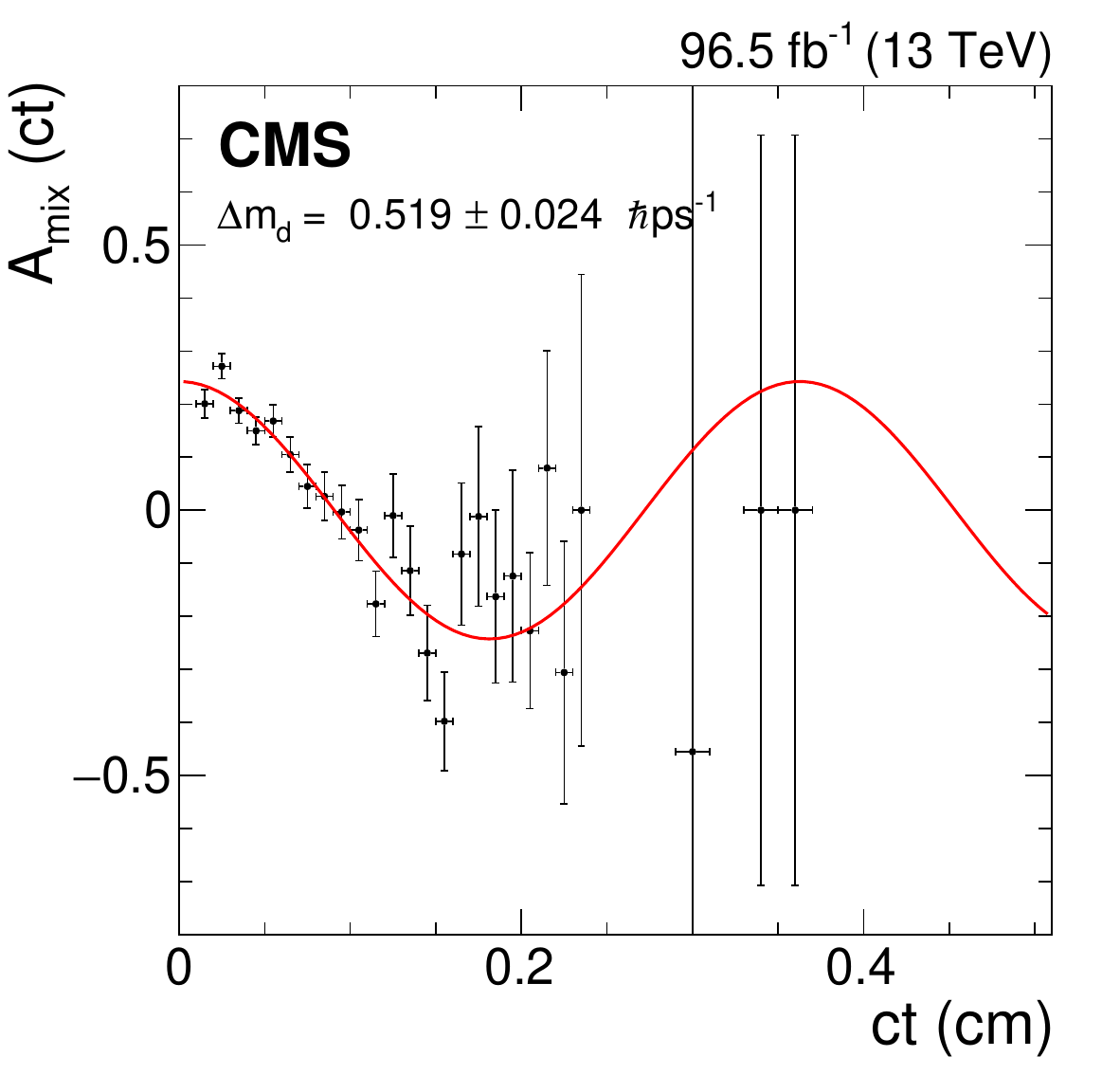} \hfil
  \includegraphics[width=0.24\textwidth]{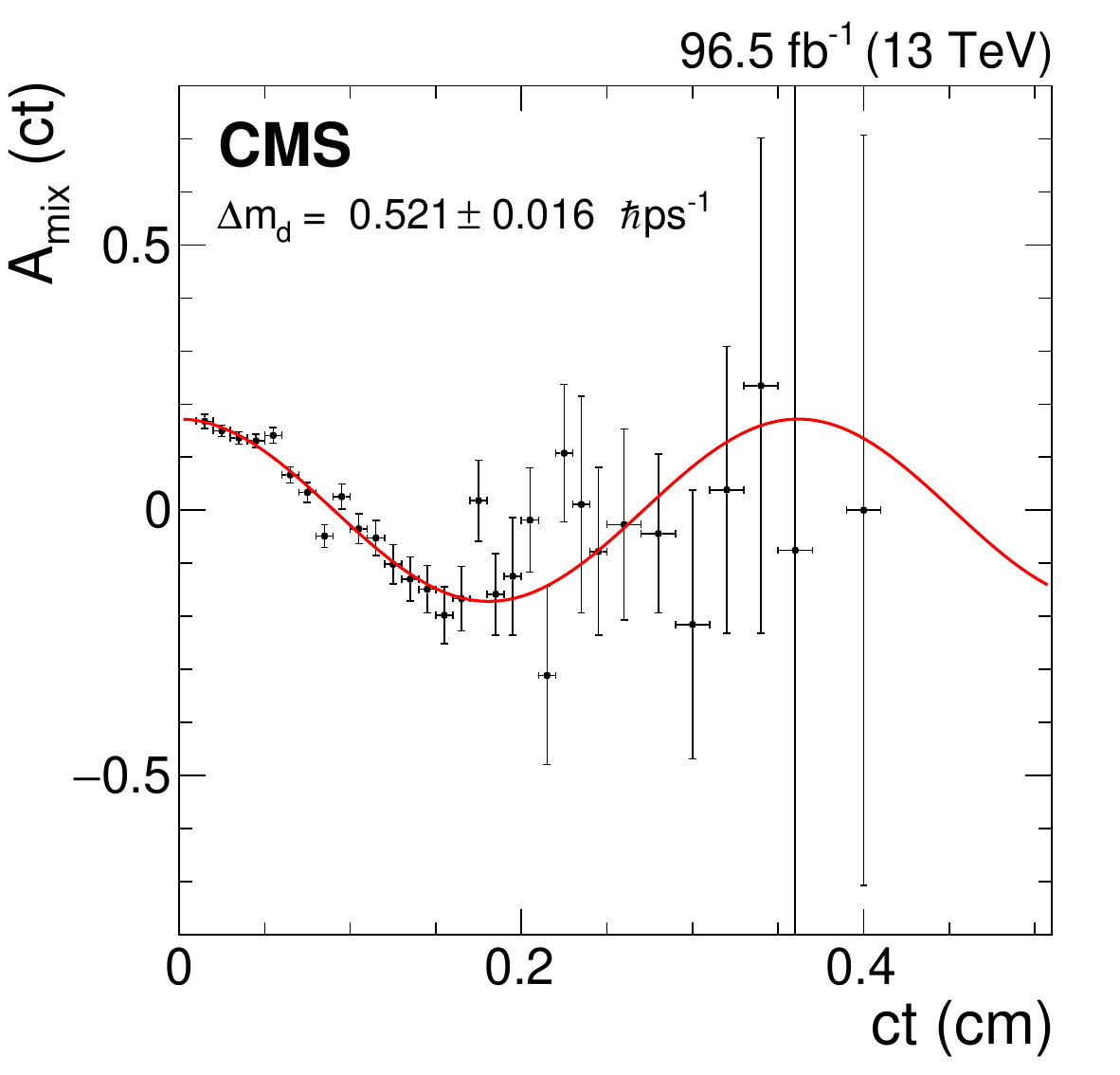} \hfil
  \includegraphics[width=0.24\textwidth]{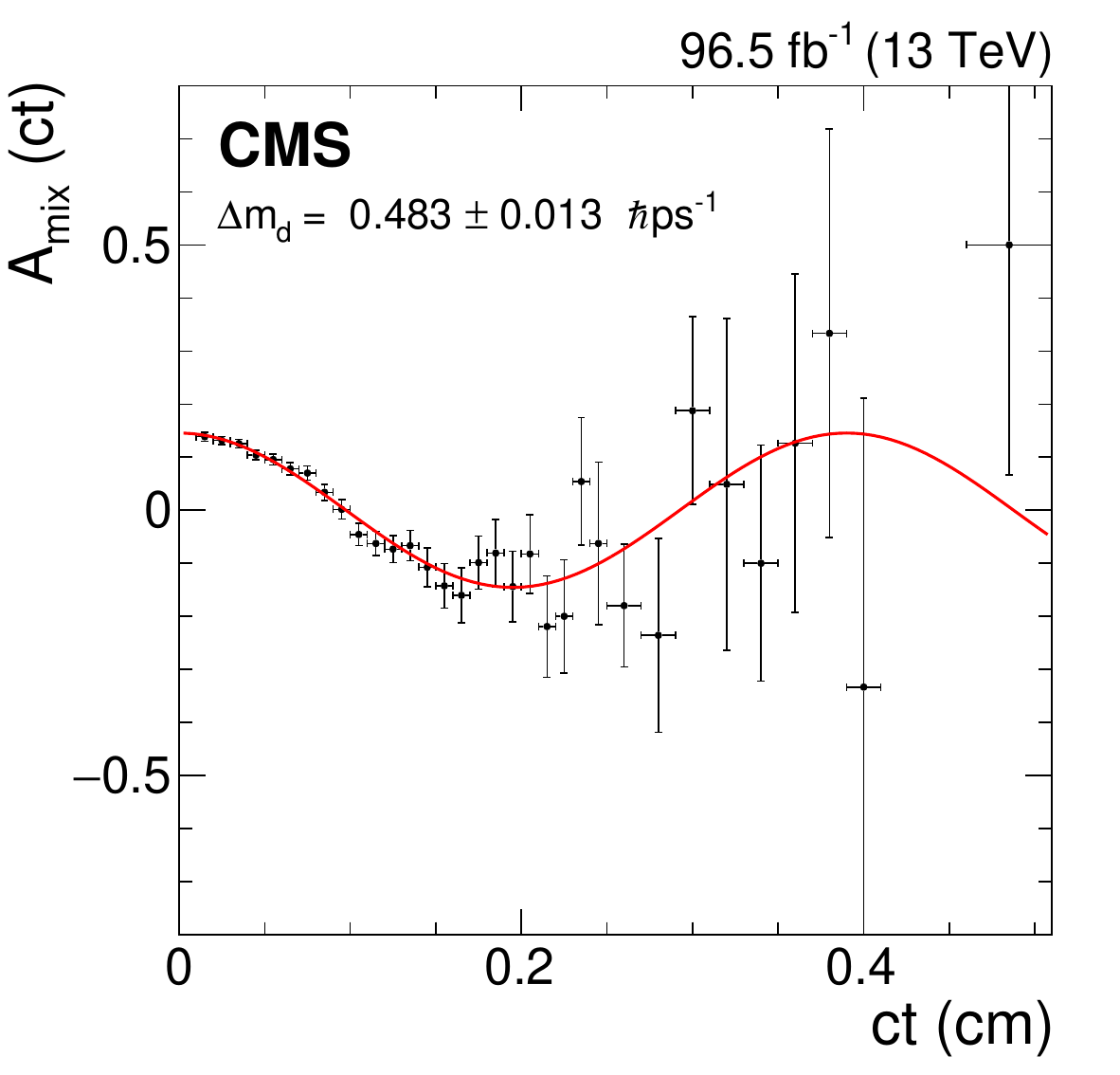} \\
  \caption{The \Bd mixing asymmetry as a function of the proper decay time as measured in the \Bdjpsikstar control data sample in the different tagging categories.
  The reported uncertainties for the \Dmd fitted value are statistical only.
  This study was conducted to validate the tagging framework and assess the consistency of information provided by various tagging techniques.
  Top row, from left to right: only OS muon (\setJpsiMu trigger category), only OS muon (\setJpsiTrk trigger category), only OS electron, and only OS jet.
  Bottom row, from left to right: only SS, SS + OS muon, SS + OS electron, and SS + OS jet.
  All tagging categories are mutually exclusive.
  The oscillation frequencies measured in each category are consistent with one another and with the world-average value $\Dmd^\text{PDG} = 0.5069 \pm 0.0019\ \hips$~\refPDG.}
  \label{fig:dmdcat}
\end{figure*}

\begin{figure*}[!ht]
  \centering
  \includegraphics[width=0.75\textwidth]{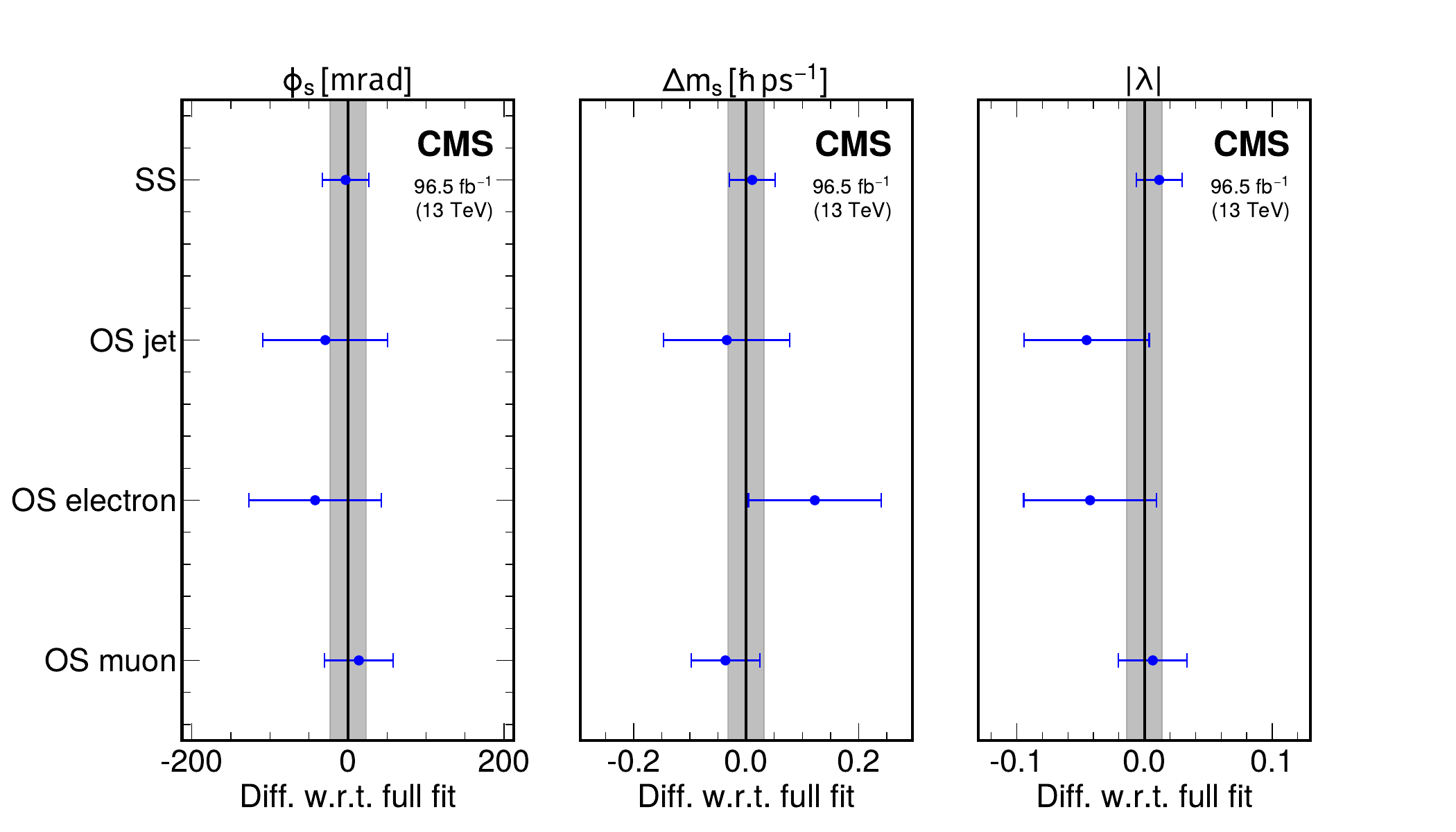}
  \caption{Comparison of CPV fit results obtained using each of the four individual flavor tagging algorithms versus the reference fit, which combines all four taggers. Only flavor-sensitive parameters are shown, and only statistical uncertainties are considered. The grey band represents the statistical uncertainty of the reference fit.}
  \label{fig:tagval}
\end{figure*}

The flavor tagging calibration fits for the OS-muon, OS-electron, OS-jet, and SS taggers are reported in Figs.~\ref{fig:calosmuon}, \ref{fig:calosele}, \ref{fig:calosjet}, and~\ref{fig:calss}, respectively.
The comparison of the SS tagger calibration curves before and after applying the corrections from simulation is shown in Fig.~\ref{fig:bubsss}.
The \Bd mixing asymmetry as a function of \ct as measured in the \Bdjpsikstar control data sample is shown in Figs.~\ref{fig:dmd} and~\ref{fig:dmdcat}.
The oscillation frequencies measured in each category are consistent with one another and with the world-average value $\Dmd^\text{PDG} = 0.5069 \pm 0.0019\ \hips$~\refPDG.
The comparison of the CP violation fit results obtained using each of the four individual flavor tagging algorithms versus the reference fit, which combines all four taggers, is reported in Fig.~\ref{fig:tagval}.
These studies were conducted to validate the tagging framework and assess the consistency of information provided by various tagging techniques.

\begin{table*}[!ht]
  \centering
  \topcaption{Matrix of the correlations of the statistical uncertainties between pairs of physics parameters.}
  \begin{tabular}{l|ccccccccccc}
    & \phis & \DGs & \Gs & \Dms & \abslambda & \sqAzero & \sqAperp & \sqAswav & \dpara & \dperp & \dswpd \\
    \hline
    \phis & $+1.00$ & $-0.05$ & $+0.02$ & $-0.14$ & $+0.21$ & $-0.04$ & $+0.04$ & $+0.03$ & $-0.04$ & $-0.16$ & $-0.02$ \\
    \DGs &  & $+1.00$ & $-0.60$ & $+0.07$ & $-0.07$ & $+0.64$ & $-0.70$ & $+0.01$ & $+0.07$ & $+0.07$ & $-0.03$ \\
    \Gs &  &  & $+1.00$ & $\hphantom{+}0.00$ & $+0.06$ & $-0.41$ & $+0.51$ & $-0.02$ & $-0.04$ & $-0.02$ & $+0.04$ \\
    \Dms &  &  &  & $+1.00$ & $-0.32$ & $+0.03$ & $-0.06$ & $-0.01$ & $+0.08$ & $+0.72$ & $-0.03$ \\
    \abslambda &  &  &  &  & $+1.00$ & $-0.06$ & $+0.06$ & $-0.02$ & $-0.29$ & $-0.41$ & $+0.05$ \\
    \sqAzero &  &  &  &  &  & $+1.00$ & $-0.58$ & $\hphantom{+}0.00$ & $+0.03$ & $+0.04$ & $-0.04$ \\
    \sqAperp &  &  &  &  &  &  & $+1.00$ & $-0.09$ & $-0.10$ & $-0.08$ & $+0.08$ \\
    \sqAswav &  &  &  &  &  &  &  & $+1.00$ & $+0.04$ & $+0.07$ & $-0.73$ \\
    \dpara &  &  &  &  &  &  &  &  & $+1.00$ & $+0.44$ & $-0.07$ \\
    \dperp &  &  &  &  &  &  &  &  &  & $+1.00$ & $-0.08$ \\
    \dswpd &  &  &  &  &  &  &  &  &  &  & $+1.00$ \\
  \end{tabular}
  \label{tab:corrStat}
\end{table*}

\begin{table*}[!ht]
  \centering
  \topcaption{Matrix of the correlations of the systematic uncertainties between pairs of physics parameters.}
  \begin{tabular}{l|ccccccccccc}
   & \phis & \DGs & \Gs & \Dms & \abslambda & \Azero & \Aperp & \Aswav & \dpara & \dperp & \dswpd \\
    \hline
    \phis & $+1.00$ & $-0.15$ & $-0.24$ & $-0.30$ & $-0.13$ & $+0.12$ & $-0.09$ & $-0.70$ & $+0.69$ & $-0.20$ & $+0.68$ \\
    \DGs &  & $+1.00$ & $+0.09$ & $+0.06$ & $-0.03$ & $+0.21$ & $-0.50$ & $+0.18$ & $-0.11$ & $+0.04$ & $-0.20$ \\
    \Gs &  &  & $+1.00$ & $+0.15$ & $-0.06$ & $-0.15$ & $+0.16$ & $+0.24$ & $-0.17$ & $+0.14$ & $-0.22$ \\
    \Dms &  &  &  & $+1.00$ & $-0.74$ & $+0.16$ & $-0.13$ & $+0.51$ & $-0.24$ & $+0.95$ & $-0.56$ \\
    \abslambda &  &  &  &  & $+1.00$ & $-0.20$ & $+0.16$ & $-0.19$ & $-0.23$ & $-0.79$ & $+0.23$ \\
    \Azero &  &  &  &  &  & $+1.00$ & $-0.87$ & $+0.02$ & $+0.06$ & $+0.17$ & $+0.02$ \\
    \Aperp &  &  &  &  &  &  & $+1.00$ & $-0.04$ & $-0.04$ & $-0.13$ & $+0.04$ \\
    \Aswav &  &  &  &  &  &  &  & $+1.00$ & $-0.63$ & $-0.03$ & $+0.60$ \\
    \dpara &  &  &  &  &  &  &  &  & $+1.00$ & $+0.43$ & $-0.49$ \\
    \dperp &  &  &  &  &  &  &  &  &  & $+1.00$ & $-0.84$ \\
    \dswpd &  &  &  &  &  &  &  &  &  &  & $+1.00$ \\
  \end{tabular}
  \label{tab:corrSyst}
\end{table*}

\begin{table*}[!ht]
  \centering
  \topcaption{Values and uncertainties of the physics parameters obtained from the combination of the CMS 8\TeV and 13\TeV results using the BLUE method. The uncertainty includes both statistical and systematic sources.}
  \begin{tabular}{l r l l l}
    Parameter & Value & Uncertainty\\
    \hline
    \phis [mrad]    & \phisResComb      & $ \pm\ \phisErrComb $       \\
    \DGs [$\ips$]   & \DGsResComb       & $ \pm\ \DGsErrComb $        \\
    \Gs [$\ips$]    & \GsResComb        & $ \pm\ \GsErrComb $         \\
    \Dms [$\hips$]  & \DmsResComb       & $ \pm\ \DmsErrComb $        \\
    \abslambda      & \abslambdaResComb & $ \pm\ \abslambdaErrComb $  \\
    \sqAzero        & \sqAzeroResComb   & $ \pm\ \sqAzeroErrComb $    \\
    \sqAperp        & \sqAperpResComb   & $ \pm\ \sqAperpErrComb $    \\
    \sqAswav        & \sqAswavResComb   & $ \pm\ \sqAswavErrComb $    \\
    \dpara [rad]    & \dparaResComb     & $ \pm\ \dparaErrComb $      \\
    \dperp [rad]    & \dperpResComb     & $ \pm\ \dperpErrComb $      \\
    \dswpd [rad]    & \dswpdResComb     & $ \pm\ \dswpdErrComb $      \\
  \end{tabular}
  \label{tab:combRes}
\end{table*}

\begin{table*}[!ht]
  \centering
  \topcaption{Matrix of the correlations between the physics parameters as obtained from the combination between the CMS 8\TeV and 13\TeV results. 
  Correlations include both statistical and systematic uncertainties.}
  \begin{tabular}{l|ccccccccccc}
   & \phis & \DGs & \Gs & \Dms & \abslambda & \Azero & \Aperp & \Aswav & \dpara & \dperp & \dswpd \\
    \hline
    \phis & $+1.00$ & $-0.06$ & $-0.06$ & $-0.17$ & $+0.11$ & $\hphantom{+}0.00$ & $+0.01$ & $-0.06$ & $+0.07$ & $-0.17$ & $+0.08$ \\
    \DGs &  & $+1.00$ & $-0.20$ & $+0.07$ & $-0.03$ & $+0.17$ & $-0.47$ & $+0.01$ & $+0.03$ & $+0.06$ & $-0.03$ \\
    \Gs &  &  & $+1.00$ & $+0.06$ & $-0.02$ & $-0.13$ & $+0.20$ & $+0.08$ & $-0.07$ & $+0.05$ & $-0.06$ \\
    \Dms &  &  &  & $+1.00$ & $-0.41$ & $+0.01$ & $-0.02$ & $+0.07$ & $+0.01$ & $+0.76$ & $-0.12$ \\
    \abslambda &  &  &  &  & $+1.00$ & $-0.12$ & $+0.08$ & $-0.05$ & $-0.24$ & $-0.52$ & $+0.08$ \\
    \Azero &  &  &  &  &  & $+1.00$ & $-0.77$ & $-0.01$ & $\hphantom{+}0.00$ & $+0.08$ & $+0.01$ \\
    \Aperp &  &  &  &  &  &  & $+1.00$ & $-0.02$ & $-0.04$ & $-0.07$ & $+0.02$ \\
    \Aswav &  &  &  &  &  &  &  & $+1.00$ & $-0.06$ & $+0.34$ & $-0.71$ \\
    \dpara &  &  &  &  &  &  &  &  & $+1.00$ & $+0.12$ & $+0.04$ \\
    \dperp &  &  &  &  &  &  &  &  &  & $+1.00$ & $-0.14$ \\
    \dswpd &  &  &  &  &  &  &  &  &  &  & $+1.00$ \\
    \end{tabular}
    \label{tab:corrComb}
\end{table*}

%% file: BPH-23-004-public-authorlist.tex
\cmsinstitute{Yerevan Physics Institute, Yerevan, Armenia}
{\tolerance=6000
A.~Hayrapetyan, A.~Tumasyan\cmsAuthorMark{1}\cmsorcid{0009-0000-0684-6742}
\par}
\cmsinstitute{Institut f\"{u}r Hochenergiephysik, Vienna, Austria}
{\tolerance=6000
W.~Adam\cmsorcid{0000-0001-9099-4341}, J.W.~Andrejkovic, L.~Benato\cmsorcid{0000-0001-5135-7489}, T.~Bergauer\cmsorcid{0000-0002-5786-0293}, S.~Chatterjee\cmsorcid{0000-0003-2660-0349}, K.~Damanakis\cmsorcid{0000-0001-5389-2872}, M.~Dragicevic\cmsorcid{0000-0003-1967-6783}, P.S.~Hussain\cmsorcid{0000-0002-4825-5278}, M.~Jeitler\cmsAuthorMark{2}\cmsorcid{0000-0002-5141-9560}, N.~Krammer\cmsorcid{0000-0002-0548-0985}, A.~Li\cmsorcid{0000-0002-4547-116X}, D.~Liko\cmsorcid{0000-0002-3380-473X}, I.~Mikulec\cmsorcid{0000-0003-0385-2746}, J.~Schieck\cmsAuthorMark{2}\cmsorcid{0000-0002-1058-8093}, R.~Sch\"{o}fbeck\cmsAuthorMark{2}\cmsorcid{0000-0002-2332-8784}, D.~Schwarz\cmsorcid{0000-0002-3821-7331}, M.~Sonawane\cmsorcid{0000-0003-0510-7010}, W.~Waltenberger\cmsorcid{0000-0002-6215-7228}, C.-E.~Wulz\cmsAuthorMark{2}\cmsorcid{0000-0001-9226-5812}
% ...
\par}
\vskip\cmsinstskip
\dag:~Deceased\\
$^{1}$Also at Yerevan State University, Yerevan, Armenia\\
$^{2}$Also at TU Wien, Vienna, Austria\\
% ...
